\documentclass[a4paper,11pt]{article}

\pdfoutput=1
\usepackage{macros}

\preprint{}
\title{Freezing and BPS jumping}

 \author{Sung-Soo Kim${}^1$, Xiaobin Li${}^2$, Satoshi Nawata${}^3$, Futoshi Yagi${}^2$}

 \affiliation[1]{School of Physics, University of Electronic Science and Technology of China, \\
 No.2006 Xiyuan Ave, West Hi-Tech Zone,
 Chengdu, Sichuan 611731, China}
 \affiliation[2]{School of Mathematics, Southwest Jiaotong University,\\
 West zone, High-tech district, Chengdu, Sichuan 611756, China}
 \affiliation[3]{Department of Physics and Center for Field Theory and Particle Physics, Fudan University,\\
 Songhu Road 2005, 200438 Shanghai, China}

\emailAdd{sungsoo.kim@uestc.edu.cn}
\emailAdd{lixiaobin@home.swjtu.edu.cn}
\emailAdd{snawata@gmail.com}
\emailAdd{futoshi\_yagi@swjtu.edu.cn}

\abstract{
We report a novel BPS jumping phenomenon of 5d $\mathcal{N}=1$ supersymmetric gauge theories whose brane configuration is equipped with an O$7$-plane. The study of the relation between O$7{}^+$-plane and O$7{}^-$-plane reveals that such BPS jumps take place when the Higgsing is triggered near the O7-plane upon a particular parameter tuning of the theories. We propose two types of gauge theories whose BPS spectra jump. One is the SU($2N+8$) gauge theory with a symmetric hypermultiplet converted to the SU($2N$) gauge theory with an antisymmetric hypermultiplet. The other is pure SO($2N+8$) gauge theory jumping to pure Sp($N$) gauge theory. We explicitly confirm our proposal through the (un)refined instanton partition functions. Furthermore, we discuss feasible generalizations involving an O$p$-plane for supersymmetric gauge theories of eight supercharges in four and three dimensions ($p=6, 5$ respectively).\\
}

\begin{document}
\maketitle
\allowdisplaybreaks
\Yboxdim5pt

\section{Introduction}\label{sec:intro}

The study of supersymmetric gauge theories in string theory has undergone significant progress with the introduction and understanding of D-branes and orientifold planes (O-planes). 
D-branes, or Dirichlet-branes, are solitonic extended objects in string theory that open strings can end \cite{Polchinski:1995mt}. Therefore, the stacking of multiple D-branes is instrumental in manifesting non-abelian gauge symmetries, which are indispensable in the formulation of gauge theories. 
Furthermore, D-branes preserve a fraction of the supersymmetry in string theory setups. Consequently, they offer a framework for the realization of supersymmetric gauge theories, enhancing our understanding of the dynamics and dualities inherent in these theories. 

In addition to D-branes, O-planes -- fixed planes where strings reflect and alter their orientation \cite{Dai:1989ua} -- play a vital role in the construction of models with specific gauge groups. They enable the realization of orthogonal and symplectic gauge groups within string theory frameworks. O-planes are also crucial for canceling the net charge and tension introduced by D-branes, ensuring the consistency and stability of the theory.

The manipulation of brane configurations with O-planes enables the exploration of incredibly rich families of supersymmetric theories. This approach provides a geometrically simple and intuitively clear perspective on these theories. Moreover, the dynamics of branes, encompassing their movement and interaction, yields profound insights into the strongly coupled dynamics of supersymmetric gauge theories. Notable phenomena like the Hanany-Witten (HW) transition, which involves brane creation and annihilation \cite{Hanany:1996ie}, and Sen's decomposition of an O7${}^-$-plane into a pair of two 7-branes \cite{Sen:1996vd}, exemplify the significance of novel brane dynamics. Utilizing the configuration and dynamics of D-branes with O-planes has repeatedly proven crucial in constructing and studying supersymmetric theories, leading to a deeper understanding of essential aspects like the moduli space of vacua, partition functions, and duality.

In this paper, we study $(p-2)$-dimensional supersymmetric gauge theories with eight supercharges constructed from D$(p-2)$-branes with an O$p^\pm$-plane. The setups give rise to not only Sp($N$)/SO($N$) gauge theories as mentioned, but also SU($N$) gauge theories with (anti-)symmetric hypermultiplet  \cite{Landsteiner:1997ei,Elitzur:1998ju,Giveon:1998sr,Bergman:2015dpa}. Our investigation focuses on the idea of \emph{freezing}: by adjusting the positions of $2^{p-4}$ D$p$-branes around an O$p^-$-plane, the effect is closely related to an O$p^+$-plane within Type II theory. Schematically, we write the freezing as
\be 
\OO p^+ ~\sim ~ \OO p^- +2^{p-4}~ \textrm{D}p\Big|_{\textrm{fixed near O$p^-$}}~.
\ee
In fact, the case of $p=7$ was recently explored in \cite{Hayashi:2023boy,Kim:2023qwh} where the effects of an O7$^+$-plane can be approximated by ``freezing'' eight D7-branes with an O7$^-$-plane (O7${}^-+$8D7) in Type IIB brane configurations.\footnote{The use of ``freezing'' in this context is inspired by the concept of a frozen singularity associated with an O7${}^+$-plane \cite{Witten:1997bs, Tachikawa:2015wka}. In this paper, ``freezing'' denotes the arrangement where $2^{p-4}$ D$p$-branes are positioned adjacent to an O$p^-$-plane, achieving an overall effect analogous to that of an O$p^+$-plane.} This process involves adjustments to the mass parameters or chemical potentials associated with the global symmetries of the D7-branes in some observables. In fact, it has been tested in the context of Seiberg-Witten curves \cite{Hayashi:2023boy} and is extendable to the level of partition functions \cite{Kim:2023qwh}.  This technique provides a useful method for calculating some partition functions for theories formulated with an O7$^+$-plane, and moreover reveals non-trivial relationships between two theories. We investigate the freezing process by examining exact partition functions, with a focus on instanton partition functions \cite{Nekrasov:2002qd}.

We must emphasize that the freezing method does not universally apply to all physical observables nor imply a duality between theories. It is primarily effective for specific observables, such as the prepotential or instanton partition functions, offering an efficient computational strategy. While it does establish identities for certain observables, there exist observables that distinguish between the two configurations. Nevertheless, the freezing method unveils curious links between seemingly distinct theories, which deserve more investigation.

Subsequently, our analysis extends to the \emph{unfreezing} process. During the freezing phase, $2^{p-4}$ D$p$-branes are utilized, but these can be removed by adjusting the positions of the ``color'' D$(p-2)$-branes and possibly adding more D$p$-branes near the O$p^-$-plane. This rearrangement is achieved through Hanany-Witten transitions \cite{Hanany:1996ie} following the Higgsing of the D$(p-2)$-branes. This process provides yet another non-trivial relationship between two different configurations with an O$p^\pm$-plane. Remarkably, we find that instanton partition functions exhibit notable discontinuities during the process of unfreezing. Since the instanton partition functions are Witten indices counting half-BPS states of instantons, this indicates a novel \emph{BPS jumping} phenomenon. This is rooted in a basic principle from elementary calculus class: integration and limit operations generally do not commute, schematically represented by
\begin{align}\label{non-comm}
    \int \lim \cZ \neq  \lim \int \cZ~.
\end{align}
The instanton partition functions are evaluated by Jeffrey-Kirwan (JK) residue integrals \cite{jeffrey1995localization}. Upon the specialization of parameters within the considered theory, these integrals encounter degenerate poles \cite{brion1999arrangement,szenes2003toric,Benini:2013xpa}. As a result, the accurate computation of residues, following this parameter specialization, differs from a naive parameter substitution in the generic integrated expressions.  This gives rise to the novel BPS jumping phenomenon. 
It is noteworthy that this phenomenon is different from wall-crossing phenomena \cite{Seiberg:1994rs} as it occurs at special points in the parameter (moduli) space of the theory where new degenerate poles emerge in JK residue integrals. This phenomenon is indeed ubiquitous when we tune chemical potentials or fugacities in JK residue integrals. Thus, this phenomenon offers a new perspective for understanding and analyzing not only instanton partition functions but also a wider range of partition functions involving JK residue methods.

In our detailed analysis, we aim to elucidate how the unfreezing processes relate two instanton partition functions with multiplicity coefficients, as recently found in  \cite{Nawata:2021dlk,Chen:2023smd}. In the context of 5-brane web configurations, D1-branes generate instantons in 5d theories so that this implies that D1-branes contribute distinct non-perturbative effects at certain configurations specifically associated with unfreezing. In other words, the configuration of branes during unfreezing leads to the creation/annihilation of unique instanton states, deviating from those observed at generic points in the parameter space. Moreover, the Witten index on the moduli spaces of instantons undergoes notable changes at particular chemical potential values, highlighting the novel dynamical nature across the parameter space.

The paper is outlined as follows. Central to our study, section \ref{sec:5d} delves into the freezing and unfreezing process involving O7$^\pm$-plane in the 5-brane webs. As a warm-up, we review 5-brane web realization of 5d $\cN=1$ supersymmetric gauge theories in Type IIB theory, and we proceed to examine the effects of the freezing and unfreezing by the cubic prepotentials. Our investigation then extends to the investigation of the (un)freezing process in terms of 5d instanton partition functions. Concretely, using the ADHM descriptions of instanton moduli spaces, we explicitly show non-trivial identities among instanton partition functions as direct outcomes of freezing and unfreezing. As an illuminating example of the identities, we deal with the relation of partition functions of E-string and M-string theory. It also elucidates the method for calculating E-string and M-string partition functions using 5d instanton partition functions. Finally, we explore the novel BPS jumping phenomenon using unrefined instanton partition functions during the unfreezing process. 

Subsequently, we study the freezing and unfreezing processes in 4d and 3d theories with eight supercharges. As emphasized before, the freezing process does not lead to the identities of all the physical observables. 
In the realm of 4d $\cN=2$ superconformal field theories, the superconformal index serves to differentiate between two theories related by the (un)freezing process with O6$^\pm$-plane. However, this process does reveal non-trivial connections among Schur indices. Similarly, in 3d $\cN=4$ theories, both the superconformal and twisted indices can distinguish theories related by the (un)freezing process involving an O5$^\pm$-plane, yet they also show noteworthy identities between 3-sphere partition functions. These aspects are examined in section \ref{sec:4d3d}.

Several appendices supplement the main text. Appendix \ref{app:notations} fixes the notations and conventions employed throughout the paper. Appendix \ref{app:ADHM} details the contributions to ADHM integrals from various gauge and matter multiplets for instanton partition functions. Appendix \ref{app:JK} delves into the detailed computations of Jeffrey-Kirwan residues for instanton partition functions, with a particular emphasis on handling degenerate poles. Through detailed calculations of JK residues, we clarify the occurrence of BPS jumping due to the appearance of degenerate poles during the unfreezing process.

%%%%%%%%%%%%%%%%%%%%%%%%%%%%%%%%%%%%%%%%%%%
%%%%%%%%%%%%%%%%%%%%%%%%%%%%%%%%%%%%%%%%%%%
\section{Freezing, unfreezing,  and BPS jumping in 5d theories}\label{sec:5d}

\subsection{5d cubic prepotential and freezing}\label{sec:cubic}
To start, we review the freezing at the level of the cubic 
prepotential, discussed in \cite{Hayashi:2023boy}, and propose yet another novel relation between theories involving an O7-plane in their brane configurations.

The cubic prepotential $\mathcal{F}$ governing Coulomb branch of 5d supersymmetric field theory of gauge group $G$ is given by \cite{Seiberg:1996bd,Morrison:1996xf,Intriligator:1997pq}
\begin{align}\label{eq:prepot}
\mathcal{F}(\phi) = \frac{1}{2g_0^2}h_{ij}\phi_i\phi_j + \frac{\kappa}{6}d_{ijk}\phi_i\phi_j\phi_k + \frac{1}{12}\bigg(\sum_{r\in\Delta}\left|r\cdot \phi\right|^3 - \sum_f\sum_{w \in R_f}\left|w\cdot \phi +m_f\right|^3\bigg ),  
\end{align}
where $g_0$ is the gauge coupling, $h_{ij} = \mathrm{Tr}(T_iT_j)$ with $T_i$ being the Cartan generators of Lie algebra $\mathfrak{g}$ associated with $G$, and $\phi$ is the vacuum expectation value of the scalar field in the vector multiplet. The one-loop contribution of the cubic prepotential is given in the last term where $\Delta$ is the root system of Lie algebra $\mathfrak{g}$, $w$ is the weight of a representation $R_f$, and $m_f$ are mass parameters for hypermultiplet in $R_f$. We note that the term with the coefficient $\frac{\kappa}{6}$ in \eqref{eq:prepot} only exists for gauge group $G=\mathrm{SU}(N\ge3)$ where $\kappa$ is the Chern-Simons (CS) level and $d_{ijk} = \frac12\mathrm{Tr}(T_i\{T_j,T_k\})$. If the cubic Casimir of the matter representation is odd, there is a parity anomaly that can be canceled by a CS term with a half-odd-integer level $\kappa$. In particular, in the case of $\SU(2N+1)$ gauge theory with (anti-)symmetric matter, a parity anomaly is present, requiring the introduction of a CS term with a half-odd-integer level. 

We also remark that the 5d $\Sp(N)$ gauge theory has a discrete $\theta$-angle due to the fact that $\pi_4(\Sp(N))=\mathbb{Z}_2$ \cite{Morrison:1996xf,Douglas:1996xp,Bergman:2013ala}. The $\theta$-angle crucially distinguishes the spectrum of the theory at the instanton level. In other words, $\Sp(N)_{\theta=0}$ gauge theory and $\Sp(N)_{\theta=\pi}$ gauge theory have distinct instanton partition functions, which will be discussed in detail in the following sections. On the other hand, as the cubic prepotential  \eqref{eq:prepot}  is one-loop exact, the prepotential is insensitive to the discrete $\theta$ angle\footnote{We note that although the cubic prepotential does not distinguish the $\theta$ angle, the complete prepotential~\cite{Hayashi:2019jvx} does distinguish the discrete $\theta$ angle and moreover it captures enhanced flavor symmetry.}. For this reason, we do not introduce the discrete $\theta$-angle when writing a prepotential $\mathcal{F}(\phi)$. 
Moreover, for Sp($N$) gauge theory with fundamental hypermultiplets, the discrete $\theta$-angle difference can be understood from the sign difference of the fundamental hypermultiplet mass and so we omit the discrete $\theta$ angle in the presence of fundamental hypermultiplet for Sp($N$) gauge theory.\footnote{As $\Sp(1)=\SU(2)$, pure $\SU(2)$ theory also has the $\theta$-angle. It was also discussed in \cite{Jia:2022dra} that the discrete $\theta$-angle for the SU(2)$_\theta$ theory can be understood as the Chern-Simons level $\kappa$ of SU($N$)$_\kappa$ for $N=2$.}

For convenience, we express the cubic prepotential in terms of the Coulomb branch parameters $a_i$, by setting all the masses of hypermultiplets to zero. For $G=\textrm{SU}(N)$, in the Weyl chamber  $a_1\ge a_2 \ge \cdots \ge a_{N-1}\ge0$ and $a_N=-\sum_{i=1}^{N-1}a_i$, the cubic prepotential for $\mathrm{SU}(N)$ gauge theory at the CS level $\kappa$ with hypermultiplets in $N_f$ fundamental representations, $N_{a}$ antisymmetric representations, and $N_{s}$ symmetric representations (for short, $\mathrm{SU}(N)_\kappa+N_f\mathbf{F}+N_{a}\AS+N_{s}\Sym$) takes the form
\begin{equation}\label{eq:Pre for SU}
\begin{aligned}
  \mathcal{F}_{\mathrm{SU}(N)_\kappa+N_f\mathbf{F}+N_{a}\AS+N_{s}\Sym}
  =&~ \frac{1}{2g_0^2}\sum_{i=1}^Na_i^2 +\frac{\kappa}{6}\sum_{i=1}^N a_i^3+  \frac{1}{6}\sum_{i<j}^N (a_i-a_j)^3\\
 & -\frac{N_f}{12}\sum_{i=1}^N |a_i|^3-\frac{N_{a}}{12}\sum_{i<j}^N |a_i+a_j|^3\cr
 &- \frac{N_{s}}{12}\Big(\sum_{i=1}^N |2a_i|^3+\sum_{i<j}^N |a_i+a_j|^3\Big)\ .
\end{aligned}
\end{equation}
For $G=\textrm{Sp}(N)$,
in the Weyl chamber $a_1\ge a_2\ge \cdots\ge a_N\ge0$, the cubic prepotential takes the form 
\begin{align}\label{eq:FforSpN}
  \mathcal{F}_{\mathrm{Sp}(N)+N_f\mathbf{F}} = \frac{1}{g_0^2}\sum_{i=1}^Na_i^2 +  \frac{1}{6}\bigg(\!\sum_{i<j}^N \big((a_i\!-\!a_j)^3
  +(a_i\!+\!a_j)^3\big)
  +(8\!-\!N_f)\!\sum_{i=1}^N a_i^3\bigg)\ .  
\end{align}
For $G=\textrm{SO}(2N)$,
in the Weyl chamber $a_1\ge a_2\ge \cdots\ge a_N\ge0$, the cubic prepotential takes the form
\begin{align}\label{eq:FforSO2N}
  \mathcal{F}_{\mathrm{SO}(2N)+N_f\mathbf{F}} = \frac{1}{g_0^2}\sum_{i=1}^Na_i^2 +  \frac{1}{6}\bigg(\sum_{i<j}^N\big( (a_i-a_j)^3+(a_i+a_j)^3\big) -N_f\sum_{i=1}^N a_i^3\bigg)\ .   
 \end{align}

We note that it is well-known that $\SU(2N)$ gauge theory with an antisymmetric hypermultiplet or a symmetric hypermultiplet has a Higgs branch to $\Sp(N)$ or $\SO(2N)$, respectively, 
\begin{align} \label{eq:HiggsBranch}
    \mathrm{SU}(2N)_\kappa+ 1\AS~&\xrightarrow{\text{Higgsing}}~ \mathrm{Sp}(N)\ , \cr
    \mathrm{SU}(2N)_\kappa+ 1\Sym~&\xrightarrow{\text{Higgsing}}~ \mathrm{SO}(2N)\ , 
\end{align}
which can be seen from the cubic prepotential by setting $a_{2N+1-i}=-a_{i}$ as well as masses of the hypermultiplets to zero, $m_{\bf AS/ Sym}=0$. 
\begin{figure}[t]
    \centering
    \includegraphics[width=15cm]{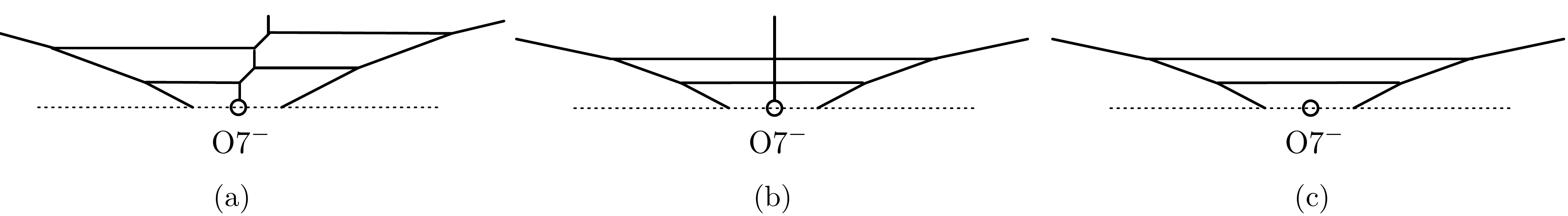}
    \caption{A Higgsing from SU($2N)+1\AS$ to Sp($N$). As an example, we choose $N=2$. (a) a 5-brane web for SU($4)+1\AS$. (b) The Coulomb branch parameters and the mass of an antisymmetric hypermultiplet are tuned so that the middle NS5-brane can be Higgsed. (c) The resulting 5-brane configuration after Higgsing the middle NS5-brane, which is a 5-brane web for Sp($2$).  }
    \label{fig-SUtoSp}
\end{figure}
\begin{figure}[t]
    \centering
    \includegraphics[width=14.cm]{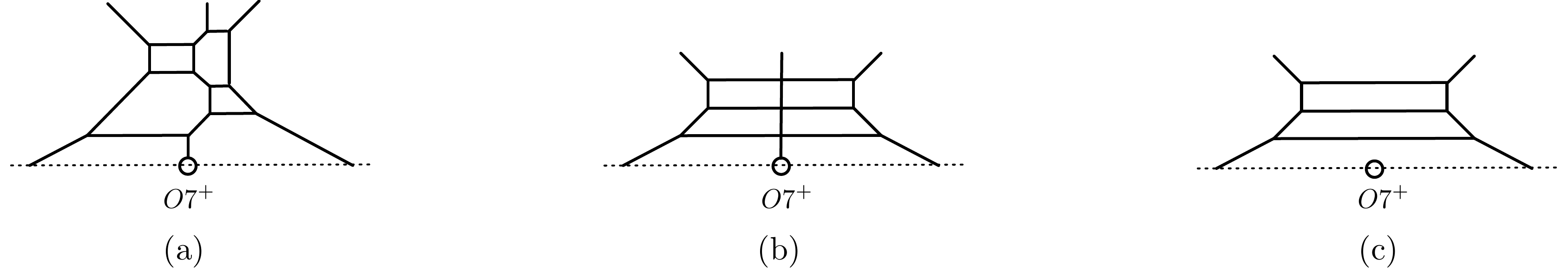}
    \caption{A Higgsing from SU($2N)+1\Sym$ to SO($2N$). As an example, we choose $N=3$. (a) a 5-brane web for SU($6)+1\Sym$. (b) The Coulomb branch parameters and the mass of a symmetric hypermultiplet are tuned so that the middle NS5-brane can be Higgsed. (c) The resulting 5-brane configuration after Higgsing the middle NS5-brane, which is a 5-brane web for SO($6$).  }
    \label{fig:SU+SymToSO}
\end{figure}
This Higgsing can be readily seen from 5-brane webs in Type IIB theory. Our convention for 5-brane webs is as follows: D5- and NS5-branes are extended in the $x^{0,1,2,3,4,6}$-directions and the $x^{0,1,2,3,4,5}$-directions, respectively so that their 5-brane configurations are constructed in the $x^{5,6}$-plane, called the $(p,q)$-plane. The fundamental hypermultiplets can be described by D7-branes which are extended in the $x^{0,1,2,3,4,7,8,9}$-directions so that a 7-brane in the $(p,q)$-plane appears as a dot on which 5-branes end. See \cite{Aharony:1997bh,Hayashi:2018lyv} for more details and examples.

Along with this convention, 5d $\SU(N)$ gauge theory with one (anti-)symmetric hypermultiplet is constructed by introducing an O7-plane on 5-brane web where a half NS5-brane is stuck \cite{Bergman:2015dpa}. An O7$^-$-plane gives rise to an antisymmetric hypermultiplet (e.g., see figure \ref{fig-SUtoSp}), while an O7$^+$-plane gives rise to a symmetric hypermultiplet (e.g., see figure \ref{fig:SU+SymToSO}). The asymptotic horizontal distance of this middle NS5-brane from the position of an O7-plane is given by $\frac{N}{2} m_{\bf AS/Sym}$, a half of the mass of (anti-)symmetric hypermultiplet multiplied by the number of color branes \cite{Hayashi:2023boy}. As an illustrative example, in figure \ref{fig:SU+SymToSO}(a), we present a 5-brane web for $\SU(6)+1\mathbf{Sym}$, where the fundamental region is given. One can imagine the covering space such that the projected image due to an O7$^+$-plane is included below the monodromy cut of an O7$^+$-plane (the dotted line in figure \ref{fig:SU+SymToSO}). 
To perform the Higgsing \eqref{eq:HiggsBranch}, the color D5-branes are aligned to realize the above Higgsing condition so that the middle NS5-brane is aligned, which is depicted in figure \ref{fig:SU+SymToSO}(b). The Higgsing takes place as follows. The parallel color D5-branes on the left and the right are reconnected and the middle half N5-brane brane is also connected through its reflected image so that the NS5-brane is taken away along the transverse directions where an O7-plane is extended. The resulting web is given in figure \ref{fig:SU+SymToSO}(c), which is a 5-brane web for pure $\SO(6)$, as expected.

\begin{figure}[t]
    \centering
    \includegraphics[width=15.1cm]{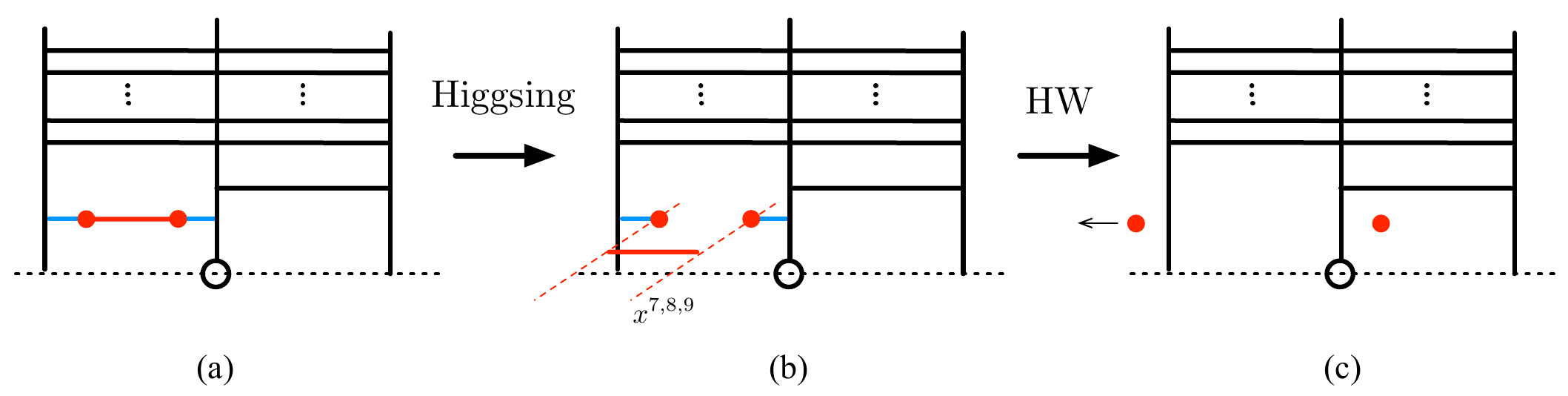}
    \caption{A Higgsing with a symmetric or antisymmetric hypermultiplet:
    $\SU(n)+1\AS/\Sym+N_f\mathbf{F} \to \SU(n-1)+1\AS/\Sym+(N_f-1)\mathbf{F}  .$
    }
    \label{fig:O7-Higgsing}
\end{figure}
\begin{figure}[t]
    \centering
    \includegraphics[width=15.1cm]{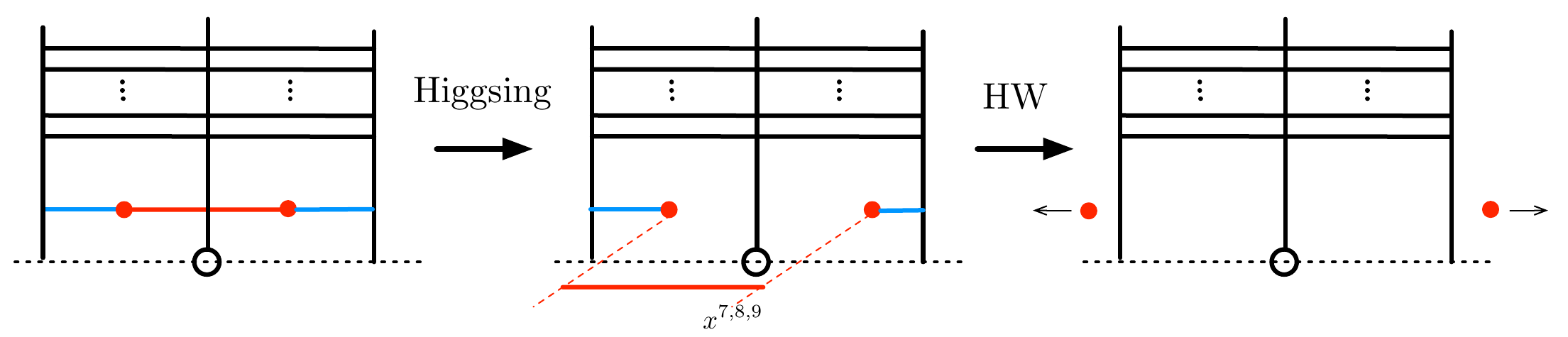}
    \caption{A Higgsing with a symmetric or an antisymmetric hypermultiplet reducing the Coulomb branch dimension by two: $\SU(n)+1\AS/\Sym+N_f\mathbf{F} \to\SU(n-2)+1\AS/\Sym+(N_f-2)\mathbf{F}.$}
    \label{fig:O7-Higgsing-rank2}
\end{figure}
When there are additional $N_f$ fundamental hypermultiplets, 
there is another Higgs branch associated with fundamental hypermultiplets, which reduces the Coulomb branch dimension and the number of fundamental hypermultiplets by one:
\begin{align}\label{eq:ASym-Higgsing}
\textrm{SU}(n)+1\AS/\Sym+N_f\mathbf{F} ~\to~ \SU(n-1)+1\AS/\Sym+(N_f-1)\mathbf{F}  \ .
\end{align}
The corresponding 5-brane configuration\footnote{The Higgsing $\textrm{SU}(n)+1\AS+1\mathbf{F} \to \SU(n-1)+1\AS$ is also possible with a special mass $m_f=0$. In the brane web, the Higgsed D5-brane is a half-D5 color brane that is connected to two half-D7-branes.} is depicted in figure \ref{fig:O7-Higgsing}, where the brane charges are neglected to highlight the relevant Higgsing procedures. In figure \ref{fig:O7-Higgsing}(a), two (flavor) D7-branes, appearing red dots, are aligned on a color D5-brane in a way that recombination can take place such that the D5-brane in red ends on two D7-branes, while the D5-branes in blue are connected to a D7-brane and a 5-brane with NS5-charge in the middle. Once this recombination happens, as illustrated in figure \ref{fig:O7-Higgsing}(b), the D5-brane in red is Higgsed away along the transverse direction to 5-brane, which is denoted by the $x^{7,8,9}$-directions. Finally, one can take the Hanany-Witten (HW) transition so that a D7-brane on the left of figure \ref{fig:O7-Higgsing}(c) is decoupled, becoming a free hypermultiplet, while a D7-brane on the right contributes to a fundamental hypermultiplet. As a result, this Higgsing described through the 5-brane web reduces the rank of the Coulomb branch and the flavor group by one, respectively. It is worth noting that this Higgsing takes out one fundamental hypermultiplet, to realize the Higgsing, one needs to tune masses of two fundamental hypermultiplets.

We note that the 5-brane configuration suggests another Higgsing process:
\begin{align}\label{eq:Higgsingwith2Fonly}
\textrm{SU}(n)_\kappa +1\AS/\Sym+N_f\mathbf{F} ~\rightarrow~ \SU(n-2)_\kappa+1\AS/\Sym+(N_f-2)\mathbf{F} \ .
\end{align}
This Higgsing, while physically akin to the one described in \eqref{eq:ASym-Higgsing}, differs in its execution. It could be viewed as applying the Higgsing process in \eqref{eq:ASym-Higgsing} twice. However, as illustrated in figure \ref{fig:O7-Higgsing-rank2}, this particular Higgsing involves the adjustment of only two Coulomb branch and two mass parameters, leaving other mass parameters and the CS level untouched. It is, hence, more convenient when we perform Higgsings successively, in particular, at the level of partition functions.  
It follows that along the Higgsings, one finds that the prepotential for SU($N+8)+1\AS+8\mathbf{F}$ becomes identical to that for SU($N)+1\AS$ by setting the Coulomb branch parameters and the hypermultiplet mass parameters as 
\begin{align}\label{eq:SUN+8to SUN}
    a_{N+1}=\cdots=a_{N+8}=m_i=0\ .
\end{align}
Here we note that as the hypermultiplet masses of the prepotential \eqref{eq:Pre for SU} were already set to zero for simplicity, the prepotential check for this Higgsing is done with zero masses. 
We shall see in the next subsection that the precise relation does not require them to zero, rather a generic value satisfying the condition \eqref{eq:SUN+8to SUN} is enough. 
\vspace{.5cm}

With these explicit forms of the prepotentials for the theories involving an O7-plane in their brane configurations, one can see the following intriguing relations among these theories.

\begin{figure}[t]
    \centering
    \includegraphics[width=10cm]{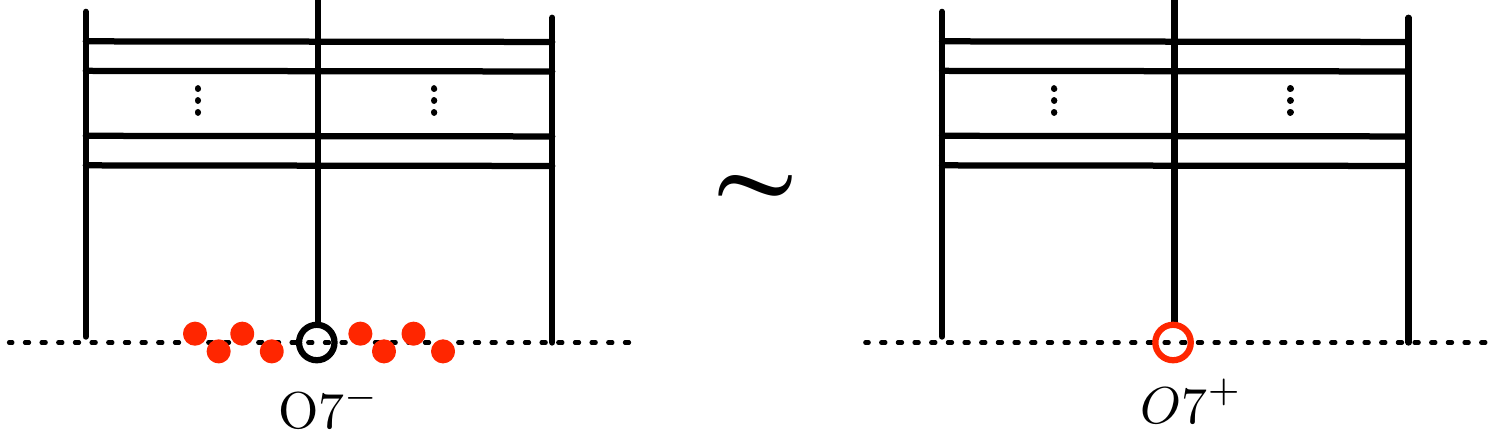}
    \caption{Freezing: O7$^-+8$D7$|_{\textrm{fixed}}\sim\ $O7$^+$. Here, D7-branes are denoted by red dots and the charges of the 5-brane are neglected to highlight the relation between an O7$^{\pm}$-plane with 8 D7-planes.}
    \label{fig:O7--to+}
\end{figure}

\paragraph{Freezing.} The cubic prepotential of SU($N$) gauge theory with an antisymmetric and eight fundamentals ($1\AS+8\mathbf{F}$) given in \eqref{eq:Pre for SU} can be reduced to that with a symmetric ($1\Sym$), as proposed in \cite{Hayashi:2023boy}:
\begin{align}\label{eq:SUNAS to SUNSym}
\mathcal{F}_{\mathrm{SU}(N)_\kappa+1\AS+(8+N_f)\mathbf{F}}
= \mathcal{F}_{\mathrm{SU}(N)_\kappa+1\Sym+N_f\mathbf{F}}\ . 
\end{align}
Though for simplicity, we set the hypermultiplet masses to zero, but this relation is generic. With $m_\mathbf{AS} =m_\mathbf{Sym}=m$ for (anti-)symmetric hypermultiplets, $m_\mathbf{F}=\frac{m}{2}$ for the eight fundamental hypermultiplets, one can easily check \eqref{eq:SUNAS to SUNSym}. We also remark that as in \cite{Hayashi:2023boy,Kim:2023qwh}, there exists a set of mass parameters giving rise to the relation even at the level of the (refined) instanton partition function which will be discussed  in the next subsection
\begin{equation}\label{eq:ASFToSym}
 \mathrm{SU}(N)_\kappa+1\AS+8\mathbf{F}~\longrightarrow ~\mathrm{SU}(N)_\kappa+1\Sym\ .
\end{equation}
We refer to this reduction procedure as the \emph{freezing} \cite{Hayashi:2023boy}.

From the perspective of Type IIB 5-brane webs, the freezing can be understood as follows.
As explained, \textrm{SU}$(N)+1\AS+8\mathbf{F}$ is constructed with eight D7-branes (flavors) and an O7$^-$-plane at which a half NS5-brane is stuck, while $\textrm{SU}(N)+1\Sym$ is constructed by a 5-brane web with an O7$^+$-plane where a half NS5-brane is stuck. 
To realize the freezing, eight D7-branes are put at the position of the O7$^-$-plane such that they are bound together to make a combination of O7$^-+8$D7 which leads to the same charge and monodromy as an O7$^+$-plane, 
\begin{align}\label{freezing}
    \textrm{O7}^-+8\,\textrm{D7}\Big|_{\textrm{fixed}} ~\sim ~ \textrm{O7}^+. 
\end{align}
 In doing so, one finds that the prepotential for $\textrm{SU}(N)+1\AS+8\mathbf{F}$ becomes identical to that for $\textrm{SU}(N)+1\Sym$. The corresponding 5-brane configuration is depicted in figure~\ref{fig:O7--to+}.

 The concept of freezing involves fixing the positions of eight D7-branes close to an O7$^-$-plane, producing an effect analogous to that of an O7$^+$-plane as if eight D7-branes and an O7$^-$-plane are ``frozen'' to form an O7$^+$-plane. 
For SU type gauge theories, the freezing is a procedure that tunes the masses of $1\AS+8\mathbf{F}$ to convert the set of hypermultiplets into $1\Sym$ while the Coulomb branch parameters are untouched.
\begin{figure}[t]
    \centering
    \includegraphics[width=15.1cm]{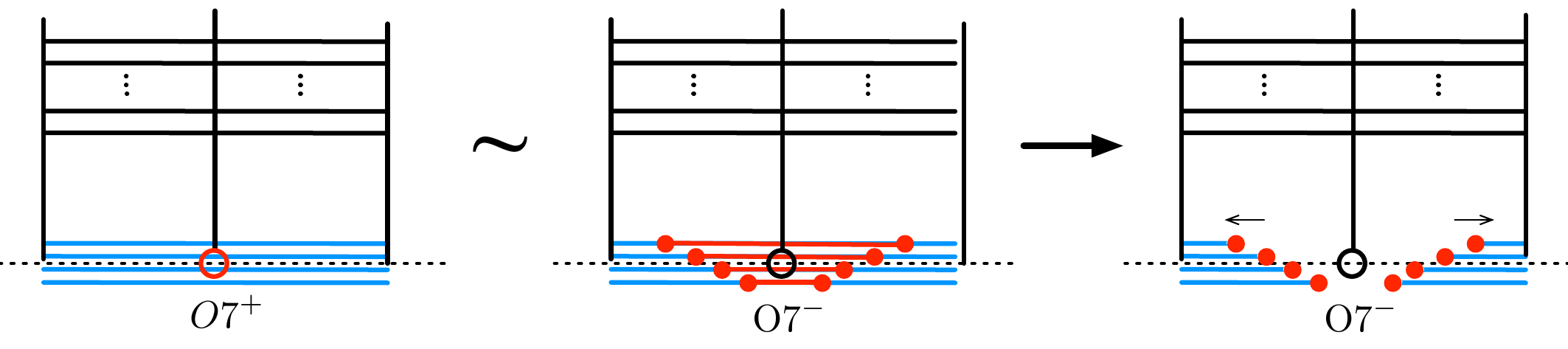}
    \caption{Unfreezing. $\mathsf{LEFT}$: A simplified 5-brane web for SU($2N+8)+1\Sym$. Here, one brings four D5-branes (in blue) to the position of an O7$^+$-plane and then one ``unfreezes'' an O7$^+$-plane so that it can be effectively regarded as an O7$^-$ and 8 D7s such that D7-branes are on the blue D5-branes. $\mathsf{MIDDLE}$: At the same time, four simultaneous Higgsing take place so that 8 D7-branes and 4 D5-branes in blue on top of O7$^-$-plane are recombined such that D5-branes in red are suspended between two D7-branes. $\mathsf{RIGHT}$: The D5-branes in red are Higgsed away along $x^{7,8,9}$-directions and the remaining D7-branes attached to D5-branes in blue are taken to infinity so that the Hanany-Witten transition makes them free hypermultiplets. As a result, one obtains a 5-brane web for SU($2N)+1\AS$.}
    \label{fig:O7-unfreezing}
\end{figure}

\paragraph{Unfreezing.} 
We now imagine a reverse procedure of the freezing. Namely, we convert $1\Sym$ into $1\AS+8\mathbf{F}$, but we also simultaneously reduce the dimension of the Coulomb branch by the Higgsing discussed in \eqref{eq:Higgsingwith2Fonly}:
\begin{align}\label{eq:SynToAS}
 \mathrm{SU}(N+8)_\kappa+1\Sym&~\longrightarrow ~\mathrm{SU}(N)_\kappa+1\AS\ .
\end{align}
By setting all the unfrozen mass parameters to zero and at the same time performing four successive Higgsings with $a_{2N+1}=\cdots =a_{2N+8}= 0$, 
one can see that the cubic prepotential yields 
\begin{align}\label{eq:SynToAS-prepot}
 \mathcal{F}_{\mathrm{SU}(N+8)_\kappa+1\Sym}&~\longrightarrow ~\mathcal{F}_{\mathrm{SU}(N)_\kappa+1\AS}\ .
\end{align}
This procedure can be viewed as ``\emph{unfreezing}'': we regard an O7$^+$-plane effectively as O7$^-$ paired with 8 D7s at fixed positions, which is at the same time followed by four consecutive Higgsings, each removing a configuration involving a color D5-brane connected to two D7-branes. Subsequent Hanany-Witten transitions \cite{Hanany:1996ie} then displace the eight D7-branes from the vicinity of the O7$^-$-plane along its monodromy cut, which makes eight D7s free hypermultiplets (or floating D7-branes). These procedures, depicted in figure \ref{fig:O7-unfreezing}, should take place at the same time.

\begin{figure}[t]
    \centering
    \begin{tikzpicture}
\node[rounded rectangle,draw] at (0,0) {$\mathcal{F}_{\mathrm{SU}(N+8)_\kappa+1\AS+8\mathbf{F}}$};
\node[rounded rectangle,draw] at (2,-5) {$\mathcal{F}_{\mathrm{SU}(N)_\kappa+1\AS}$};
\node[rounded rectangle,draw] at (-2,-2.5) {$\mathcal{F}_{\mathrm{SU}(N+8)_\kappa+1\Sym}$};
\draw[->] (-1,-.5) to node [left,scale=0.8] {Freezing~
} (-2,-2);
\draw[->] (1,-.5) to node [right,scale=0.8] { Successive Higgsings} (2,-4.5);
\draw[->] (-1.5,-3) to node [left=0.25,scale=0.8] {Unfreezing
} (1.5,-4.5);
\end{tikzpicture}
    \caption{Freezing, unfreezing, and Higgsings for SU gauge theories.}
    \label{fig:SU_triangle_prepot}
\end{figure}
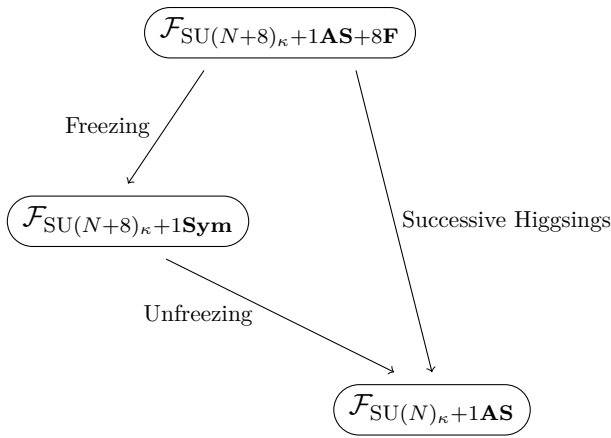

We note that having discussed the freezings and unfreezings, one finds an intriguing relation between theories involving an O7-plane, which can be summarized as in figure \ref{fig:SU_triangle_prepot}.

We also remark that a similar relation arises for Sp/SO theories, which follows from the Higgsings of antisymmetric or symmetric hypermultiplet given in \eqref{eq:HiggsBranch}. More explicitly, one can see the relations as follows. Starting from the cubic prepotential for Sp($N+4$) theory with eight fundamental hypermultiplets. We freeze the theory by setting the masses of eight flavors to zero, 
\begin{align} \label{eq:massfreezingSptoSO}
 m_{N+1} =m_{N+2}=\cdots=m_{N+8}= 0\ ,   
\end{align}
which leads to the form of the prepotential given in \eqref{eq:FforSpN}, which is nothing but the prepotential for pure SO($2N+8$) theory: 
\begin{align}\label{eq:FreezeSptoSO}
\mathcal{F}_{\mathrm{Sp}(N+4)+8\mathbf{F}}\xrightarrow[\eqref{eq:massfreezingSptoSO}]{\textrm{freezing}}~&\frac{1}{g_0^2}\sum_{i=1}^{N+4}a_i^2 +  \frac{1}{6}\sum_{i<j}^{N+4}\big( (a_i-a_j)^3+(a_i+a_j)^3\big)
=\mathcal{F}_{\mathrm{SO}(2N+8)}\ .   
 \end{align}

We then do the unfreezing by setting $a_{N+j}=0$ ($j=1,2,3,4$) which leads to the cubic prepotential for pure Sp($N$) theory:
\begin{align}\label{eq:UnfreezeSOtoSp}
\mathcal{F}_{\mathrm{SO}(2N+8)}\xrightarrow{\textrm{unfreezing}}&~\frac{1}{g_0^2}\sum_{i=1}^{N}a_i^2 +  \frac{1}{6}\sum_{i<j}^{N}\!\big( (a_i\!-\!a_j)^3+(a_i\!+\!a_j)^3\big) \!+\! \frac{8}{6} \sum_{i=1}^{N}a_i^3
=\mathcal{F}_{\mathrm{Sp}(N)}\ .   
 \end{align}

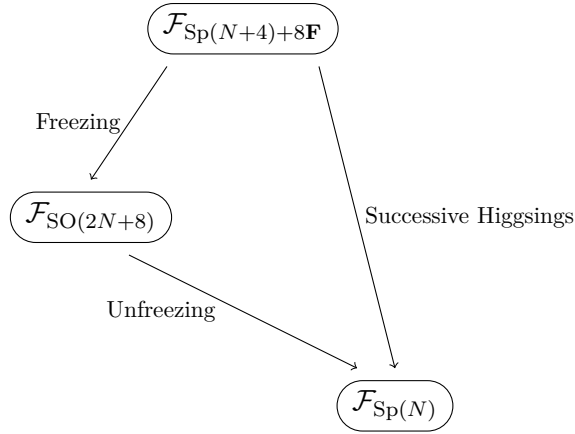
\begin{figure}[H]
    \centering
\begin{tikzpicture}
\node[rounded rectangle,draw] at (0,0) {$\mathcal{F}_{\mathrm{Sp}({N+4})+8\mathbf{F}}$};
\node[rounded rectangle,draw] at (2,-5) {$\mathcal{F}_{\mathrm{Sp}({N})}$ };
\node[rounded rectangle,draw] at (-2,-2.5) {$\mathcal{F}_{\mathrm{SO}({2N+8})}$};
\draw[->] (-1,-.5) to node [left, scale=0.8] {Freezing~} (-2,-2);
\draw[->] (1,-.5) to node [right,scale=0.8] {Successive Higgsings
} (2,-4.5);
\draw[->] (-1.5,-3) to node [left=0.25, scale=0.8] {Unfreezing} (1.5,-4.5);
\end{tikzpicture}
    \caption{Freezing, unfreezing, and Higgsings for Sp/SO gauge theories.}
    \label{fig:Sp/So_triangle_prepot}
\end{figure}

Together with a successive application of conventional Higgsing, $\Sp(N+4)+8F\to\Sp(N)$, one can summarize the relation between $\Sp/\SO$ gauge theories as depicted in figure \ref{fig:Sp/So_triangle_prepot}. 

Because these various intriguing relations between theories involving an O7$^\pm$-plane are presented based on the cubic prepotentials, one is tempted to speculate whether the same relation holds even at the non-perturbative level. It is suggestive that though the cubic prepotential only captures perturbative aspects, the existence of the corresponding brane configurations would imply that a similar relation can be extended to the non-perturbative level. We now consider 5d instanton partition functions on the $\Omega$-background and generalize the freezing with refined $\Omega$-deformation parameters, $\epsilon_{1,2}$. 

%%%%%%%%%%%%%%%%%%%%%%%%%%%

\subsection{Identities of 5d instanton partition functions}\label{sec:instanton}

In 5d supersymmetric theories, one of the most important observables is an instanton partition function \cite{Nekrasov:2002qd} on $S^1\times \mathbb{R}^4$ with the $\Omega$-background. It is defined as an index that counts BPS states within a 5d theory:
\be\label{inst-PF}
Z_{\textrm{inst}}\left(\epsilon_1, \epsilon_2, a_i, m\right)=\operatorname{Tr}\left[(-1)^\mathsf{F} e^{-\beta\left\{Q, Q^{\dagger}\right\}} e^{-\epsilon_1\left(J_1+J_R\right)} e^{-\epsilon_2\left(J_2+J_R\right)} e^{-a_i C_i} e^{-m \cdot F}\right] \ .
\ee
Here, $J_1, J_2$ are defined by the Cartans $J_l, J_r$ of the spacetime symmetry $\SO(4)\cong \SU(2)_l \times \SU(2)_r$ of $\mathbb{R}^4$ via $J_1=\frac{J_r+J_l}{2}, J_2=\frac{J_r-J_l}{2}$. The operator $J_R$ represents the Cartan generator of the $\SU(2)_R$ R-symmetry.
The Coulomb branch parameters are denoted as $a_i$, where the index $i$ ranges from 1 to the rank of the gauge group $G$. All additional flavor symmetries are collectively labeled as $F$, which are conjugate to the mass parameters $m$.

Instantons, representing non-perturbative effects in gauge theories,
play a crucial role in understanding the behavior of gauge theories in strong coupling regimes, where traditional perturbative methods fail. 
The method employed by Nekrasov \cite{Nekrasov:2002qd} involved localization techniques, which allowed the path integral of the gauge theory to be exactly computed by localizing it to the moduli space of instantons. Consequently, the instanton partition function can be understood as a character of the equivariant action 
\be 
\bC_{\e_1}^\times\times\bC_{\e_2}^\times\times \prod_i^{\rk G} \bC_{a_i}^\times \times \prod_j^{\rk F} \bC_{m_j}^\times~.
\ee 
on the instanton moduli spaces. A more physical perspective interprets it that a 5d theory on the $\Omega$-background effectively localizes to supersymmetric quantum mechanics on the instanton moduli spaces, and \eqref{inst-PF} is its partition function. As exact results, encoding all non-perturbative effects in a gauge theory, we make use of the instanton partition functions to see the consequence of freezing and unfreezing. 

First, let us recall the Higgsing of $\SU(N)_\kappa+1\Sym$ to $\SO(N)$ at the level of instanton partition functions \cite{Chen:2023smd}. Turning on the $\Omega$-background, a brane may change its position by a certain linear combination of $\e_{1,2}$ from the absence of the background. As in figure  \ref{fig:SU+SymToSO}, to Higgs $\SU(N)_\kappa+1\Sym$ to $\SO(N)$, all the D5-branes are aligned and the middle NS5-brane is positioned at $\e_+/2$ so that we tune the parameters of the theory 
\be\label{SU-SO-Higgsing}
m=\e_+~, \qquad a_{N-i+1}=-a_i~,
\ee
where $m$ is the mass of the symmetric hypermultiplet. Note that when $N$ is odd, we set $a_{\frac{N+1}{2}}=0$. At this specialization, the instanton partition function behaves
\be\label{SU-SO}
Z_{2k,\kappa=\frac12 (N \bmod 2)}^{\SU(N)+1\Sym} \ \xrightarrow[\eqref{SU-SO-Higgsing}]{\textrm{Higgsing}} \  (-1)^{k (N+ 1)}Z_{k}^{\SO(N)}~,
\ee
for $N\ge4$.
It is important to note that $2k$-instanton on the left-hand side coincides with $k$-instanton on the right-hand side, and the odd-instanton partition functions of $\SU(N)_\kappa+1\Sym$ vanish at the specialization \eqref{SU-SO-Higgsing}.

We can apply a similar procedure to the brane configuration involving an O7${}^-$-plane ($\SU(N)_\kappa+1\AS$) to obtain the brane configuration for pure $\Sp(N)$ gauge theory, as given in figure \ref{fig-SUtoSp}. This involves the specialization 
\be\label{SU-Sp-Higgsing}
m=\e_+~, \qquad  a_{2N-i+1}=-a_i~,
\ee
manipulating the instanton partition functions, which yields the following identities:
\begin{equation}\label{SU-Sp}
\begin{aligned}
Z_{k,\kappa\equiv N\bmod 2}^{\SU(2N)+1\AS} \ \xrightarrow[\eqref{SU-Sp-Higgsing}]{\textrm{Higgsing}} &\  
 (-1)^{k(N+1)+\lceil\frac{k}{2}\rceil}Z_{k,\theta=0}^{\Sp(N)}\ ,\cr
Z_{k,\kappa\equiv N+1 \bmod 2}^{\SU(2N)+1\AS}\ \xrightarrow[\eqref{SU-Sp-Higgsing}]{\textrm{Higgsing}} &\ (-1)^{k(N+1)+\lceil\frac{k}{2}\rceil}Z_{k,\theta=\pi }^{\Sp(N)}\ .
\end{aligned}
\end{equation}
The value of the discrete $\theta$-angle in the resulting $\Sp(N)$ theory is determined by both the Chern-Simons level and the rank of the gauge group.

\subsection*{SU(\texorpdfstring{$N$}{N}) gauge theories with (anti-)symmetric hypermultiplet}

In the context of instanton partition functions, we apply the Higgsing process as depicted in figure \ref{fig:O7-Higgsing-rank2} to reduce the rank of a gauge group. It is important to note that, on the $\Omega$-background, the positions of branes, which correspond to the Coulomb branch and mass parameters, might be shifted by a linear combination of $\e_{1,2}$. Particularly, the Higgsing of the $\SU(N)_\kappa+1\AS+N_f\bfF$ instanton partition function \`a la figure \ref{fig:O7-Higgsing-rank2} can be implemented using two distinct sets of parameterizations. One such set of parameters is defined as
\begin{equation}\label{Higgsing-parameters}
a_N=-a_{N-1}-m-\e_+,  \quad m_{N_f}=-a_{N-1}+\e_+, \quad  m_{N_f-1}=a_{N-1}+m+2\e_+ ~,
\end{equation}
and an alternate set is obtained by changing $\e_i\to -\e_i$ above. Under these conditions, one can verify that the instanton partition function of $\SU(N)_\kappa+1\AS+N_f\bfF$ reduces both the gauge group rank and the number of fundamental hypermultiplets by two:
\be \label{SU-Higgsing}
Z^{\SU(N)_\kappa+1\AS+N_f\textbf{F}}_{\textrm{inst}}\ \xrightarrow[\eqref{Higgsing-parameters}]{\textrm{Higgsing}}\ 
Z^{\SU(N-2)_\kappa+1\AS+(N_f-2)\textbf{F}}_{\textrm{inst}}~,
\ee
which can be checked at the level of the integrand:
\begin{align}\label{Higgsing-k}
\cZ_{\SU(N)_\kappa,k}^{\textrm{vec}}\cZ_{\SU(N),k}^{\textrm{anti}}(m)\cZ_{\SU(N),k}^{N_f}\Big|_{\eqref{Higgsing-parameters}}=
\cZ_{\SU(N-2)_\kappa,k}^{\textrm{vec}}\cZ_{\SU(N-2),k}^{\textrm{anti}}(m)\cZ_{\SU(N-2),k}^{N_f-2}
\ .
\end{align}
The various contributions to the ADHM integrals are summarized in appendix \ref{app:ADHM}.

\bigskip

Now let us consider the effect of freezing \eqref{freezing} on the instanton partition functions for $\SU(N)_\kappa+1\AS+8\textbf{F}$ whose brane setting is explained in the previous section.
On a generic $\Omega$-background, we apply the freezing $8\DD7$ to O7$^-$ by setting the mass parameters $m_{\ell=1,\cdots, 8}$ of eight fundamental hypermultiplets to
\be \label{AS-to-S} 
 m_\ell=\frac{m\pm\e_{\pm}}{2}(+\pi i)\ ,
 \ee 
where $m$ is the mass of an antisymmetric hypermultiplet. (See also our notation convention in Appendix \ref{app:notations}.)  Here, we consider all possible combinations of signs, and $(+\pi i)$ indicates both the inclusion and exclusion of this shift. Upon the freezing, the instanton partition function reduces to that of $\SU(N)_\kappa+1\Sym$ 
\be \label{AS-to-S-inst}
Z^{\SU(N)_\kappa+1\AS+8\textbf{F}}_{\textrm{inst}}\ \xrightarrow[\eqref{AS-to-S}]{\textrm{freezing }}\ Z^{\SU(N)_\kappa+1\Sym}_{\textrm{inst}}~.
\ee 
 This can be shown by comparing the integrands of the instanton partition functions at each instanton level, without performing JK-residues. Using the identity \eqref{eq:ch-sh-identity}, one can verify that
\begin{align}\label{AS-to-S-k}
\cZ_{\SU(N)_\kappa,k}^{\textrm{vec}}\cZ_{\SU(N),k}^{\textrm{anti}}(m)\cZ_{\SU(N),k}^{N_f=8}\Big|_{\eqref{AS-to-S}}
=\cZ_{\SU(N)_\kappa,k}^{\textrm{vec}}\cZ_{\SU(N),k}^{\textrm{sym}}(m)\ .
\end{align}
Therefore, the identity \eqref{AS-to-S-inst} holds when we introduce the same 5d Chern-Simons term to the two theories.

For unfreezing, there are notable differences compared to the prepotential case \eqref{eq:SynToAS-prepot}. While we tune the positions of only eight color branes for prepotential, we need more color branes and flavor branes for unfreezing at the level of instanton partition functions. This is because the positions \eqref{AS-to-S} of 8D7 are different from each other. To Higgs these 8D7 branes, we require the introduction of eight more color branes and an equal number of additional D7 branes, leading to the following 
\be \label{S-to-AS-inst}
Z^{\SU(N+16)_\kappa+1\Sym+8\textbf{F}}_{\textrm{inst}}\ \xrightarrow[\eqref{S-to-AS}]{\textrm{Unfreezing}} \ Z^{\SU(N)_\kappa+1\AS}_{\textrm{inst}}~,
\ee 
which can be executed by two different parameter specializations. One such set of parameters is given by
\begin{align}\label{S-to-AS}
&a_s=-\frac{m}{2}-\epsilon_{+} \pm \frac{\epsilon_{ \pm}}{2}(+ \pi i),\quad  -\frac{m}{2}+2 \epsilon_{+} \pm \frac{\epsilon_{ \pm}}{2}(+\pi i)\cr 
&m_l= \frac{m}{2}-3 \epsilon_{+} \pm \frac{\epsilon_{ \pm}}{2}(+\pi i) \ ,
\end{align}
and the other set is obtained by changing $\e_i\to-\e_i$ above. 
The other set is obtained by changing $\e_i\to-\e_i$ above. 
As detailed in \eqref{Higgsing-parameters}, there are two parametrizations for Higgsing of the instanton partition function, leading to the above two transformations.  This can be confirmed, in a similar manner to \eqref{AS-to-S-k}, by examining the identity of the ADHM integrands at each instanton level.

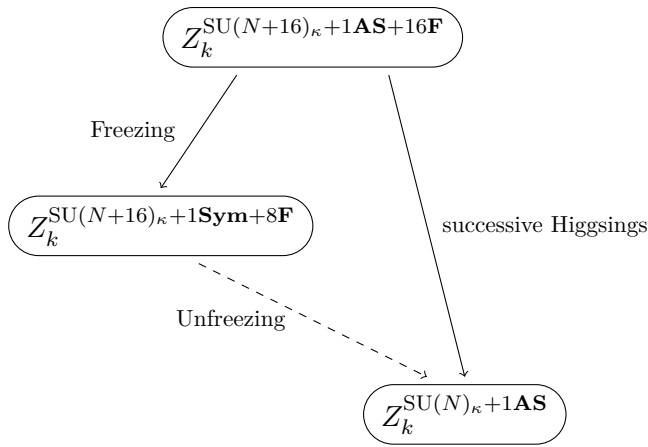
\begin{figure}[t]
    \centering
    \begin{tikzpicture}
\node[rounded rectangle,draw] at (0,0) {$Z^{\mathrm{SU}({N+16})_\kappa+1\AS+16\textbf{F}}_k$};
\node[rounded rectangle,draw] at (2,-5) {$Z^{\mathrm{SU}({N})_\kappa+1\AS}_k$};
\node[rounded rectangle,draw] at (-2,-2.5) {$Z^{\mathrm{SU}({N+16})_\kappa+1\Sym+8\textbf{F}}_k$};
\draw[->] (-1,-.5) to node [left=0.2,scale=0.8] {Freezing} (-2,-2);
\draw[->] (1,-.5) to node [right=0.1,scale=0.8] {successive Higgsings} (2,-4.5);
\draw[->,dashed] (-1.5,-3) to node [left=0.25,scale=0.8] {Unfreezing} (1.5,-4.5);
\end{tikzpicture} 
    \caption{This figure illustrates the relationships between refined instanton partition functions for SU gauge groups under specific parameter specializations. It shows that the partition function $Z^{\mathrm{SU}(N+16)_\kappa+1\AS+16\textbf{F}}_k$, following the freezing process \eqref{AS-to-S}, becomes identical to $Z^{\mathrm{SU}(N+16)_\kappa+1\Sym+8\bfF}_k$. Additionally, when the unfreezing procedures \eqref{S-to-AS} are applied, $Z^{\mathrm{SU}(N+16)_\kappa+1\Sym+8\bfF}_k$ corresponds to $Z^{\mathrm{SU}(N)_\kappa+1\AS}_k$. The dashed line in the figure represents BPS jumping.}
    \label{fig:Refined_triangles}
\end{figure}

We remark that if one combines the two processes, Freezing and Unfreezings, together, the total effect is equivalent to eight sequential Higgsings of \eqref{SU-Higgsing}, which is illustrated in figure \ref{fig:Refined_triangles}. Moreover, during the unfreezing process, the instanton partition function exhibits a sudden jump. This novel phenomenon will be treated in section \ref{sec:jump} in detail.

\subsection*{Sp(\texorpdfstring{$N$}{N}) and SO(\texorpdfstring{$2N$}{2N}) gauge theories}

As demonstrated in \eqref{SU-SO} and \eqref{SU-Sp}, the Higgsing applied to $\SU(2N)+1\Sym$ or $\SU(2N)+1\AS$ transitions into pure Yang-Mills theory for either $\SO(2N)$ or $\Sp(N)$ gauge group. Consequently, this Higgsing process, as outlined in \eqref{SU-Sp-Higgsing}, alongside the subsequent steps of freezing and unfreezing, establishes a connection between the instanton partition functions of the $\Sp(N)$ and $\SO(2N)$ gauge theories. Let us look at this relationship more closely.

We start the $\Sp(N)_\theta+8\mathbf{F}$ setup, where eight mass parameters are specifically arranged for the freezing process as follows:
\be \label{Sp-to-SO-para}
m_\ell~=~ 0,~~\e_+,~~\frac{\e_{1,2}}2,~~\pi i,~~\e_++\pi i,~~ \frac{\e_{1,2}}2+\pi i\ .
\ee 
These parameterizations can be seen as the $m=\e_+$ specialization of \eqref{AS-to-S}. When we apply this freezing process, the $2k$-instanton partition function of $\Sp(N)_\theta+8\mathbf{F}$ becomes identical to the $k$-instanton partition function of the pure $\SO(2N)$ Yang-Mills
\be\label{Sp-to-SO}
Z^{\Sp(N)_\theta+8\textbf{F}}_{2k}\ \xrightarrow[\eqref{Sp-to-SO-para}]{\textrm{freezing }}\ Z^{\SO(2N)}_{k}~.
\ee
Interestingly, this process of freezing, particularly the setting of one fundamental mass to zero, eliminates the minus sector from the $\Sp(N)$ instanton partition function,  as indicated in \eqref{Sp-fund}. Consequently, this freezing process is independent of a choice of the discrete $\theta$-angle, and we can verify it  from the identity of the ADHM integrands
\be 
\mathcal{Z}_{\mathrm{Sp}(N), 2k}^{\mathrm{vec},+}\mathcal{Z}_{\mathrm{Sp}(N), 2k}^{N_f=8,+}\Big|_{\eqref{Sp-to-SO-para}}=\mathcal{Z}_{\mathrm{SO}(2N), k}^{\mathrm{vec}}~.
\ee 

The subsequent stages of unfreezing become more subtle. As mentioned above, the freezing process \eqref{Sp-to-SO-para} removes the minus sector of the $\Sp(N)$ instanton partition function because one of the fundamental masses is set to be zero $m_\ell=0$. The processes of unfreezing then revive the plus sector with even instanton numbers:
\begin{equation}
 Z_k^{\mathrm{SO}(2 N+16)+8\bfF} \xrightarrow[\eqref{SO-to-Sp}]{\textrm{unfreezing}}  Z_{2 k}^{\mathrm{Sp}(N)^+}\ ,
\end{equation}
adjusting the Coulomb branch ($a_s$) and mass parameters ($m_\ell$) as follows:
\begin{align}\label{SO-to-Sp}
a_s=&~0,~~ \epsilon_{+},~~   \frac{\epsilon_{1,2}}{2},~~   
\pi i,~~ \epsilon_{+}
+\pi i,~~  \frac{\epsilon_{1,2}}{2}+\pi i,\cr 
m_l= &~a_s+\epsilon_{+}.
\end{align}

By increasing the rank of the gauge group and the number of fundamental masses, other sectors of the $\Sp(N)$ instanton partition function can be accessed through unfreezing. The details are committed, but these can be also verified at the level of the integrand (up to some factor):
\begin{align}
 Z_k^{\mathrm{SO}(2 N+18)+10\bfF}& \xrightarrow{\textrm{unfreezing}}  Z_{2 k+1 }^{\mathrm{Sp}(N)^+} \cr 
 &a_s=0(+\pi i)~,\  \epsilon_{+}(+\pi i)~, \  \frac{\epsilon_{1.2}}{2}(+\pi i)~,\  \epsilon_{-}\cr 
&m_l= \epsilon_{+}(+\pi i)~,\ 2 \epsilon_{+}(+\pi i)~,\ \epsilon_{+}+\frac{\epsilon_{1 \cdot 2}}{2}(+\pi i)~, \ 0,\, \ 2\e_+.\cr 
 Z_k^{\mathrm{SO}(2 N+18)+10\bfF}& \xrightarrow{\textrm{unfreezing}}  Z_{2 k+1 }^{\mathrm{Sp}(N)^-} \cr 
 &a_s=0(+\pi i)~,\  \epsilon_{+}(+\pi i)~, \  \frac{\epsilon_{1.2}}{2}(+\pi i)~,\  \epsilon_{-}+\pi i\cr 
&m_l= \epsilon_{+}(+\pi i)~,\ 2 \epsilon_{+}(+\pi i)~,\ \epsilon_{+}+\frac{\epsilon_{1 \cdot 2}}{2}(+\pi i)~, \ \pi i,\, \ 2\e_++\pi i.\cr 
 Z_{k-1}^{\mathrm{SO}(2 N+20)+12\bfF}& \xrightarrow{\textrm{unfreezing}}  Z_{2 k }^{\mathrm{Sp}(N)^-} \cr 
 &a_s=0(+\pi i)~,\  \epsilon_{+}(+\pi i)~, \  \frac{\epsilon_{1.2}}{2}(+\pi i)~,\  \epsilon_{-}(+\pi i)\cr 
&m_l= \epsilon_{+}(+\pi i)~,\ 2 \epsilon_{+}(+\pi i)~,\ \epsilon_{+}+\frac{\epsilon_{1 \cdot 2}}{2}(+\pi i)~, \ 0(+\pi i),\, \ 2\e_+(+\pi i).\cr 
\end{align}

\paragraph{Comments on 4d instanton partition functions} While we have focused on 5d instanton partition functions so far, let us make brief comments on 4d instanton partition functions. In fact, from 5d to 4d instanton partition functions, we can simply make the following replacements
\be \sh(\alpha) \rightarrow \alpha~, \qquad \ch(\alpha) \rightarrow 2~.\ee  
To implement the freezing process in the NS5-D4-D6-O6 system within Type IIA, we position four D6-branes at the location of the O6$^-$-plane. These D6-branes are bound together, forming a composite structure of O6$^-+4$D6. The resulting effect can be compared with that of an O6$^+$-plane:
\be \label{freezing-O6}
    \textrm{O6}^-+4\,\textrm{D6} \Big|_{\textrm{fixed}} ~\sim ~ \textrm{O6}^+. 
\ee 
(For a more detailed explanation, refer to section \ref{sec:4d}.) This leads to the identity of 4d instanton partition functions of $\SU(N)+1\AS+4\bfF$ and $\SU(N)+1\Sym$ where the mass parameters are as in \eqref{AS-to-S} without $(+\pi i)$ shifts. Similarly, the unfreezing process in this setting provides the identity of 4d instanton partition functions of $\SU(N+8)+1\Sym+4\bfF$ and $\SU(N)+1\AS$ where the parameter specializations are as in \eqref{S-to-AS} again without $(+\pi i)$ shifts.  It becomes evident that the (un)freezing process also implicates identities between 4d $\SO(N)$ and $\Sp(N)$ instanton partition functions, although we will omit the details here.

\subsection{E-string \texorpdfstring{$\to$}{to} M-string}\label{sec:EtoM}
Up to this point, our study revealed a striking correspondence in the instanton partition functions. Specifically, we have found that, as a result of freezing, the partition function for $\text{SU}(N)_\kappa+1\AS+8\bfF$ coincides with that of $\text{SU}(N)_\kappa+1\Sym$, upon the eight masses of the fundamentals are appropriately tuned. This finding not only enhances our understanding of instanton dynamics but also bridges different configurations in string theory.

A particularly intriguing manifestation of this correspondence is found in the context of E-string \cite{Witten:1995gx,Ganor:1996mu,Seiberg:1996vs} and M-string \cite{Haghighat:2013gba} theories. Although these theories have 6d origins, they admit 5d realizations, offering a unique playground to explore the above correspondence. In this respect, the E-string theory is effectively disguised as a 5d SU(2) theory with eight fundamental hypermultiplets, while M-string theory reveals itself as a 5d SU(2) theory of the discrete theta $\theta=0$ with one symmetric (adjoint) hypermultiplet:
\bea 
\textrm{E-string on }S^1 =& \textrm{ 5d SU(2)+8\textbf{F}} \ ,\cr 
\textrm{M-string on }S^1 =& \textrm{ 5d SU(2)$_0$+1\Sym}\ .
\eea 
Therefore, while the partition functions of E-string and M-string have been computed by various methods 
\cite{Haghighat:2013gba, Haghighat:2014pva,Kim:2014dza,Cai:2014vka, Kim:2015jba,Kim:2017jqn,Kim:2020hhh,Kim:2022dbr},
 we can make use of 5d instanton partition functions to evaluate them here as in \cite{Hwang:2014uwa}. Since the antisymmetric representation of SU(2) is trivial, the $N=2$ specialization \eqref{AS-to-S-inst} of the freezing effect can be understood as the relation between E-string and M-string through this approach. Namely, upon specializing the eight mass parameters in E-string, the partition function agrees remarkably with that of M-string.

\begin{figure}
    \centering
    \includegraphics[width=12.1cm]{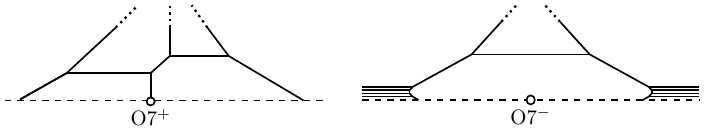}
\caption{$\mathsf{LEFT}$: the brane diagram for SU(2) with one symmetric hypermultiplet. $\mathsf{RIGHT}$: the brane diagram for Sp(1) with eight fundamental hypermultiplets. The external 5-branes meet at a certain point in both diagrams, which indicates the presence of a higher-order pole at infinity in ADHM integrands.}
 \label{fig:leg-close}
\end{figure}

\paragraph{Poles at infinity and brane diagrams}
Before delving into the partition functions of E-string and M-string theory, it is important to discuss a pole at infinity in the ADHM integrand and its interpretation in a brane diagram. As figure \ref{fig:leg-close} depicts, the external 5-branes meet at a certain point in the 5-brane web diagram for SU(2)+1\Sym, and a D1-brane suspended by these external 5-branes is bounced back. In such scenarios, the ADHM instanton integrals suffer from higher-order poles at infinity. For example, the expression for the 1-instanton case from \eqref{SU-instanton},
\be
\cZ^{\SU(2)_0+1\Sym}_{k=1}=\frac{\sh(\phi+m\pm a) \sh(\pm({m}+2 \phi)-\e_-)  \sh(2\epsilon_+)}{\sh(\epsilon_{1,2})  \sh(\e_+\pm{a}\pm\phi) }~\overset{\phi\to \infty}{\sim}~ \cO(e^{\phi})\ ,
\ee 
shows the presence of a higher-order pole as $\phi\to \infty$. This issue is present even for higher instantons. A similar issue arises in the instanton partition function for SU(2)+1\AS+8\textbf{F}. As a result, a naive application of the JK residue method fails to yield the correct partition function.

\bigskip

\paragraph{1-instanton} For SU(2)$_0$+1\Sym, we can use the expression for the adjoint hypermultiplet in \eqref{SU-instanton} since $\Sym\cong\textbf{Adj}$ for SU(2). For instance, the 1-instanton part behaves as
\be
\cZ_{k=1}^{\SU(2)_0+1\textbf{Adj}}=\frac{\sh(m\pm\phi\pm a)\sh(\epsilon_-\pm m) \sh(2\epsilon_+)}{\sh(\epsilon_{1,2})  \sh(\e_+\pm{a}\pm\phi) } ~\overset{\phi\to \infty}{\sim}~\cO(1)\ .
\ee
Therefore, the JK residue integral provides 
\bea 
Z_{k=1}^{\SU(2)_0+1\textbf{Adj}}=&~\frac{\sh(m\pm(2a-\e_+))\sh(m\pm\epsilon_-)}{\sh(\epsilon_{1,2}) \sh(2{a}) \sh(2\e_+-2{a}) } \cr 
&+\frac{\sh(m\pm(2a+\e_+))\sh(m\pm\epsilon_-) }{\sh(\epsilon_{1,2}) \sh(-2{a}) \sh(2\e_++2{a}) }\cr 
=&~ Z_{\textrm{extra},k=1}^{\SU(2)_0+1\textbf{Adj}}+Z_{\textrm{QFT},k=1}^{\SU(2)_0+1\textbf{Adj}}\ ,
\eea 
where $Z_{\textrm{extra}}$ is the part independent of the Coulomb branch parameter, $a$,
\bea \label{M-1}
Z_{\textrm{extra},k=1}^{\SU(2)_0+1\textbf{Adj}}=&~2 \frac{\sh(m\pm\epsilon_{-})}{\sh(\epsilon_{1,2})}~,\cr 
Z_{\textrm{QFT},k=1}^{\SU(2)_0+1\textbf{Adj}}=&~ \frac{
\sh(\epsilon_{\pm}\pm m)
\ch(2\e_+)}{\sh(\epsilon_{1,2})\sh(2\epsilon_{+}\pm2a)} \ .
\eea

Now let us turn to E-string theory. The ADHM integrand of SU(2)+8\textbf{F} obtained from \eqref{SU-instanton} exhibits a higher-order pole $\cO(e^{2\phi})$ at infinity. In contrast, the ADHM integrand of $\Sp(1)_{\theta=0}+8\textbf{F}$,\footnote{When hypermultiplets are introduced, the discrete theta angle $\theta$ does not lead to a topologically distinct theory. Here we specify the value of $\theta$ to keep track of what ADHM integral is used for the computation.} derived from the formulas in appendix \ref{app:Sp}, presents a less severe singularity $\cO(e^{\phi})$ at infinity (see the right of figure \ref{fig:leg-close}) although it provides the same partition function as SU(2)+8\textbf{F}. This makes the latter more manageable. To mitigate the singularity at infinity, we can make use of an antisymmetric hypermultiplet \cite{Hwang:2014uwa,Gaiotto:2015una,Chen:2021ivd}. For the $\SU(2)\cong\Sp(1)$ gauge group, the antisymmetric representation is trivial, meaning the addition of an antisymmetric hypermultiplet does not fundamentally change the physics. However, this addition is beneficial as the contribution of the antisymmetric hypermultiplet tends towards $\cO(e^{-\phi})$ at infinity, therefore softening the singularity. Thus, to derive the E-string partition function, we will examine the ADHM instanton partition function of $\Sp(1)_0$+1\AS+8\textbf{F} in detail because it behaves as $\cO(1)$ at infinity.\footnote{The pole at infinity could provide two distinct contributions: one that yields a physically relevant contribution and the other that leads to the extra/decoupled factor. If the pole at infinity can be appropriately handled before introducing the antisymmetric hypermultiplet, the outcomes are expected to be identical. However, to the best of the authors' knowledge, the proper method of dealing with the pole at infinity has not been established yet.}

At 1-instanton, the ADHM instanton partition function of $\Sp(1)_0$+1\AS+8\textbf{F} does not involve an integral \cite{Shadchin:2005mx,Kim:2012gu} so that the evaluation is straightforward:
\bea 
Z_{k=1}^{\Sp(1)_0+1\AS+8\textbf{F}}=&~Z_{k=1}^{\Sp(1)_0+1\AS+8\textbf{F},+}+Z_{k=1}^{\Sp(1)_0+1\AS+8\textbf{F},-}\cr 
=&~\frac{\sh(\pm a+m)\prod_{\ell=1}^8 \sh(m_\ell)}{\sh(\e_{1,2})\sh(\pm m -\!\e_+)\sh(\pm a +\e_+)}\!+\!\frac{\ch(\pm a+m)\prod_{\ell=1}^8 \ch(m_\ell)}{\ch(\e_{1,2})\ch(\pm m -\!\e_+)\ch(\pm a +\e_+)}\cr 
=&~Z_{\textrm{extra},k=1}^{\Sp(1)_0+1\AS+8\textbf{F}}+Z_{\textrm{QFT},k=1}^{\Sp(1)_0+1\AS+8\textbf{F}}\ ,
\eea 
where, in the last line, we separate the parts dependent (labelled by QFT) and independent (labelled by extra) of the Coulomb branch parameter:
\bea \label{E-1}Z_{\textrm{QFT},k=1}^{\Sp(1)_0+1\AS+8\textbf{F}}=-\frac{\prod_{\ell=1}^8 \sh(m_\ell)}{\sh(\e_{1,2})\sh(\pm a +\e_+)}-\frac{\prod_{\ell=1}^8 \ch(m_\ell)}{\ch(\e_{1,2})\ch(\pm a +\e_+)}~,\cr 
Z_{\textrm{extra},k=1}^{\Sp(1)_0+1\AS+8\textbf{F}}=\frac{\prod_{\ell=1}^8 \sh(m_\ell)}{\sh(\e_{1,2})\sh(\pm m -\e_+)}+\frac{\prod_{\ell=1}^8 \ch(m_\ell)}{\ch(\e_{1,2})\ch(\pm m -\e_+)}~.
\eea 
Upon specializing the eight masses $m_\ell$ as in \eqref{AS-to-S}, we obtain
\be 
Z_{\textrm{QFT},k=1}^{\Sp(1)_0+1\AS+8\textbf{F}}\Big|_{\eqref{AS-to-S}}= -\frac{
\sh(\epsilon_{\pm}\pm m)
\ch(2\e_+)}{\sh(\epsilon_{1,2})\sh(2\epsilon_{+}\pm2a)}\ .
\ee 
Comparing this with \eqref{M-1}, we see a relationship between the 1-instanton partition functions of the E-string and M-string,
\be 
Z_{\textrm{QFT},k=1}^{\Sp(1)_0+1\AS+8\textbf{F}} \ \xrightarrow[]{\eqref{AS-to-S}} \ -Z_{\textrm{QFT},k=1}^{\SU(2)_0+1\textbf{Adj}}~.
\ee

\paragraph{Higher-instanton} Let us move on to higher-instanton. The instanton partition function of SU(2)$_0$+1\textbf{Adj} \cite{Nekrasov:2002qd,Bruzzo:2002xf} can be expressed in terms of sums over Young diagrams as
\begin{align}
Z^{\SU(2)_0+1\textbf{Adj}}=\sum_{k\ge0}e^{-m_0k}\!\!\!\!\sum_{\sum_s|\lambda^{(s)}|=k}\, \prod_{s,t=1}^2 \prod_{x \in \lambda^{(s)}}\, \frac{\sh( N_{st}+m-\epsilon_{+} )\sh( N_{st}-m-\epsilon_{+})}{\sh( N_{st} )\sh( N_{st}-2 \epsilon_{+})} ,
\end{align}
where $m_0$ is the inverse coupling squared (or instanton mass), and we define the Nekrasov factor as
\be \label{Nij}
N_{st}(x)=a_s-a_t-\epsilon_1 L_{\lambda^{(s)}}(x)+\epsilon_2(A_{\lambda^{(t)}}(x)+1)~.
\ee
(See \eqref{arm-leg} for the definitions of the arm $A_\lambda(x)$ and leg $L_\lambda(x)$ length for a content $x\in \lambda$.)
As seen above, this can be factored into two parts
\be 
Z^{\SU(2)_0+1\textbf{Adj}}=Z_{\textrm{extra}}^{\SU(2)_0+1\textbf{Adj}}
Z_{\textrm{QFT}}^{\SU(2)_0+1\textbf{Adj}}~,
\ee 
where $Z_{\textrm{extra}}^{\SU(2)_0+1\textbf{Adj}}$ is expressed in terms of a plethystic exponential
\be 
Z_{\textrm{extra}}^{\SU(2)_0+1\textbf{Adj}}=\textrm{PE}\left[\frac{e^{-m_0}}{1-e^{-m_0}}\frac{2\sh(m\pm\e_-)}{\sh(\e_{1,2})}\right]~.
\ee

On the other hand, a closed-form expression of the refined instanton partition function of $\Sp(1)_0$+1\AS+8\textbf{F} is unavailable at this moment because it is difficult to classify the JK pole structure. Therefore, we perform the JK residue integral at each instanton level. To extract a genuine instanton contribution,  it is necessary to eliminate the extra factor that is independent of the Coulomb branch parameter. For the sake of simplicity, we focus on the extra factor specifically at the specialization \eqref{AS-to-S}, as a computation for generic mass parameters is computationally intensive:
\be 
Z^{\Sp(1)_0+1\AS+8\textbf{F}}\Big|_{\eqref{AS-to-S}}=Z_{\textrm{extra}}^{\Sp(1)_0+1\AS+8\textbf{F}}\cdot
Z_{\textrm{QFT}}^{\Sp(1)_0+1\AS+8\textbf{F}}\Big|_{\eqref{AS-to-S}}~,
\ee 
where $Z_{\textrm{extra}}^{\Sp(1)_0+1\AS+8\textbf{F}}$ is expressed in terms of a plethystic exponential
\begin{equation}
Z_{\textrm{extra}}^{\Sp(1)_0+1\AS+8\textbf{F}}
=\textrm{PE}\left[\frac{e^{-m_0}}{1+e^{-m_0}}\frac{\sh(\e_-\pm m)}{\sh(\e_{1,2})}\bigg(1-\frac{e^{-m_0}}{1-e^{-m_0}}\frac{\ch(2\e_+)}{2\ch(m\pm\e_+)}\bigg)\right]~.
\end{equation}
We have verified the equivalence of the partition functions up to $k=4$ at the refined level (up to sign\footnote{This sign difference can be absorbed into the redefinition of the instanton factor. %Therefore, this would not imply any significant mismatch.
}):
\be
Z_{\textrm{QFT},k}^{\Sp(1)_0+1\AS+8\textbf{F}}\Big|_{\eqref{AS-to-S}}=(-1)^k Z_{\textrm{QFT},k}^{\SU(2)_0+1\textbf{Adj}}~.
\ee

We remark that in a similar fashion, one can also take another freezing limit on the E-string partition function which leads to the partition function for $\SU(2)_\pi+1\mathbf{Adj}$. The map is based on \cite{Kim:2023qwh} and given by:
\be \label{E-to-SU2pi} 
m_{1,\cdots,8}=\pm\frac{ m+\epsilon_+}{2},~~\frac{m\pm\epsilon_-}{2},~~  
\frac{m\pm\epsilon_\pm}{2}+\pi i
\ ,
\ee 
where $m$ is the mass of the adjoint hypermultiplet.

The two different freezings on E-string were also checked in \cite{Kim:2024klw}.

\subsection{BPS jumping in instanton spectra}\label{sec:jump}

Now we turn to the part of the unfreezing procedure, and we delve into the BPS jumping phenomenon briefly mentioned in the preceding subsection through the partition function. The unfreezing requires the specialization of parameters in the theory. 
As we fine-tune parameters in the theory, the Jeffrey-Kirwan integral encounters degenerate poles, as explained in appendix \ref{app:JK}. Interestingly, these degenerate poles can give rise to the appearance of  multiplicity coefficients in \cite{Nawata:2021dlk,Chen:2023smd}.  Consequently, we observe jumps in the instanton partition functions. In other words, the BPS spectrum of one theory jumps to that of another upon a particular tuning of the physical parameters (masses and Coulomb branch moduli) of the theory, which we will discuss in detail. For the sake of simplicity, we will concentrate on \emph{unrefined} instanton partition functions to illustrate this remarkable BPS jumping phenomenon.

\begin{figure}[ht]
    \centering
    \begin{tikzpicture}
\node[rounded rectangle,draw] at (0,0) {$Z^{\mathrm{SU}({N+8})_\kappa+1\AS+8\textbf{F}}_k$};
\node[rounded rectangle,draw] at (2,-5) {$Z^{\mathrm{SU}({N})_\kappa+1\AS}_k$};
\node[rounded rectangle,draw] at (-2,-2.5) {$Z^{\mathrm{SU}({N+8})_\kappa+1\Sym}_k$};
\draw[->] (-1,-.5) to node [left=0.1,scale=0.8] {Freezing} (-2,-2);
\draw[->] (1,-.5) to node [right,scale=0.8] {Higgsing} (2,-4.5);
\draw[->,dashed] (-1.5,-3) to node [left=0.25,scale=0.8] {Unfreezing}(1.5,-4.5);
\end{tikzpicture}
\quad  \begin{tikzpicture}
\node[rounded rectangle,draw] at (0,0) {$Z^{\mathrm{Sp}({N+4})+8\textbf{F}}_{2\ell+\chi}$};
\node[rounded rectangle,draw] at (2,-5) {$Z^{\mathrm{Sp}({N}),\pm}_{2\ell+\chi}$ };
\node[rounded rectangle,draw] at (-2,-2.5) {$Z^{\mathrm{SO}({2N+8})}_{\ell}$};
\draw[->] (-1,-.5) to node [left=0.1,scale=0.8] {Freezing} (-2,-2);
\draw[->,dashed] (-1.5,-3) to node [left=0.25,scale=0.8] {Unfreezing} (1.5,-4.5);
\end{tikzpicture}
    \caption{$\mathsf{LEFT}$: Illustration of the identities among unrefined instanton partition functions for SU gauge groups upon specific parameter specializations. The partition function $Z^{\mathrm{SU}({N+8})_\kappa+1\AS+8\textbf{F}}_k$, after the freezing process as in \eqref{AS-to-S-unref}, is equal to $Z^{\mathrm{SU}({N+8})_\kappa+1\Sym}_k$. Furthermore, applying the unfreezing procedures as in \eqref{S-to-AS-unref} identifies $Z^{\mathrm{SU}({N+8})_\kappa+1\Sym}_k$ with $Z^{\mathrm{SU}(N)_\kappa+1\AS}_k$ where the dashed line indicates BPS jumping. \\
    $\mathsf{RIGHT}$: Demonstrates the identities of unrefined instanton partition functions for SO/Sp gauge groups upon parameter specializations. The partition function $Z^{\mathrm{Sp}({N+4})+8\textbf{F}}_{2\ell+\chi}$, after the freezing process as in \eqref{Sp-to-SO-unref}, is equal to $Z^{\mathrm{SO}({2N+8})}_{\ell}$. Furthermore, applying the unfreezing procedures as in \eqref{SO-to-Sp-unref} identifies $Z^{\mathrm{SO}({2N+8})}_{\ell}$ with $Z^{\mathrm{Sp}({N}),\pm}_{2\ell+\chi}$ where the dashed line indicates BPS jumping.}
    \label{fig:Unrefined_triangles}
\end{figure}
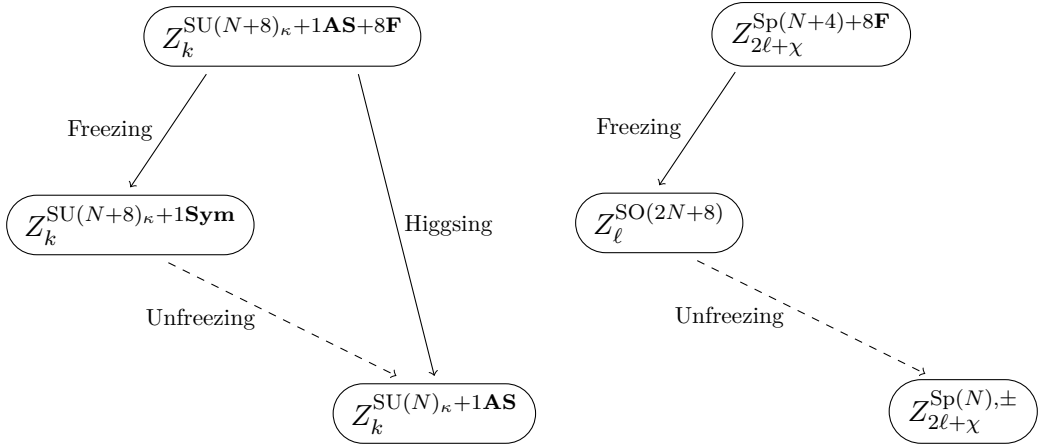

The unrefined limit $\epsilon_1=-\epsilon_2=\hbar$ offers further simplification since brane positions become more degenerate compared to the refined case. This effect notably reduces the necessary number of color and flavor branes for the processes of unfreezing. At the unrefined level, the fundamental hypermultiplets are indeed unnecessary, and the ranks of gauge groups are the same as the analysis of the prepotential, as in figure \ref{fig:Unrefined_triangles}. Thus, let us first investigate the unfreezing process for the SU gauge groups using the unrefined instanton partition functions.

\subsection*{SU(\texorpdfstring{$N$}{N}) gauge theories with (anti-)symmetric hypermultiplet}
In this subsection, we adopt the same notation for the unrefined limit of the Nekrasov factor \eqref{Nij} as follows:
\begin{equation}
  N_{s,t}(x):= a_s- a_t-\hbar(A_{\lambda^{(s)}}(x)+L_{\lambda^{(t)}}(x)+1)\ .
\end{equation}
Then, the formula for $\mathrm{SU}({N})_\kappa+1\Sym$ ($N\ge3$)\footnote{Since the JK residue integral for the SU(2) gauge group suffers from the higher-order pole at infinity as discussed around figure \ref{fig:leg-close}, the formulas \eqref{SU-sym} is \emph{not} valid for $N=2$. This remark is also applied to \eqref{SU-AS-8F}.} is expressed as a sum over $N$-tuples $\vec\lambda$ of Young diagrams 
\begin{align}\label{SU-sym}
& Z_{k}^{\mathrm{SU}({N})_\kappa+1\Sym} \cr
= & (-1)^{k N+\left\lceil\frac{k}{2}\right\rceil} \sum_{|\vec\lambda|=k} \prod_{s=1}^N \prod_{x \in \lambda^{(s)}} e^{\kappa \phi_s(x)} \frac{\operatorname{sh}\left(2 \phi_s(x)+m \pm \hbar\right) \cdot \prod_{t=1}^N \operatorname{sh}\left(\phi_s(x)+a_t+m\right)}{\prod_{t=1}^N \operatorname{sh}^2\left(N_{s, t}(x)\right)} \cr
& \times \prod_{s \leq t}^N \prod_{\substack{x \in \lambda^{(s)}, y \in \lambda^{(t)} \\
x<y}} \frac{\operatorname{sh}\left(\phi_s(x)+\phi_t(y)+m \pm \hbar\right)}{\operatorname{sh}^2\left(\phi_s(x)+\phi_t(y)+m\right)}\\
=: & \sum_{|\vec\lambda|=k}  Z_{\vec\lambda}^{\mathrm{SU}({N})_\kappa+1\Sym} \nonumber\ ,
\end{align}
where $\phi_s(x)$ indicates a pole location associated to a content $x=\left(i,j\right) \in \lambda^{(s)}$ as
\be \label{pole-location}
\phi_s(x)=a_s+(i-j) \hbar \ .
\ee 
It is important to emphasize that this formula is valid only under the assumption that all the Coulomb branch parameters $a_s$ take \emph{generic values}. The reason for this remark becomes evident later. When the Coulomb branch parameters take special values, an additional phenomenon arises, and the partition function differs from the expression \eqref{SU-sym}.

On the other hand, the formula for $\mathrm{SU}({N})_\kappa+1\AS$ is expressed as a sum over $(N+8)$-tuples $\vec\lambda$ of Young diagrams\footnote{The expression \eqref{SU-AS} is indeed different from \cite[Eq. (2.12)]{Chen:2023smd}, where only four effective Coulomb branch parameters are present. The expression \eqref{SU-AS} takes into account the freezing/unfreezing of O7${}^+$$\sim$O7${}^-+$8D7, resulting in the emergence of eight additional effective Coulomb branch parameters \eqref{S-to-AS-unref}.}
\begin{equation}\label{SU-AS}
\begin{aligned}
 Z_{k}^{\mathrm{SU}({N})_\kappa+1\AS} = \sum_{|\vec\lambda|=k}C_{\vec{\lambda}, \vec{a}}^{\operatorname{anti}}  Z_{\vec\lambda}^{\mathrm{SU}({N+8})_\kappa+1\Sym} \Big|_{\eqref{S-to-AS-unref}} \ ,
\end{aligned}
\end{equation}
where we set eight additional effective Coulomb branch parameters as 
\be \label{S-to-AS-unref}
a_{N+j}= -\frac{m}{2} (+\pi i )~, \ -\frac{m}{2} (+\pi i )~, \   \frac{\pm\hbar-m}{2} (+\pi i )~.
\ee
The poles coming from the eight additional Young diagrams are generically \emph{degenerate} \cite{brion1999arrangement,szenes2003toric,Benini:2009gi}, meaning that more than $k$ factors in the denominator of the ADHM integrand simultaneously become zero at these poles. Consequently, the residues at these degenerate poles give rise to the multiplicity coefficients $C_{\vec{\lambda}, \vec{A}}^{\operatorname{anti}}$. The conjectured values for these constants are as follows:
$C^{\textrm{anti}}_{\lambda^{(s)}=\emptyset,a_s}=1$, and
\bea \label{C-anti}
C^{\textrm{anti}}_{\lambda^{(s)},a_s=\frac{m}2(+\pi i)}=&-1&\textrm{for } s=N+1,N+2\ ,\cr 
C^{\textrm{anti}}_{\lambda^{(s)},a_s=\frac{m}2(+\pi i)}=&~(-1)^{\alpha(\lambda^{(s)})}&\textrm{for } s=N+3,N+4\ ,\cr 
C^{\textrm{anti}}_{\lambda^{(s)},a_s=\frac{\hbar-m}{2}(+\pi i)}=&~ \beta(\lambda^{(s)}) \bmod 2 &\textrm{for } s=N+5,N+6\ , \cr 
C^{\textrm{anti}}_{\lambda^{(s)},a_s=\frac{-\hbar-m}{2}(+\pi i)}=&~ \beta((\lambda^{(s)})^t) \bmod 2 
&\textrm{for } s=N+7,N+8\ .\eea
See \eqref{alpha-beta} for the definitions of $\alpha(\lambda)$ and $\beta(\lambda)$.

The unrefined instanton partition function of $\mathrm{SU}({N})_\kappa+1\AS+8\textbf{F}$ is given by
\begin{equation}\label{SU-AS-8F}
\begin{aligned}
 Z^{\mathrm{SU}({N})_\kappa+1\AS+8\textbf{F}}_k = \sum_{|\vec\lambda|=k}C_{\vec{\lambda}, \vec{a}}^{\operatorname{anti}}  Z_{\vec\lambda}^{\mathrm{SU}({N+8})_\kappa+1\Sym} \prod_{s=1}^{N+8}\prod_{x \in \lambda^{(s)}} \prod_{l=1}^{8}\sh(\phi_s(x)+ m_l)  \Big|_{\eqref{S-to-AS-unref}} \ .
\end{aligned}
\end{equation}
To obtain the partition function $Z^{\mathrm{SU}({N})_\kappa+1\Sym}_k$ from $Z^{\mathrm{SU}({N})_\kappa+1\AS+8\textbf{F}}_k$, the masses of the eight fundamentals in $\mathrm{SU}({N})_\kappa+1\AS+8\textbf{F}$ must be adjusted as 
\be \label{AS-to-S-unref}
m_{j}= \frac{m}{2} , ~~ \frac{m}{2} , ~~  \frac{m\pm \hbar}{2} ,~~~
\frac{m}{2} +\pi i , ~~ \frac{m}{2} +\pi i , ~~   \frac{m\pm \hbar}{2} +\pi i ~,
\ee
for $j=1,\ldots,8$. Like the refined case, this can be verified at the level of the ADHM integrand:
\begin{align}
\cZ_{\SU(N)_\kappa,k}^{\textrm{vec}}\cZ_{\SU(N),k}^{\textrm{sym}}(m)=\cZ_{\SU(N)_\kappa,k}^{\textrm{vec}}\cZ_{\SU(N),k}^{\textrm{anti}}(m)\cZ_{\SU(N),k}^{N_f=8}\Big|_{\eqref{AS-to-S-unref}}\ .
\end{align}
In fact, when the masses of the fundamentals are set to \eqref{AS-to-S-unref}, the residues at the degenerate poles become zero. This is evident from the fact that the contributions $\phi_s(x)$ ($s=N+1,\ldots,N+8$) from the effective Coulomb branch parameters \eqref{S-to-AS-unref} vanish due to the presence of $\sh(\phi_s(x)+m_l)$ factors arising from $8\textbf{F}$. Consequently, we arrive at the following identity
\begin{equation}\label{SU-S-8F}
 Z^{\mathrm{SU}({N})_\kappa+1\AS+8\textbf{F}}_k\Big|_{\eqref{AS-to-S-unref}} =\sum_{|\vec\lambda|=k} Z_{\vec\lambda}^{\mathrm{SU}({N})_\kappa+1\Sym} = Z^{\mathrm{SU}({N})_\kappa+1\Sym}_k~.
\end{equation}

To obtain the partition function $Z_k^{\mathrm{SU}({N})_\kappa+1\AS}$ from $Z_k^{\mathrm{SU}({N+8})_\kappa+1\Sym}$, we need to adjust the eight Coulomb branch parameters in $\mathrm{SU}({N+8})_\kappa+1\Sym$ as \eqref{S-to-AS-unref}
for $j=1,\ldots,8$. Once again, verifying the equality of the partition functions at the level of the ADHM integrand is straightforward. However, we observe an intriguing phenomenon. A naive substitution of \eqref{S-to-AS-unref} into \eqref{SU-sym} does not yield \eqref{SU-AS} due to the presence of the multiplicity coefficients. These multiplicity coefficients arise due to two reasons.  Firstly, they emerge because the adjustment of some Coulomb branch parameters by a difference of $\hbar$ in \eqref{S-to-AS-unref} disrupts the conventional classification of poles by Young diagrams. Secondly, when tuning the eight Coulomb branch parameters of $\mathrm{SU}({N+8})_\kappa+1\Sym$ to \eqref{S-to-AS-unref}, the JK residue integrals encounter degenerate poles. For a detailed analysis of these phenomena, we refer to appendix \ref{app:JK}. Consequently, at the specific value of the Coulomb branch parameters, the partition function suddenly \emph{jumps}, implying the corresponding jump in the BPS spectra:
\begin{equation}\label{SU-S-jump}
\begin{aligned}
 Z_{k}^{\mathrm{SU}({N+8})_\kappa+1\Sym} = &\sum_{|\vec\lambda|=k} Z_{\vec\lambda}^{\mathrm{SU}({N+8})_\kappa+1\Sym}\cr 
\xrightarrow[\textrm{BPS jump}]{a_{N+j} \textrm{ at } \eqref{S-to-AS-unref}}\quad & \sum_{|\vec\lambda|=k}C_{\vec{\lambda}, \vec{a}}^{\operatorname{anti}}  Z_{\vec\lambda}^{\mathrm{SU}({N+8})_\kappa+1\Sym} \Big|_{\eqref{S-to-AS-unref}} \ .
\end{aligned}
\end{equation}
Consequently, the unrefined instanton partition function of  $\mathrm{SU}({N+8})_\kappa+1\Sym$ at the special value of Coulomb branch parameters \eqref{S-to-AS-unref} agree with that \eqref{SU-AS} of $\mathrm{SU}({N})_\kappa+1\AS$.

In this discussion, we primarily explore unrefined instanton partition functions. However, it is important to note that can also occur during the unfreezing processes at the refined level as illustrated in figure \ref{fig:Refined_triangles}. 
In fact, we analyze the BPS jumping at the refined level in appendix \ref{app:JK}. The jumping phenomenon indeed becomes more complicated in refined settings. Due to the lack of a closed-form expression for refined instanton partition functions, we avoid delving into these complexities in this paper.

\subsection*{Sp(\texorpdfstring{$N$}{N}) and SO(\texorpdfstring{$2N$}{2N}) gauge theories}
A similar phenomenon can be observed among Sp($N$) and SO($n$) gauge theories. 
The formula for the pure $\mathrm{SO}({n})$ gauge theory ($n\ge 4$) is expressed as a sum over $\lfloor \frac{n}{2}\rfloor$-tuples $\vec\lambda$ of Young diagrams 
\begin{equation}\label{SO}
\begin{aligned}
 Z_{k}^{\mathrm{SO}(n)}= &\sum_{|\vec\lambda|=k} \prod_{s=1}^N \prod_{x \in \lambda^{(s)}}\frac{\operatorname{sh}^4\left(2\phi_s(x)\right) } {\operatorname{sh}^{2\chi}\left(\phi_s(x)\right)\prod_{t=1}^N \operatorname{sh}^2\left(N_{s, t}(x)\right) \cdot\operatorname{sh}^2\left(\phi_s(x)+a_t\right)} \\
& \times \prod_{s \leq t}^N \prod_{\substack{x \in \lambda^{(s)}, y \in \lambda^{(t)} \\
x<y}} \frac{\operatorname{sh}^4\left(\phi_s(x)+\phi_t(y)\right)}{\operatorname{sh}^2\left(\phi_s(x)+\phi_t(y)\pm \hbar\right)}\cr 
=: & \sum_{|\vec\lambda|=k}  Z_{\vec\lambda}^{\mathrm{SO}(n)}\ ,
\end{aligned}
\end{equation}
where $n=2N+\chi$ $(n\equiv\chi \mod 2)$. Again, we emphasize that this formula remains valid when the Coulomb branch parameters are \emph{generic}. 

As is well-known from \cite{Kim:2012gu}, for 5d $\Sp(N)$ gauge theory, the supersymmetric quantum mechanics on the $k$-instanton moduli spaces is described by $\OO(k)$ gauge group, which consists of two connected components $\OO(k)^\pm$. 
Consequently, the instanton partition function receives two contributions, denoted by $\pm$ sectors corresponding to the sign of the $\OO(k)^\pm$ determinant. As explained in appendix \ref{app:Sp}, the sum (resp. difference) of these two contributions gives the partition function at $\theta=0$ (resp. $\theta=\pi$). At the unrefined level, the formulas as sums over $(N+4)$-tuples $\vec\lambda$ of Young diagrams  are provided in \cite{Nawata:2021dlk}, and moreover the relation to the SO($2N+8$) instanton partition function is pointed out, which is given as follows:
\begin{equation}\label{Sp}
Z_{k}^{\Sp(N),\pm}=\mathrm{pref}_{k}^\pm\sum_{|\vec{\lambda}|=\ell} C^{\Sp}_{\vec{\lambda},\vec{a}}Z_{\vec{\lambda}}^{\SO(2N+8)} \Big|_{\eqref{SO-to-Sp-unref}} \ ,
\end{equation}
where the prefactor functions are given by
\begin{equation}\label{pref}
\mathrm{pref}_{k}^\pm=\begin{cases}
1  &(k=2\ell,+)\ \textrm{sector}\\
\frac{1}{\operatorname{sh}^2\left(\hbar\right) \prod_{s=1}^N \operatorname{sh}^2\left(a_s\right)}  &(k=2\ell+1,+)\ \textrm{sector} \\
2(-1)^N\frac{1}{\operatorname{sh}^2\left(\hbar\right) \operatorname{sh}^2\left(2\hbar\right) \prod_{s=1}^N \operatorname{sh}^2\left(2a_s\right)}&(k=2\ell+2,-)\ \textrm{sector}\\
(-1)^N\frac{1}{\operatorname{sh}^2\left(\hbar\right) \prod_{s=1}^N \operatorname{ch}^2\left(a_s\right)} &(k=2\ell+1,-)\ \textrm{sector}\ .
    \end{cases}
\end{equation}
Here $a_{N+j}$ ($j=1,\ldots,4$) are four additional effective Coulomb branch parameters that take specific values at each sector as 
\begin{equation}\label{SO-to-Sp-unref}
a_{N+j}=\begin{cases}
    \frac{\hbar}{2}(+\pi i), \  0(+\pi i)   &(2\ell,+)\ \textrm{sector}\\
        \frac{\hbar}{2}(+\pi i), \ \hbar, \ \pi i&(2\ell+1,+)\ \textrm{sector} \\
        \frac{\hbar}{2}(+\pi i), \hbar(+\pi i)&(2\ell+2,-)\ \textrm{sector}\\
   \frac{\hbar}{2}(+\pi i), 0, \hbar+\pi i     &(2\ell+1,-)\ \textrm{sector}
    \end{cases}
\end{equation}
The multiplicity coefficients are given by 
\be\label{C-Sp}
C^{\Sp}_{\lambda^{(s)},a_s=0(+\pi i), \frac{\hbar}{2}(+\pi i)}=\frac{2^{2\alpha(\lambda^{(s)})-1}}{\binom{2\alpha(\lambda^{(s)})-1}{\alpha(\lambda^{(s)})-1}}~,\qquad 
C^{\Sp}_{\lambda^{(s)},a_s=\hbar(+\pi i)}=\frac{2^{2\beta(\lambda^{(s)})}}{\binom{2\beta(\lambda^{(s)})+1}{\beta(\lambda^{(s)})}}\ ,\ee
with $C^{\Sp}_{\lambda^{(s)}=\emptyset,a_s}=1$.
See \eqref{alpha-beta} for the definitions of $\alpha(\lambda)$ and $\beta(\lambda)$.

The unrefined instanton partition function of $\mathrm{Sp}({N+4})+8\textbf{F}$ in the plus sector is given by
\begin{align}\label{Sp--8F}
& Z^{\mathrm{Sp}({N+4})+8\textbf{F}}_{k=2\ell+\chi,+}\\ =&\left(\frac{ \prod_{l=1}^{8}\sh( m_l)  }{\operatorname{sh}^2\left(\hbar\right) \prod_{s=1}^N \operatorname{sh}^2\left(a_s\right)}\right)^\chi \sum_{|\vec\lambda|=\ell}C_{\vec{\lambda}, \vec{a}}^{\Sp}  Z_{\vec\lambda}^{\mathrm{SO}({N+4})} \prod_{l=1}^{8}  \prod_{s=1}^{N+4}\prod_{x \in \lambda^{(s)}}\sh(\pm\phi_s(x)+ m_l)  \Big|_{\eqref{SO-to-Sp-unref}}.\nonumber
\end{align}
The expression in the minus sector for $\mathrm{Sp}({N+4})+8\textbf{F}$ can be written in a similar way. 
To obtain the partition function $Z^{\mathrm{SO}({2N+8})}_\ell$ from $Z^{\mathrm{Sp}({N+4})+8\textbf{F}}_{2\ell+\chi}$ ($\chi=0,1$), 
we must freeze the masses of the eight fundamentals in $\mathrm{Sp}({N+4})+8\textbf{F}$ as
\be \label{Sp-to-SO-unref}
m_{j}= \begin{cases}
  ~0,~~ 0,~\pi i ,~ \pi i ,~ \pm\frac{\hbar}{2}, ~ \pm\frac{\hbar}{2}+\pi i &~~  k=2\ell \ ,\\ 
 ~\hbar,-\hbar,~\pi i ,~ \pi i ,~ \pm\frac{\hbar}{2},~ \pm\frac{\hbar}{2}+\pi i &~~ k=2\ell+1\ .

\end{cases}
\ee 
At this specialization, the minus sector of the partition function drops because $\mathcal{Z}_{\mathrm{Sp}(N), 2\ell+\chi}^{N_f=8,-}$ vanishes due to \eqref{Sp-fund}. 
Then, it is straightforward to verify that the integrands agree (up to a factor) at this specialization
\begin{equation}
\mathcal{Z}_{\mathrm{SO}({2N+8})),\ell}^{\mathrm{vec}}=\left(\frac{ \prod_{l=1}^{8}\sh( m_l)  }{\operatorname{sh}^2\left(\hbar\right) \prod_{s=1}^N \operatorname{sh}^2\left(a_s\right)}\right)^{-\chi} \mathcal{Z}_{\mathrm{Sp}(N+4), 2 \ell+\chi}^{\mathrm{vec},+} \mathcal{Z}_{\mathrm{Sp}(N+4), 2 \ell+\chi}^{N_f=8,+}\bigg|_{\eqref{Sp-to-SO-unref}}\ .
\end{equation}
Note that the residues of the degenerate pole become zero when we tune the mass parameters as \eqref{Sp-to-SO-unref} so that no multiplicity coefficients are involved 
\be 
\left(\frac{ \prod_{l=1}^{8}\sh( m_l)  }{\operatorname{sh}^2\left(\hbar\right) \prod_{s=1}^N \operatorname{sh}^2\left(a_s\right)}\right)^{-\chi}  Z^{\mathrm{Sp}({N+4})+8\textbf{F}}_{k=2\ell+\chi,+} \Bigg|_{\eqref{Sp-to-SO-unref}}=\sum_{|\vec\lambda|=\ell}
Z_{\vec\lambda}^{\mathrm{SO}({2N+8})}=Z_{\ell}^{\mathrm{SO}({2N+8})}~.
\ee 

To derive $Z^{\mathrm{Sp}({N}),\pm}_{2\ell+\chi}$ from $Z^{\mathrm{SO}({2N+8})}_{\ell}$, we specialize the four Coulomb branch parameters in $\mathrm{SO}({2N+8})$ as in \eqref{SO-to-Sp-unref}. It is straightforward to confirm the equivalence of the partition functions at the level of the ADHM integrand up to the factor \eqref{pref}. Nonetheless, we note an interesting occurrence. Naively substituting \eqref{SO-to-Sp-unref} into \eqref{SO} does not lead to \eqref{Sp} because of the multiplicity coefficients. As previously highlighted, these coefficients originate from degenerate poles.  When tuning the four Coulomb branch parameters of pure $\mathrm{SO}({2N+8})$ Yang-Mills to \eqref{SO-to-Sp-unref}, the JK poles become degenerate, resulting in the presence of multiplicity coefficients. As a result, at particular values of the Coulomb branch parameters, there is a \emph{notable jump} in the partition function, suggesting the corresponding jump in the BPS spectra:
\begin{equation}\label{SO-jump}
\begin{aligned}
Z_{\ell}^{\mathrm{SO}({2N+8})} =& \sum_{|\vec\lambda|=\ell}
Z_{\vec\lambda}^{\mathrm{SO}({2N+8})}\cr 
\xrightarrow[\textrm{BPS jump}]{a_{N+j} \textrm{ at } \eqref{SO-to-Sp-unref}}{}\quad &  \sum_{|\vec{\lambda}|=\ell} C^{\Sp}_{\vec{\lambda},\vec{a}}Z_{\vec{\lambda}}^{\SO(2N+8)} \Big|_{\eqref{SO-to-Sp-unref}}\ .
\end{aligned}
\end{equation}
Thus, exactly at the specific Coulomb branch parameter values \eqref{S-to-AS-unref}, the unrefined instanton partition function for pure $\mathrm{SO}({2N+8})$ Yang-Mills agrees with the one \eqref{SU-AS} for the corresponding sector of the pure $\mathrm{Sp}({N})$ Yang-Mills, up to the factor \eqref{pref}. In turn, this explains the finding in \cite{Nawata:2021dlk}, demonstrating that each sector of the  $\mathrm{Sp}({N})$ unrefined instanton partition function can be expressed in terms of that of $\mathrm{SO}({2N+8})$, augmented by multiplicity coefficients. 
It is also noteworthy that the BPS jumping occurs in a more intricate manner at the refined level although a detailed analysis is beyond the scope of this paper.

\section{Freezing and unfreezing in 4d and 3d theories}\label{sec:4d3d}
In this section, we extend the freezing and unfreezing processes to lower dimensions, 4d and 3d. We investigate a lower-dimensional version of such relation between an O$p^+$-plane and O$p^-+2^{p-4}\textrm{D}p$ and compute relevant physical observables that clearly show the relation. 

\subsection{4d \texorpdfstring{$\cN=2$}{N=2} SCFT and Schur index}\label{sec:4d}
In this section, we shall see the effects of freezing and unfreezing phenomena of an O6-plane with D6-branes in 4d $\cN=2$ Schur indices \cite{Gadde:2011ik}.

The 4d $\mathcal{N}=2$ superconformal  algebra $\SU(2,2|2)$ is generated by supercharges $(Q_{\alpha}^I,\overline Q_{\dot\alpha}^I)$ and their superconformal partners $(S^{\alpha}_I,\overline S^{\dot\alpha}_I)$. Relevant bosonic symmetry generators of  $\SU(2,2|2)$ here are $\left(E, j_1, j_2, R, r\right)$ where  $j_{1,2}$ signifies the angular momentum of $\mathrm{SO}(4) \simeq \mathrm{SU}(2)_1 \times \mathrm{SU}(2)_2$, while $(R, r)$ correspond to Cartan generators associated to the $\cN=2$ superconformal $R$-symmetry $\mathrm{SU}(2)_R\times \mathrm{U}(1)_r$.

The 4d  $\cN=2$  superconformal index counts the 1/8-BPS states annihilated by one supercharge and its superconformal partner, say $\overline Q_{1\dot{-}}$ and $\overline S^{1\dot{-}}$.
In other words, it counts $\overline Q_{1\dot{-}}$-cohomology states, which saturate the bound
$$
\bar\delta_{1\dot{-}}:=\{\overline S^{1\dot{-}},\overline Q_{1\dot{-}} \}=E-2j_2-2R+r=0~.
$$
Then, the 4d $\cN=2$ superconformal index is defined as
\begin{align}
\mathcal I(\frakp, \frakq, \frakt) = \Tr(-1)^\sfF\,
e^{-\beta\,\bar \delta_{{1}\dot{-}}}\,
\frakp^{j_1+j_2-r}\,
\frakq^{-j_1+j_2-r}\,
\frakt^{R+r}\, \prod_{j}\frakm_j^{f_j}\ , 
\end{align}
where  $\frakm_j$ are flavor fugacities and $f_j$ are Cartan generators of flavor symmetry in an $\cN=2$ SCFT.
It is straightforward to evaluate the 4d $\cN=2$ superconformal index for an $\cN=2$ SCFT with Lagrangian description \cite{Gadde:2009kb,Gadde:2010te}.
In particular, the contribution of a half-hypermultiplet with representation $\lambda$ to the multi-particle index gives rise to the elliptic gamma function, given by
\be
\cI^{\frac12H}(z;\frakp,\frakq,\frakt)=\prod_{i,j=0}^{\infty}\frac{1-z^{-w} \frakp^{i+1} \frakq^{j+1}/\sqrt{\frakt}}{1-z^{w}\sqrt{\frakt}\, \frakp^i \frakq^j}=: \Gamma(z^w\sqrt{\frakt})~,
\ee
where $w$ runs over the weights of the representation $\lambda$.
The 4d $\cN=2$ vector multiplet contributes
\be
\cI^{\textrm{vec}}(z;\frakp,\frakq,\frakt)=\frac{\kappa^{\textrm{rk}G} \Gamma(\tfrac{\frakp\frakq}{\frakt})^{\textrm{rk}G}}{|W_G|}\prod_{\a \in \Delta} \frac{\Gamma\left(z^\alpha \frac{\frakp \frakq}{\frakt} \right)}{\Gamma\left(z^\alpha\right)}~,\qquad \textrm{where}\qquad \kappa=(\frakp;\frakp)(\frakq;\frakq)~,
\ee
where $\Delta$ represents the set of roots associated with the gauge group $G$, and $|W_G|$ is the order of the Weyl group of $G$. 
Then, for a superconformal theory with a gauge group $G$ and half-hypermultiplets carrying representations $\lambda_i$, the full index is schematically expressed by
\begin{equation}
{I}= \oint_{\mathrm{T}^{\textrm{rk}G}} \frac{d\boldsymbol{z}}{2\pi i\boldsymbol{z}}~ \cI^{\textrm{vec}}(\boldsymbol{z};\frakp,\frakq,\frakt) \prod_{\textrm{matter}} \cI^{\frac12H}(\boldsymbol{z};\frakp,\frakq,\frakt)\ , 
\end{equation}
where the integral is performed over the maximal torus $\mathrm{T}^{\textrm{rk}G}$ of the gauge group $G$.

In the context of 4d $\cN=2$ superconformal index, there are several intriguing specializations \cite{Gadde:2011ik,Gadde:2011uv}. Of particular relevance in this paper is the Schur index, obtained by setting $\frakt=\frakq$, which enumerates the 1/4 BPS operators consisting of Higgs branch operators annihilated by $(Q_{-}^1, S^{-}_1)$ and $(\overline{Q}_{\dot{-}}^1,\overline{S}^{\dot{-}}_1)$. In the Schur limit, the contribution from the hypermultiplet simplifies to
\be
\cI^{H}=\Gamma(z^{\pm} \sqrt{\mathfrak{t}})\quad \xrightarrow{\frakt\to \frakq} \quad   \cI^{H}_{\textrm{Schur}}=\frac{1}{\theta(z \sqrt{\frakq};\frakq)}~,
\ee
and the vector multiplet contribution is reduced to
\be
\cI^{\textrm{vec}}=\frac{\kappa^{\textrm{rk}G} \Gamma(\tfrac{\frakp\frakq}{\frakt})^{\textrm{rk}G}}{|W_G|}\prod_{\alpha \in \Delta} \frac{\Gamma\left(z^{\alpha} \frac{\frakp\frakq}{\frakt}\right)}{\Gamma\left(z^\alpha\right)}\quad \xrightarrow{\frakt\to \frakq} \quad  \cI^{\textrm{vec}}_{\textrm{Schur}}=\frac{(\frakq;\frakq)^{2\textrm{rk}G}}{|W_G|}\prod_{\alpha \in \Delta}  \theta(z^{\alpha};\frakq)~.
\ee

\subsection*{SU(\texorpdfstring{$N$}{N}) gauge theories with (anti-)symmetric hypermultiplet}
Let us consider 4d $\cN=2$ SCFTs with SU($N$) gauge group and one (anti-)symmetric hypermultiplet. For an $\cN=2$ theory to be conformal, the $\beta$-function must vanish:
\be 
\beta\propto 2T(\mathbf{adj})-T(R)\ ,
\ee 
where $R$ is the representation of a half-hypermultiplet in a theory. The Dynkin indices for the representations of SU($N$) are given by
\be T(\square)=\frac12, \quad T(\mathbf{adj})=N, \quad T(\Sym)=\frac{N}2+1, \quad  T(\AS)=\frac{N}2-1~. \ee
Therefore, for $\SU(N)$ gauge theory with one symmetric (resp. antisymmetric) hypermultiplet to be superconformal, we have to add $N-2$ (resp. $N+2$) fundamental hypermultiplets. 

\begin{figure}
    \centering
    \includegraphics[width=10.1cm]{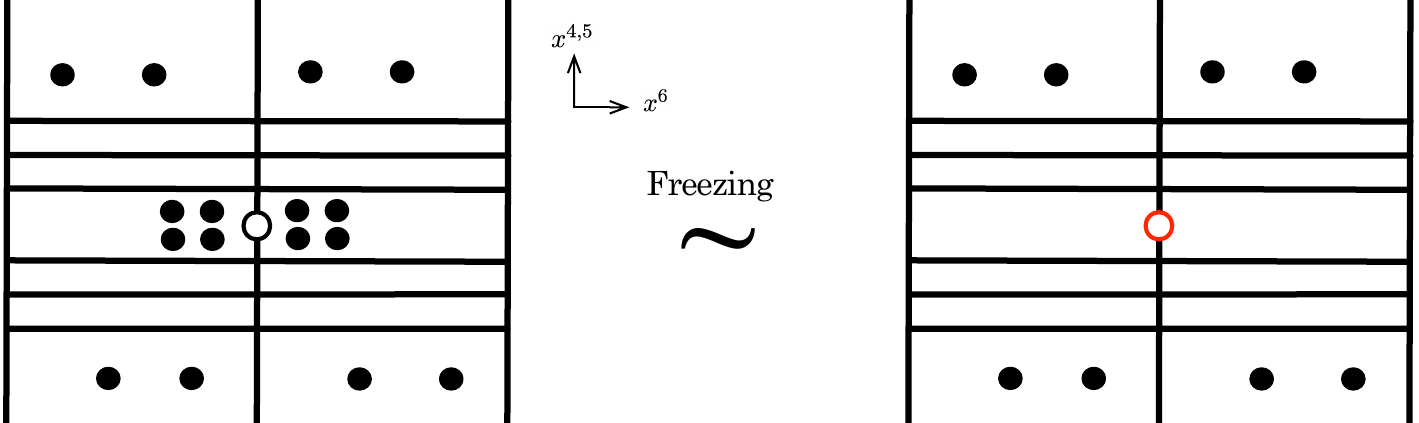}
\caption{$\mathsf{LEFT}$: A IIA brane configuration for 4d SU($N$)+1$\AS+(N+2)\mathbf{F}$ ($N=6$ in this case). $\mathsf{RIGHT}$: A brane configuration for 4d SU($N$)+1$\Sym+(N-2)\mathbf{F}$. Here, (color) D4 branes are extended along the $x^{6}$-direction, NS5-branes are extended along the $x^{4,5}$-directions, and (flavor) D6-branes  ($x^{7,8,9}$) are denoted by black dots. Also, an O6$^-$-plane is denoted by a black circle and an O6$^+$-plane is by a red circle.}
 \label{fig:4dIIAbraneO6}
\end{figure}
\begin{figure}
    \centering
    \includegraphics[width=14.1cm]{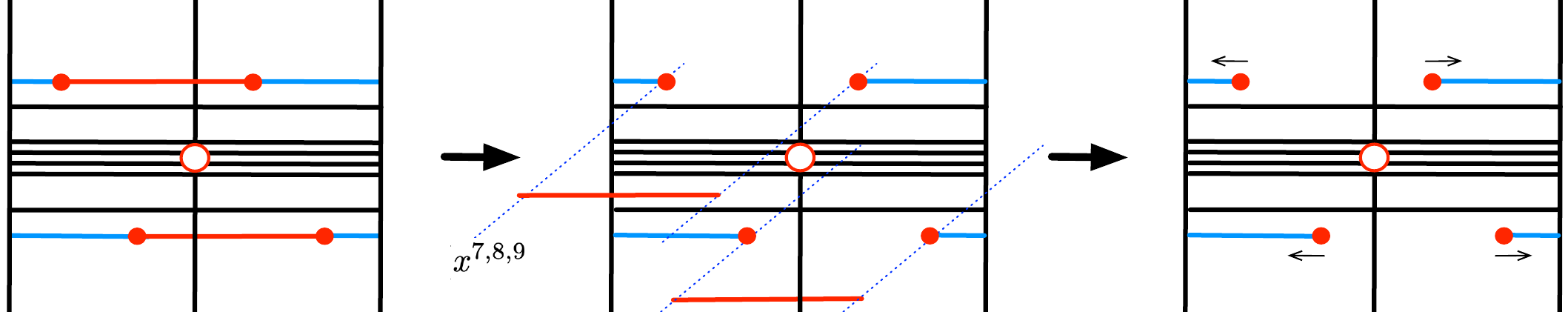}
\caption{Higgsing with $\AS/\Sym$ in IIA brane setup: $\SU(N+2)+1\AS/\Sym+2\mathbf{F}\to \SU(N)+1\AS/\Sym$. In the last figure, Hanany-Witten transitions are performed to two D6 branes along the direction of the arrows. }
 \label{fig:4dHiggsingAS}
\end{figure}

Similar to the construction of 5d gauge theories with (anti-)symmetric hypermultiplet in IIB 5-brane web diagrams, 4d SU$(N)$ gauge theories can be constructed from IIA brane setup with
NS5-, D4-, and D6-branes/O6-planes, which is a T-dual of IIB brane configuration. More precisely, SU($N$) gauge theories with $N_f$ fundamental hypermultiplets are represented by $N$ D4-branes stretched between two NS5-branes, accompanied by $N_f$ D6-branes to account for the fundamental hypermultiplets. When one (anti-)symmetric hypermultiplet is introduced, the corresponding brane setup requires an O6$^\pm$-plane, respectively.  This setup is illustrated in figure \ref{fig:4dIIAbraneO6}, where the brane arrangement is shown in the covering space, highlighting the D4/D6-branes mirrored by an O6-plane. 
The freezing here is the process in \eqref{freezing-O6}, $\textrm{O6}^-+4\textrm{D6}|_{\textrm{fixed}}\sim  \textrm{O6}^+.$ 
One then clearly sees that if the middle NS5-brane is Higgsed away, then the resulting brane configuration becomes that for Sp/SO gauge theories with only fundamental hypermultiplets. Analogous to the case of the IIB brane configuration discussed earlier, the same kind of Higgsing reducing ranks of gauge and flavor groups by two is naturally realized on this IIA brane configuration as depicted in figure \ref{fig:4dHiggsingAS}. 

\begin{figure}
    \centering
    \includegraphics[width=14.1cm]{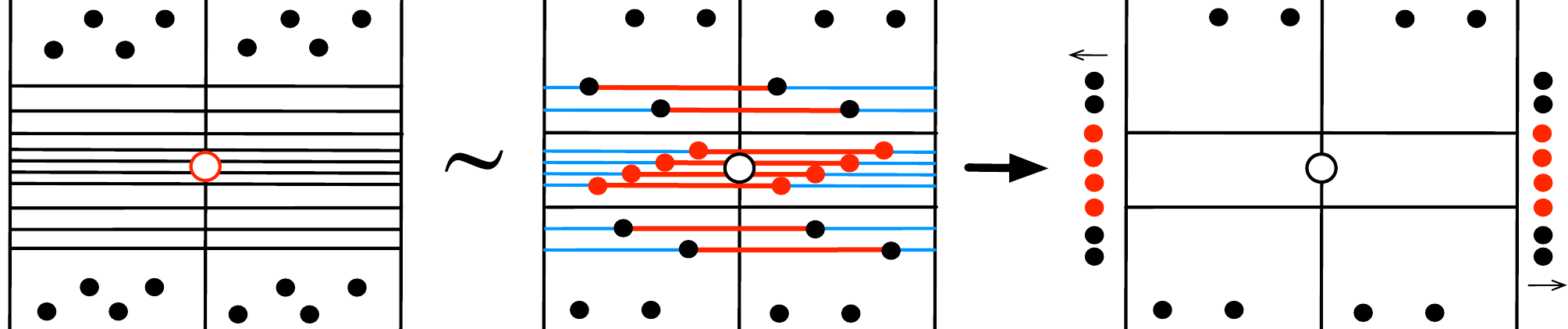}
\caption{Unfreezings for on IIA brane setup, suggesting $\SU(N+8)+1\Sym+(N+6)\mathbf{F}\to \SU(N)+1\AS+(N+2)\mathbf{F}$. Here, $N=2$. {\sf LEFT}: a brane configuration for $\SU(10)+1\Sym+8\mathbf{F}$ with a red circle denoting an O6$^+$. {\sf MIDDLE}: an $\textrm{O6}^+\sim \textrm{O6}^-+$4D6 (unfreezing) and color D4-branes are align with D6-branes to be Higgsed. {\sf RIGHT:} 8 D6-branes become free hypermultiplets by Hanany-Witten moves. The brane setup describes $\SU(2)+1\AS+4\mathbf{F}$}
 \label{fig:4dunHiggsing}
\end{figure}

The process of unfreezing in the IIA brane setup is also evident. In figure \ref{fig:4dunHiggsing}, we illustrate a brane configuration for SU$(10)+1\Sym+8\mathbf{F}$ with an O6$^+$-plane for $1\Sym$. Through unfreezing, we utilize  $\textrm{O6}^+\sim\textrm{O6}^-+4\textrm{D6}|_{\textrm{fixed}}$,  with the four D6 branes highlighted in red. At the same time, Higgsing is applied to eight color D4-branes arranged alongside eight D6-branes, subsequently relocated in the $x^{7,8,9}$ direction.  Following the Hanany-Witten transitions, one finds that the resulting configuration is a IIA brane configuration  SU$(2)+1\AS+4\mathbf{F}$, as shown on the right of the figure \ref{fig:4dunHiggsing}, where red dots represent free hypermultiplets.

Although the brane configuration illustrated in figure \ref{fig:4dunHiggsing} suggests the relation between $\SU(N+8)+1\Sym+(N+6)\mathbf{F}$ and $\SU(N)+1\AS+(N+2)\mathbf{F}$, capturing this relation within the complete BPS spectrum in 4d is far from straightforward. Our analysis reveals that only a select few BPS observables truly reflect this relation as depicted by the brane setup. The Schur index stands out as one such observable. We will now study a detailed demonstration of how the unfreezing process manifests at the level of the Schur index.

\paragraph{Schur Index. }
Based on the discussion of the Schur limit of the superconformal index earlier, we find that the Schur index for $\SU({N+8})+1\AS+({N+10})\mathbf{F}$ can be expressed as follows:
\begin{align}\label{SU-1AS-Schur}
{I}^{\SU(N+8)+1\AS+(N+10)\textbf{F}}_{\textrm{Schur}}=& \oint_{\mathrm{T}^{N+7}}\frac{d\boldsymbol{z}}{2\pi i\boldsymbol{z}} {\cI}^{\SU(N+8)+1\AS+(N+10)\textbf{F}}_{\textrm{Schur}}\\ 
=&\frac{(\frakq;\frakq)^{2(N+7)}}{(N+8) !} \oint_{\mathrm{T}^{N+7}}\frac{d\boldsymbol{z}}{2\pi i\boldsymbol{z}}\prod_{i=1}^{N+8} \prod_{j > i} \frac{\theta((z_j/z_i)^\pm)}{\theta(z_iz_j\mathfrak{m} \sqrt{\frakq})} \prod_{k=1}^{N+10} \frac{1}{\theta(z_i\mathfrak{m}_k \sqrt{\frakq})} ~ .\nonumber
\end{align}
where $\frakm$ is U(1) flavor fugacity of one antisymmetric hypermultiplet while $\frakm_k$ are the $\U(N+10)$ flavor fugacities for the fundamental hypermultiplets. 

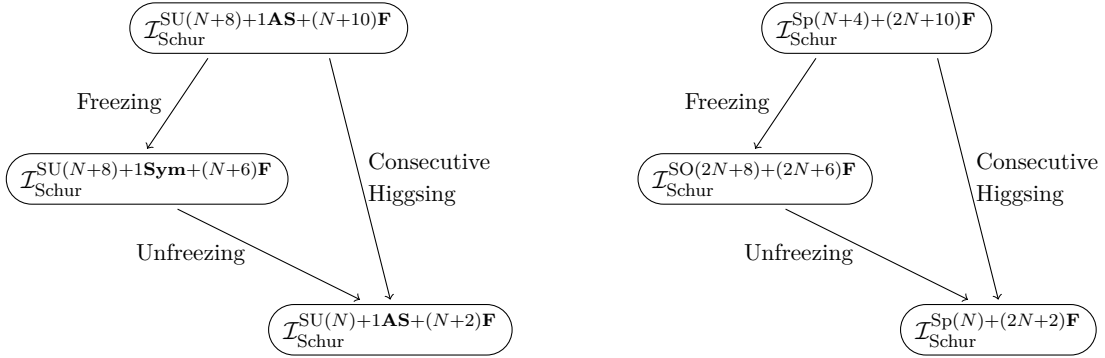
\begin{figure}[ht]
    \centering
    \begin{tikzpicture}[scale=0.8]
\node[rounded rectangle,draw,scale=0.85] at (0,0) {$\cI^{\mathrm{SU}({N+8})+1\AS+({N+10})\textbf{F}}_{\textrm{Schur}}$};
\node[rounded rectangle,draw,scale=0.85] at (2,-5) {$\cI^{\mathrm{SU}({N})+1\AS+({N+2})\textbf{F}}_{\textrm{Schur}}$};
\node[rounded rectangle,draw,scale=0.85] at (-2,-2.5) {$\cI^{\mathrm{SU}({N+8})+1\Sym+({N+6})\textbf{F}}_{\textrm{Schur}}$};
\draw[->] (-1,-.5) to node [left=0.1,scale=0.8] {Freezing} (-2,-2);
\draw[->] (1,-.5) to node [right,scale=0.8,text width=2.8cm] {Consecutive Higgsing} (2,-4.5);
\draw[->] (-1.5,-3) to node [left=0.2,scale=0.8] {Unfreezing} (1.5,-4.5);
\begin{scope}[xshift=10cm]
\node[rounded rectangle,draw,scale=0.85] at (0,0) {$\cI^{\mathrm{Sp}({N+4})+({2N+10})\textbf{F}}_{\textrm{Schur}}$};
\node[rounded rectangle,draw,scale=0.85] at (2,-5) {$\cI^{\mathrm{Sp}({N})+({2N+2})\textbf{F}}_{\textrm{Schur}}$ };
\node[rounded rectangle,draw,scale=0.85] at (-2,-2.5) {$\cI^{\mathrm{SO}({2N+8})+({2N+6})\textbf{F}}_{\textrm{Schur}}$};
\draw[->] (-1,-.5) to node [left=0.1,scale=0.8] {Freezing} (-2,-2);
\draw[->] (1,-.5) to node [right,scale=0.8,text width=2.8cm] {Consecutive Higgsing} (2,-4.5);
\draw[->] (-1.5,-3) to node [left=0.2,scale=0.8] {Unfreezing} (1.5,-4.5);
\end{scope}
\end{tikzpicture}
    \caption{Freezing, unfreezing and Higssing procedures for Schur indices. $\mathsf{LEFT}$: The Schur index of ${\mathrm{SU}({N+8})+1\AS+({N+10})\textbf{F}}$, after the freezing process as in \eqref{SCI-flavor-SU}, is equal to that of ${\mathrm{SU}({N+8})+1\Sym+({N+6})\textbf{F}}$. Furthermore, applying the unfreezing as in \eqref{mass-tuning} and \eqref{poles} identifies the  Schur index of ${\mathrm{SU}({N+8})+1\Sym+({N+6})\textbf{F}}$ with that of ${\mathrm{SU}({N})+1\AS+({N+2})\textbf{F}}$, up to a factor.\\
$\mathsf{RIGHT}$: The Schur index for $\mathrm{Sp}(N+4)+ (2N+10)\mathbf{F}$, following the freezing process as described in \eqref{SCI-flavor-SpSO}, matches the Schur index for $\mathrm{Sp}(N) + (2N+2)\mathbf{F}$. Moreover, by implementing the unfreezing processes, as detailed in \eqref{mass-tuning2} and \eqref{poles2}, we can equate the Schur index of $\mathrm{Sp}(N) + (2N+2)\mathbf{F}$ to that of $\mathrm{SO}(2N+8) + (2N+6)\mathbf{F}$, up to a factor. }
    \label{fig:SCI_triangles}
\end{figure}

In this setup, two Higgs branch operators acquire the vacuum expectation values (VEVs), which Higgses the gauge group by rank two:
\begin{equation}
\mathrm{SU}(N+2)+1 \mathbf{A S}+(N+4) \mathbf{F}\ \xrightarrow{\textrm{Higgsing}} \ \mathrm{SU}(N)+1 \mathbf{A S}+(N+2) \mathbf{F}~.
\end{equation}
The Higgsing procedure can be seen at the level of the superconformal index \cite{Gaiotto:2012xa} as follows. To move two D6 branes on a D4-brane, we tune the flavor fugacities $\mathfrak{m}_{N+9}=\frac{\mathfrak{m}}{\mathfrak{m}_{N+10}\sqrt{\frakq}}$, and take residues at the particular values of the gauge fugacities
\begin{multline}
\Big[(N+8)(N+7) (\frakq;\frakq)^{2}\prod_{i=1}^{N+8}\theta(\frakm_{N+10}/\frakm_{i})\theta(\frakm_{i}\frakm_{N+10}\sqrt{\frakq}/\frakm)\Big]^{-1}\mathcal{I}^{\SU(N+6)+1\AS+(N+8)\textbf{F}}_{\textrm{Schur}}\\
=\Res^\prime_{z_{N+7}=\frac{\mathfrak{m}_{N+10}}{\mathfrak{m}}} \Res_{z_{N+8}=\frac{1}{\mathfrak{m}_{N+10}\sqrt{\frakq}}}\mathcal{I}^{\SU(N+8)+1\AS+(N+10)\textbf{F}}_{\textrm{Schur}}(\mathfrak{m}_{N+9}=\frac{\mathfrak{m}}{\mathfrak{m}_{N+10}\sqrt{\frakq}}) ~.
\end{multline}
Here, $\Res^\prime$ indicates that we remove the divergent zero modes by hand, which are typically present in the Higgsing process, when taking the residue. As illustrated on the left side of figure \ref{fig:SCI_triangles}, by employing this procedure four times, we can reduce the gauge rank by eight through the appropriate assignment of VEVs to eight hypermultiplets.

Now let us see the freezing effect 
\eqref{freezing-O6}
at the level of the Schur index.
To fix the position of four D6-branes near the O6${}^-$-plane, we set the flavor fugacities in the Schur index \eqref{SU-1AS-Schur} as
\be \label{SCI-flavor-SU}\mathfrak{m}_{N+j}=\pm\mathfrak{m}^{\frac{1}{2}}\frakq^{\pm\frac{1}{4}}~, \ee
where $j$ ranges from 7 to 10, and here we take all possible sign combinations. 
With this specialization, it reduces to the Schur index of $\SU(N+8)$ with one symmetric and $(N+6)$ fundamental hypermultiplets
\be \label{Schur-freeze}
\cI^{\mathrm{SU}({N+8})+1\AS+({N+10})\textbf{F}}_{\textrm{Schur}} \ \xrightarrow[\eqref{SCI-flavor-SU}]{\textrm{Freezing}} \ \cI^{\mathrm{SU}({N+8})+1\Sym+({N+6})\textbf{F}}_{\textrm{Schur}}~,
\ee 
where
\begin{multline}\label{SU-1S-Schur}
{I}^{\SU(N+8)+1\Sym+(N+6)\textbf{F}}_{\textrm{Schur}}\\
=\frac{(\frakq;\frakq)^{2(N+7)}}{(N+8)!} \oint_{\mathrm{T}^{N+7}}\frac{d\boldsymbol{z}}{2\pi i\boldsymbol{z}}\prod_{i=1}^{N+8} \frac{1}{\theta(z_i^{2}\mathfrak{m} \sqrt{\frakq})}\prod_{j > i} \frac{\theta((z_j/z_i)^\pm)}{\theta(z_iz_j\mathfrak{m} \sqrt{\frakq})} \prod_{k=1}^{N+6} \frac{1}{\theta(z_i\mathfrak{m}_k \sqrt{\frakq})}  ~.
\end{multline}

To transition from the Schur index of $\SU(N+8)+1\Sym+(N+6)\textbf{F}$ to $\SU(N)+1\AS+(N+2)\textbf{F}$, we position an additional four D6-branes close to the O6${}^-$-plane. This is achieved by adjusting four flavor fugacities as follows:
\be\label{mass-tuning}
\mathfrak{m}_{N+j}=\pm\mathfrak{m}^{-\frac{1}{2}}\frakq^{\pm\frac{1}{4}}~,
\ee
where $j$ ranges from 3 to 6, and we take all possible sign combinations here. 
Then, to implement the unfreezing, we bring eight color D4-branes near the O6${}^-$-plane. This is done by taking residues in the integrand  of the Schur index for $\mathrm{SU}({N+8})+1\Sym+({N+6})\textbf{F}$ at
\be \label{poles}
z_{N+j}=\pm \mathfrak{m}^{\pm \frac{1}{2}}\frakq^{\pm \frac{1}{4}}~,\qquad  (j=1\ldots,8)
\ee 
where we remove the divergent zero modes appropriately. Consequently, we obtain the integrand of the Schur index for $\mathrm{SU}({N})+1\AS+({N+2})\textbf{F}$ up to some factor:
\be 
\cI^{\mathrm{SU}({N+8})+1\Sym+({N+6})\textbf{F}}_{\textrm{Schur}}(\mathfrak{m}_{N+j}=\pm\mathfrak{m}^{-\frac{1}{2}}\frakq^{\pm\frac{1}{4}}) \ \xrightarrow[\eqref{poles}]{\textrm{Unfreezing}} \ \cI^{\mathrm{SU}({N})+1\AS+({N+2})\textbf{F}}_{\textrm{Schur}}~.
\ee 
The entire procedure is illustrated in the left of figure \ref{fig:SCI_triangles}.

\bigskip

\paragraph{Comment on the full superconformal indices:} 
Unlike the Schur index, when we extend our analysis to the full superconformal index with  $\frakp,\frakq,\frakt$, we find that no specific flavor fugacities provide the identity for the SCFTs involving the antisymmetric hypermultiplet and the one with the symmetric hypermultiplet. Moreover, while the Macdonald index \cite{Gadde:2011uv} receives the same contributions from Higgs branch operators as the Schur index does, these theories cannot be related by the identities of the Macdonald indices. Specifically, the fugacity  $\frakt$ distinguishes these theories. It is desirable to discern why this overlap appears exclusively at the Schur index level.

\subsection*{Sp(\texorpdfstring{$N$}{N}) and SO(\texorpdfstring{$2N$}{2N}) gauge theories}
We will now compare the Schur indices between the superconformal QCD of $\Sp(N)$ and $\SO(2N)$. For $\Sp(N)$, the superconformal condition is achieved with $2N+2$ fundamental half-hypermultiplets while the $\SO(2N)$ group achieves this with $2N-2$ fundamental half-hypermultiplets.\footnote{A Higgsing of an antisymmetric/symmetric full-hypermultiplet on SU($2N$) gauge theories with (anti-) symmetric matter leads to the superconformal conditions for Sp($N$) and SO($2N$) gauge theories.} Hence, the Schur index for the superconformal QCD of $\Sp(N+4)$ is determined as follows:
\begin{equation}\label{Sp-Schur}
{I}^{\Sp({N+4})+({2N+10})\textbf{F}}_{\textrm{Schur}}=\frac{(\frakq;\frakq)^{2(N+4)}}{(N+4)!2^{N+4}} \oint_{\mathrm{T}^{N+4}}\frac{d\boldsymbol{z}}{2\pi i\boldsymbol{z}}\prod_{i=1}^{N+4}\theta( z_i^{\pm2}) \prod_{j > i}\theta( z_j^\pm z_i^\pm) \prod_{k=1}^{2N+10} \frac{1}{\theta(z_i^\pm \mathfrak{m}_k \sqrt{\frakq})} ~ .
\end{equation}

Let us Higgs the gauge group by rank one via the Hanany-Witten process after placing two D6-branes onto a D4-branes:
\begin{equation}
\Sp(N+4)+({2N+10})\textbf{F}\ \xrightarrow{\textrm{Higgsing}} \ \Sp(N+3)+({2N+8})\textbf{F}~.
\end{equation}
 Regarding the Schur index, we equate the flavor fugacities $\frakm_{2N+10}$ and $\frakm_{2N+9}$, and set the gauge fugacity to $z_{N+4} = \frakm_{2N+9}\frakq^{1/2}$:
\begin{multline}
\Big[2(N+4)(\frakq;\frakq)^2\prod_{i=1}^{2N+8}\theta(\frakm_{2N+9}^\pm/\frakm_{i})\Big]^{-1}{\cI}^{\Sp(N+3)+({2N+8})\textbf{F}}_{\textrm{Schur}}\\
=\Res^\prime_{z_{N+4}=\frakm_{2N+9}\frakq^{1/2}}{\cI}^{\Sp(N+4)+({2N+10})\textbf{F}}_{\textrm{Schur}}(\frakm_{2N+10}=\frakm_{2N+9})
\end{multline}
where $\Res^\prime$ implies the exclusion of the divergent zero modes when taking residues. Sequentially applying this Higgsing process four times ultimately leads to the $\Sp(N)+(2N+2)\textbf{F}$ theory.

The freezing effect from \eqref{freezing-O6} can be seen at the level of the Schur index. To set the positions of four D6-branes close to the O6${}^-$-plane, we specialize the four flavor fugacities in the Schur index \eqref{Sp-Schur} as follows:
\be \label{SCI-flavor-SpSO}
\mathfrak{m}_{2N+6+j}=\pm1,\pm\frakq^{\frac{1}{2}}~,  
\ee 
it reduces to the Schur index of the ${\SO(2N+8)+({2N+6})\textbf{F}}$ theory up to a factor of 2:
\be 
\cI^{\Sp({N+4})+({2N+10})\textbf{F}}_{\textrm{Schur}} \ \xrightarrow[\eqref{SCI-flavor-SpSO}]{\textrm{Freezing}} \ \cI^{\SO(2N+8)+({2N+6})\textbf{F}}_{\textrm{Schur}}~,
\ee 
where
\begin{equation}
\frac12{I}^{\SO(2N+8)}_{\textrm{Schur}}=\frac{(\frakq;\frakq)^{2(N+4)}}{(N+4)!2^{N+4}} \oint_{\mathrm{T}^N}\frac{d\boldsymbol{z}}{2\pi i\boldsymbol{z}}\prod_{i=1}^{N+4}\prod_{j > i}\theta( z_j^\pm z_i^\pm) \prod_{k=1}^{2N+6} \frac{1}{\theta(z_i^\pm \mathfrak{m}_k \sqrt{\frakq})}~ .
\end{equation}

To transition from the Schur index of $\mathrm{SO}(2N+8)+(2N+6) \mathbf{F}$ to $\mathrm{Sp}(N)+(2N+2) \mathbf{F}$, we relocate an additional four D6-branes close to the $\mathrm{O}6^{-}$-plane, and Higgs four D4-branes out. This relocation is achieved by setting four flavor fugacities as follows:
\be \label{mass-tuning2}
\mathfrak{m}_{2N+2+j}= \pm 1,\pm \mathfrak{q}^{\frac{1}{2}}~,
\ee
where $j=1,\ldots,4$.
Subsequently, we take residues in the Schur index formula of $\mathrm{SO}(2N+8)+(2N+6) \mathbf{F}$ at 
\be \label{poles2}
z_{N+j}=\pm 1,\pm \mathfrak{q}^{\frac{1}{2}}~,\qquad (j=1,\ldots,4)
\ee 
where we appropriately remove the divergent zero modes in the denominator. Consequently, we obtain the integrand of the Schur index for $\mathrm{Sp}(N)+(2N+2) \mathbf{F}$ up to an unimportant factor:
\be 
\cI^{\mathrm{SO}(2N+8)+(2N+6) \mathbf{F}}_{\textrm{Schur}}(\mathfrak{m}_{2N+2+j}= \pm 1,\pm \mathfrak{q}^{\frac{1}{2}}) \ \xrightarrow[\eqref{poles2}]{\textrm{Unfreezing}} \ \cI^{\mathrm{Sp}({N})+({2N+2})\textbf{F}}_{\textrm{Schur}}~.
\ee 
The whole relationship is depicted in the right of figure \ref{fig:SCI_triangles}.

As noted earlier, the relation holds only at the Schur indices. We emphasize that no specific tuning of flavor and gauge fugacities can extend this relation to encompass the full superconformal index involving $\frakp,\frakq,\frakt$, or its various other limits.

\subsection{3d \texorpdfstring{$\cN=4$}{N=4} sphere partition functions}\label{sec:3d}
In the context of 3d setups, the brane configurations for 3d $\cN=4$ theories are closely related to those used for 4d $\cN=2$ gauge theories. For example, as depicted in figure \ref{fig:4dIIAbraneO6} with O6-planes and D6-branes, the concept of freezing can be adapted to 3d by substituting O6-planes with O5-planes and D6-branes with D5-branes, as illustrated by 
\be\label{freezing-O5}
\OO5^-+ 2\, \textrm{D5} \Big|_{\textrm{fixed}} ~\sim ~\OO5^+~.
\ee
A notable distinction is the extension of NS5-branes along the $x^{4,5,6}$-directions in 3d configurations.
Thus, we will now turn our attention to the sphere partition function, which effectively represents the concept of freezing in 3d. Given the similarity of this procedure to those previously described, we will provide a concise overview here.
In this setting, the (un)freezing can be captured by the three-sphere partition function \cite{Kapustin:2009kz}. Since the procedure is very similar to the previous cases, our discussion here will be succinct for simplicity.

\subsection*{U(\texorpdfstring{$N$}{N}) gauge theories with (anti-)symmetric hypermultiplet}

\paragraph{Freezing} An $S^3$ partition function is evaluated by supersymmetric localization in the pioneering work \cite{Kapustin:2009kz}.
The $S^3$ partition function for $\U(N)$ gauge group with one antisymmetric and $(N_f+2)$ fundamental hypermultiplets is
\begin{equation}\label{eq:SU-1AS-S3}
Z_{S^3}^{\U(N)+1\AS+(N_f+2)\bfF}(m)=\frac{1}{N!}\int [d\mathrm{s}]\prod_{i=1}^{N}\frac{e^{2\pi i\xi  s_i}\prod_{j> i}\sh^22\pi(s_i- s_j)}{\prod_{j> i}\ch2\pi(s_i+ s_j-m) \prod_{k=1}^{N_f+2}\ch2\pi( s_i-m_k)}~,
\end{equation}
where $\xi$ is the Fayet-Iliopoulos term, $m$ is the mass of the antisymmetric and $m_k$ are those of the fundamentals. 
To incorporate the effect of freezing two D5-branes with an O5$^-$-plane \eqref{freezing-O5}, we specialize two mass parameters as 
\be \label{S3-freezing}
m_{N_f+1}=\frac{m}{2}-\frac{i}{4}~, \quad m_{N_f+2}=\frac{m}{2}+\frac{i}{4}
\ee
in the aforementioned partition function. This adjustment yields the partition function for $\U(N)$ gauge group with one symmetric and $N_f$ fundamental hypermultiplets (up to a factor of $i$):
\be 
Z_{S^3}^{\U(N)+1\AS+(N_f+2)}\ \xrightarrow[\eqref{S3-freezing}]{\textrm{Freezing}} \ Z_{S^3}^{\U(N)+1\Sym+N_f\bfF}~.
\ee 
where
\begin{multline}
    \label{eq:SU-1S-S3}
Z_{S^3}^{\U(N)+1\Sym+N_f\bfF}(m)\\=\frac{1}{N!}\int [d\mathrm{s}]\prod_{i=1}^{N}\frac{e^{2\pi i\xi  s_i}\prod_{j> i}\sh^22\pi(s_i- s_j)}{\ch2\pi(2 s_i-m)\prod_{j> i}\ch2\pi(s_i+ s_j-m) \prod_{k=1}^{N_f}\ch2\pi(s_i-m_k)}~.
\end{multline}
Here we use the addition formula
\be
\sh2\pi(2s)=\ch2\pi(s)\sh2\pi(s)~.
\ee

\paragraph{Unfreezing} Now let us turn to the process of unfreezing. To separate the two D5-branes used in the freezing process \eqref{freezing-O5}, we move two color D3-branes to the position of the D5-branes. Subsequently, through Higgsing and the Hanany-Witten transition, the D5-branes are moved away. To apply this change to the $S^3$ partition function, we can take the residues at $\frac{m}{2}\pm \frac{i}{4}$ 
\begin{align}
&\cZ_{S^3}^{\U(N-2)+1\AS+N_f\bfF}(m)\\
=&\Big[2N(N-1)ie^{-2\pi i m \xi}\prod_{k=1}^{N_f}\ch(m-2m_k)\Big] \ \Res_{z_N=\frac{m}{2}- \frac{i}{4}}\Res_{z_{N-1}=\frac{m}{2}+\frac{i}{4}}\cZ_{S^3}^{\U(N)+1\Sym+N_f\bfF}(m)~.\nonumber 
\end{align}
In this way, we can see the unfreezing at the level of the $S^3$ partition functions
\be 
\cZ_{S^3}^{\U(N)+1\Sym+N_f\bfF} \ \xrightarrow{\textrm{unfreezing}} \ \cZ_{S^3}^{\U(N-2)+1\AS+N_f\bfF}~.
\ee

\subsection*{Sp(\texorpdfstring{$N$}{N}) and SO(\texorpdfstring{$2N$}{2N}) gauge theories}

\paragraph{Freezing}  The $S^3$ partition function of 3d $\cN=4$ SQCD with a simple gauge group $G$ and $N_f$ half-hypermultiplets carrying representations $w$ can be computed by
\be
Z_{S^3}^{G+N_f \bfF}=\frac{1}{|W_G|}\lim_{\xi\to0} \int [d\mathrm{s}] \frac{e^{2\pi i\xi\sum_{i=1}^{\textrm{rk}G} s_i}\prod_{\alpha\in \Delta}\sh2\pi(\alpha\cdot s)}{\prod_{j=1}^{N_h}\prod_{w\in \cR}\ch2\pi(w\cdot s-m_j)}\ ,\ee
where $\Delta$ and $\cR$ refer to, respectively, the root system of $G$ and the weights of the representation associated with the hypermultiplet. Although a simple gauge group does not admit the Fayet-Ilioupolous term, we introduce a regulator $\xi$ and take the limit $\xi\to0$ to calculate the $S^3$ 
 partition functions \cite{Benvenuti:2011ga}. 

 The $S^3$ partition function of $\Sp(N)$ SQCD with $(N_f+2)$ fundamental  hypermultiplets is given by
\begin{align}
Z_{S^3}^{\Sp(N)+(N_f+2)\bfF}(m)=&\frac{1}{N!2^N}\lim_{\xi\to0}\int [d\mathrm{s}]\prod_{i=1}^{N} \frac{e^{2\pi i\xi  s_i}\sh^22\pi(2s_i)\prod_{j>i}\sh^22\pi(s_i\pm s_j)}{\prod_{j=1}^{N_f+2}\ch2\pi(\pm s_i-m_j)}\label{S3-Sp-int}\\
 =&\sum_{I \in C_N^{N_f+2}} \prod_{j=1}^N \frac{m_{I_j} \sh2\pi(2 m_{I_j})}{\prod_{\ell \not\in I} \sh2\pi(m_{\ell} \pm m_{I_j})} ~,\label{S3-Sp-sum}
\end{align}
where $I$ denotes a subset of $N$ distinct integers selected from $\{1, \ldots, N_f+2\}$, representing all possible combinations $C_N^{N_f+2}$.
The transition from the integral expression in the first line to the sum in the second line is facilitated by isolating the poles at
$$
s_j= \pm m_{I_j}+i \frac{2k_j+1}{2}, \quad k_j \in \mathbb{Z}_{\geq 0}
$$
and considering the symmetries of permutations $S_N$ of the set $I$. The limit $\xi \rightarrow 0$ is then applied using L'Hospital's rule to derive the summation form as shown in the second line \cite[appendix D]{Nawata:2021nse}.

Now to perform the freezing \eqref{freezing-O5}, we specialize the two mass parameters in the $\Sp(N)$ partition function \eqref{S3-Sp-int} as
\be \label{S3-freezing2}
m_{N_f+1}=0~, \qquad m_{N_f+2}=\frac{i}{2}~.
\ee 
This adjustment brings the partition function into the one for $\SO(2N)$ SQCD with $N_f$ fundamental hypermultiplets up to a factor of 2:
\begin{align}
Z_{S^3}^{\SO(2N)+N_f\bfF}(m)
=&\,\frac{1}{N!2^{N-1}}\lim_{\xi\to0}\int [d\mathrm{s}]\prod_{i=1}^N \frac{e^{2\pi i\xi  s_i}\prod_{j>i}\sh^22\pi(s_i\pm s_j)}{\prod_{j=1}^{N_f}\ch2\pi(\pm s_i-m_j)}\label{S3-SO-int}\\
=&\ 2\sum_{I \in C_N^{N_f}}\prod_{j=1}^N \frac{m_{I_j}}{ \sh2\pi(2m_{I_j}) \prod_{\ell \notin I} \sh2\pi(m_{\ell} \pm m_{I_j})} ~.
\label{S3-SO-sum}
\end{align}
This freezing can be verified directly in the integrands and also through the integrated expressions:
\be 
Z_{S^3}^{\Sp(N)+(N_f+2)\bfF}\ \xrightarrow[\eqref{S3-freezing2}]{\textrm{Freezing}} \ Z_{S^3}^{\SO(2N)+N_f\bfF}~.
\ee

\paragraph{Unfreezing} The subsequent steps are also analogous. To undo the freezing process described in \eqref{freezing-O5}, we introduce two additional D5-branes alongside two color branes in the vicinity of the $\OO5$-plane. This is followed by the processes of Higgsing and Hanany-Witten transition. For the $S^3$ partition functions, we then adjust the mass parameters to
\be \label{S3-mass2}
m_{N_f}=0~, \quad m_{N_f-1}=\frac{i}{2}~,
\ee
and fine-tune the gauge fugacities to
\be\label{S3-gauge2}
z_{N}=0~, \quad z_{N-1}=\frac{i}{2}~.
\ee 
Through these adjustments, the unfreezing processes become evident at the level of the $S^3$ partition functions, leading to the transformation
\be 
\cZ_{S^3}^{\SO(2N)+N_f\bfF} \ \xrightarrow[(\ref{S3-mass2},\ref{S3-gauge2})]{\textrm{unfreezing}} \ \cZ_{S^3}^{\Sp(N-2)+(N_f-2)\bfF}.
\ee

\paragraph{Comments on other partition functions}
In 3d, other exact partition functions are available. 
The 3d $\mathcal{N}=4$ superconformal index \cite{Razamat:2014pta} can be understood as a partition function on $S^1\times S^2$ with appropriate background fields introduced. However, there is no fine-tuning of the flavor symmetries to match the superconformal indices of the two theories compared above. Since 3d $\cN=4$ theories are endowed with $\SU(2)_H\times \SU(2)_C$ $R$-symmetry, performing topological twists \cite{Rozansky:1996bq}, one can also consider a twisted partition function on a general three-manifold. Notably, A-twisted and B-twisted indices on a Riemann surface, $S^1\times \Sigma_g$, have been studied in \cite[\S6]{Closset:2016arn} where A-twist (resp. B-twist) corresponds to a topological twist with $\SU(2)_H$ (resp. $\SU(2)_C$). Specifically,
A-twisted and B-twisted indices on $S^1\times S^2$ provide the Hilbert series of the Coulomb and Higgs branch of the 3d $\cN=4$ theory, respectively. Once more,  the twisted indices show no specialization in flavor fugacities that leads the indices of the two compared theories to be equal. This particularly indicates that the two theories have distinct Coulomb and Higgs branch moduli spaces. This distinction demonstrates that the two theories are not dual to each other, even when considering specific mass parameters and flavor fugacities. Hence, the correspondence in the $S^3$ partition function remains observational.

%%%%%%%%%%%%%%%%%%%%%%%%%%%%%
%%%%%%%%%%%%%%%%%%%%%%%%%%

\section{Conclusion}\label{sec:concl}
In this study, we introduce a fascinating link between the BPS partition functions for a family of supersymmetric gauge theories equipped with eight supercharges. These theories are characterized by brane configurations that incorporate an O$p^+$-plane across dimensions $d=5,4,3$, with $p=d+2$.

In particular, in five dimensions, we show that the partition functions tied to an O7$^+$-plane emerge from configurations with an O7$^-$-plane and eight D7-branes through a process we term ``freezing.'' This process is applicable to theories such as $\SU(N)_\kappa+1\Sym+N_f\mathbf{F}\leftarrow \SU(N)_\kappa+1\AS+(N_f+8)\mathbf{F}$ and $\SO(2N)+N_f\mathbf{F}\leftarrow \Sp(N)+(N_f+8)\mathbf{F}$, respectively for $N\ge2$ and $N_f$ up to the flavor limit of 5d Kaluza-Klein (KK) theories. We find that their refined instanton partition functions match when we apply mass specializations to the hypermultiplets, a direct outcome of freezing. A notable example is the application of freezing to the 5d KK theory $\SU(2)+8\mathbf{F}$, commonly referred to as E-string theory on a circle. At the level of partition functions, this results in the KK theory $\SU(2)_\theta+1\mathbf{Adj}$, where $\theta=0$ aligns with the 6d M-string on a circle, and $\theta=\pi$ links to the 6d $A_2$ theory on a circle with a $\mathbb{Z}_2$ outer automorphism twist.

Furthermore, we identify a novel phenomenon where the BPS instanton spectrum jumps from one theory to another through ``unfreezing,'' a reversal of the freezing process, in 5d theories. Unfreezing necessitates precise adjustments of mass and Coulomb branch parameters, followed by successive Higgsings in the partition function calculation. This leads to the emergence of degenerate poles in the JK residue integrals, subsequently resulting in the appearance of multiplicity coefficients that facilitate shifts in the partition functions. We have concretely verified that 
$\SU(N+8)_\kappa+1\Sym+8\mathbf{F}\dashrightarrow \SU(N)_\kappa+1\AS$ and $\SO(2N+8)\dashrightarrow \Sp(N)$ for the unrefined case. 

The processes of freezing and unfreezing can be applied to both 4d and 3d supersymmetric gauge theories. While the brane configurations in these dimensions may not be as visually intuitive as in the five-dimensional case, the charges associated with orientifold planes connect distinct theories. Specifically, in four dimensions, this relationship is represented by $\mathrm{O}6^+\sim\mathrm{O}6^-+4\mathrm{D}6$, and in three dimensions, by $\mathrm{O}5^+\sim\mathrm{O}5^-+2\mathrm{D}5$. We have found that, in these lower-dimensional theories, only certain physical observables agree under the processes of freezing and unfreezing. For example, under the freezing, the Schur index for a 4d $\mathcal{N}=2$ SU($N$) gauge theory with a symmetric hypermultiplet matches that of a theory with an antisymmetric hypermultiplet and $(N+2)$ fundamental hypermultiplets. Similarly, the relationship of these theories through unfreezing is depicted in figure \ref{fig:SCI_triangles}. In three dimensions, the equivalence of $S^{3}$ partition functions between 3d $\mathcal{N}=4$ U($N$) gauge theories with symmetric and antisymmetric hypermultiplets, alongside two fundamental hypermultiplets, is explicitly verified. The $S^{3}$ partition functions for Sp/SO gauge theories have also been confirmed.

\begin{figure}
    \centering
    \includegraphics[width=15.1cm]{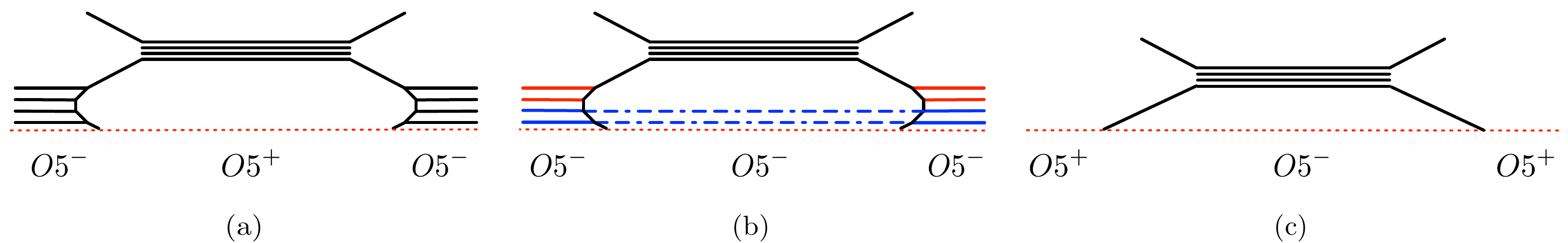}
\caption{5-brane webs with an O5-plane and freezing. (a) 5-brane web for $\Sp(N)+8\mathbf{F}$ theory.  (b) Freezing: ${\rm O}5^+\sim {\rm O}5^-+2\mathbf{F}$ denoted by blue dashed lines. (c) Together with virtual Higgsings with blue ``color'' branes and flavor branes, the freezing takes place on O5$^-$-plane with flavor D5-brane (in red), leading to an O5$^+$-plane, which results in a 5-brane web for $\SO(2N)$ theory.}
 \label{fig:O5}
\end{figure}

We remark that it is well-known that 5-brane webs for SO/Sp gauge theories can be constructed using an O5-plane instead of an O7-plane. For a single gauge group, naively a 5-brane web with an O7-plane is not much distinct from that with an O5-plane. For instance, SW curves obtained from 5-brane webs with an O7- and an O5-plane are equivalent \cite{Hayashi:2017btw, Hayashi:2023boy}. Hence, the freezing and unfreezing associated with an O5-plane can also be understood similarly. For instance, a freezing process with an O5-plane is depicted in figure  \ref{fig:O5} relating the 5d Sp($N)+8\mathbf{F}$ and the 5d SO($2N$) theory.

The study of the freezing and unfreezing in this paper raises interesting questions and several promising directions to consider. First of all, we observe that certain physical observables agree upon freezing though this agreement is not universal across all observables, as demonstrated in the 4d and 3d contexts. Given that freezing does not constitute a duality, discrepancies in observables are to be expected. Nonetheless, it is crucial to uncover any underlying principles that explain why specific observables match under freezing.

Moreover, we encounter the BPS jumping phenomenon during unfreezing. This behavior is anticipated to occur more generally when degenerate poles appear in the JK residue integrals upon varying chemical potentials.  This suggests that observing BPS jumping through JK residue integrals becomes feasible when chemical potentials, serving as physical parameters, are adjustable. A systematic examination of the conditions leading to BPS jumping and the physical mechanisms driving this phenomenon would be essential next steps.

Expanding this analysis to other dimensions presents another compelling direction for research. While our attention has primarily been on 5d instanton partition functions, it would be interesting to extend the investigation to 6d elliptic genera and Little String Theory partition functions through the viewpoint of $\mathrm{O}8^+\sim \mathrm{O}8^-+16 \mathrm{D}8$. Exploring a 2d extension, where freezing implies a relation $\mathrm{O}4^+\sim \mathrm{O}4^-+1 \mathrm{D}4$, may offer another valuable perspective. Additionally, the study of the freezing in the context of string dualities such as T-duality invites further investigation.

Another aspect worth exploring is whether the process of (un)freezing can be applied to theories and configurations with less supersymmetry.  Since the freezing process $\OO p^+ \sim \OO p^- +2^{p-4}~ \textrm{D}p$ can be implemented locally within brane configurations, it is feasible to decrease the amount of supersymmetry by altering global configurations or incorporating additional branes. Investigating the process of (un)freezing in more general brane setups would also be an interesting direction for future research.

\acknowledgments
We thank Jin Chen, Hirotaka Hayashi, Chiung Hwang, Jiaqun Jiang, Hee-Cheol Kim, Minsung Kim, Kimyeong Lee, Yiwen Pan, Marcus Sperling, Kaiwen Sun, Hao Wang, Yi-Nan Wang, Xin Wang, Piljin Yi, Gabi Zafrir, Xinyu Zhang, Jiahao Zheng, Zhenghao Zhong and Rui-Dong Zhu for valuable discussions. 
SK thanks KIAS for hosting ``Autumn Symposium on String Theory 2022,'' ``Recent Trends in Supersymmetric Field Theories 2023'' where part of the work is presented and the organizers of ``2023 East Asia Joint Workshop on Fields and Strings'' for the invitation and hospitality. SK, XL, and FY thank Fudan university for hosting ``String Theory and Quantum Field Theory 2024'' where part of the work is done. SK is supported by NSFC grant No. 12250610188. XL is supported by NSFC grant No. 11501470, No. 11426187, No. 11791240561, the Fundamental Research Funds for the Central Universities 2682021ZTPY043 and partially supported by NSFC grant No. 11671328.
The work of SN was supported by NSFC Grant No.11850410428 and Shanghai Foreign Expert grant No. 22WZ2502100.  FY is supported by the NSFC grant No. 11950410490 and by Start-up research grant YH1199911312101 and also partially supported by the Fundamental Research Funds for the Central Universities 2682021ZTPY043. 

\appendix

\section{Notation and convention}\label{app:notations} 

In this appendix, we outline the specific notations and conventions employed throughout this paper. 

First of all, throughout the paper, we denote the product (resp. sum) of all possible sign combinations for multiplicative (resp. additive) parameters in the following expressions:
\be
\begin{aligned}
f(\fraka^{\pm} \frakb^{\pm}) & =f(\fraka\frakb) f(\fraka^{-1}\frakb) f(\fraka\frakb^{-1}) f(\fraka^{-1}\frakb^{-1}), \\
g(\pm a \pm b) & =g(a+b) g(-a+b) g(a-b) g(-a-b) ,
\end{aligned}
\ee
For mass or Coulomb branch specialization, we use the following shorthand notation 
\begin{align}
    a = \pm\e_\pm (+ \pi i)\quad \Leftrightarrow\quad a= \e_+,\, \e_+ +\pi i ,\,- \e_+,\,- \e_+ +\pi i ,\, \e_-,\, \e_- +\pi i ,\, -\e_-,\, -\e_- +\pi i  \ . 
\end{align}

For various SU, Sp, and SO gauge theories, we frequently use the following concise notations for matter hypermultiplet representations:
        \begin{itemize}[nosep]
            \item Fundamental representation: $\mathbf{F}$
            \item Antisymmetric representation: $\AS$
            \item Symmetric representation: $\Sym$
        \end{itemize}

    \paragraph{Young diagrams}:
    
A Young diagram $\lambda$ is defined by a non-increasing sequence of non-negative integers, $\lambda = (\lambda_1 \geq \lambda_2 \geq \ldots \geq \lambda_{\ell(\lambda)}>0)$. The length of a Young diagram, denoted by $\ell(\lambda)$, is the number of its non-zero rows. The total number of boxes in the Young diagram is denoted by \begin{equation}
|\lambda|=\sum_{i=1}^{\ell(\lambda)} \lambda_i~.
\end{equation} The transpose of the Young diagram $\lambda$, denoted by $\lambda^t$, is obtained by reflecting $\lambda$ along its main diagonal. For a box $x = (i, j)\in \lambda$, where $i$ is the row index and $j$ is the column index, the arm length $A_{\lambda}(x)$ and the leg length $L_{\lambda}(x)$ are defined as \be \label{arm-leg}A_{\lambda}(x) = \lambda_i - j \qquad L_{\lambda}(x) = \lambda^t_j - i\ee 

\begin{figure}[ht]
    \centering
    \includegraphics[width=5cm]{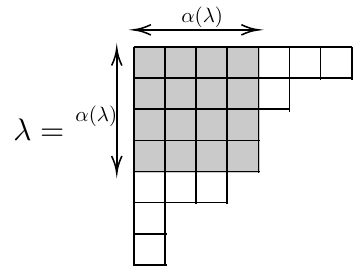}\qquad 
    \includegraphics[width=5cm]{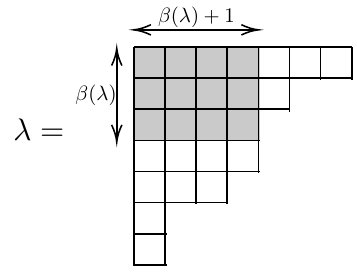}
    \caption{$\alpha(\lambda)$ is the number of rows with $\lambda_i\ge i$ while
 $\beta(\lambda)$ is the number of rows with $\lambda_i\ge i+1$. For $\lambda=(7,5,4,4,3,1,1)$, $\alpha(\lambda)=4$ and $\beta(\lambda)=3$.}
    \label{fig:alpha-beta}
\end{figure}

We also denote that $\alpha(\lambda)$ is the number of rows with $\lambda_i\ge i$ while
 $\beta(\lambda)$ is the number of rows with $\lambda_i\ge i+1$:
\be \label{alpha-beta}
\alpha(\lambda)=\max(i\mid \lambda_i\ge i)~,\qquad \beta(\lambda)=\max(i\mid \lambda_i\ge i+1)~.
\ee 
(See figure \ref{fig:alpha-beta} for an example.)

For an expression of an instanton partition function, we often use the following notation of the $P$-tuple Young diagrams with $k$ a total number of boxes:
        \[
        \vec{\lambda}=(\lambda^{(1)}, \ldots, \lambda^{(P)})~,
        \]
and
        \[
        |\vec{\lambda}|:=\sum_{s=1}^{P}|\lambda^{(s)}|=k~.
        \]

    \paragraph{Partition Functions}:

In our notation, the integrand of a partition function is denoted in calligraphic font, while the resultant partition function is represented in standard lettering. For example, a $k$-instanton partition function is expressed as 
\be 
Z_k^{\cT}=\oint_\textrm{JK} \prod_I \frac{d\phi_I}{2\pi i}~ \cZ^{\cT}_k(\boldsymbol{\phi})\ , 
\ee 
where $\cT$ symbolizes the theory under consideration. 
An instanton partition function is defined on the $\Omega$-background \cite{Nekrasov:2002qd}, and we use the standard notation $\e_{1,2}$ to represent the $\Omega$-deformation parameters. We also introduce the notation $2 \epsilon_{ \pm}=\epsilon_1 \pm \epsilon_2$, with the unrefined limit being defined as $\e_1=-\e_2=\hbar$ (therefore $\epsilon_{+}=0,~ \epsilon_{-}=\hbar$) For the 5d instanton partition function, it is also helpful to introduce the notations
\be
        \sh(x):=e^{\frac{x}{2}}-e^{-\frac{x}{2}},\qquad \ch(x):=e^{\frac{x}{2}}+e^{-\frac{x}{2}}~,
 \ee
which satisfy the relation
        \be\label{eq:ch-sh-identity}
        \ch(x) = -i \sh( x+\pi i) ~, \qquad \sh(2x)=\sh(x)\ch(x)\ .
        \ee

For a superconformal index, the integrand is also denoted in calligraphic type, and the index post-integration in standard lettering:
\begin{equation} 
I^\cT=\oint_\mathrm{T} \prod_j \frac{dz_j}{2\pi i z_j}~ \cI^{\cT}(\boldsymbol{z}).
\end{equation} 
We use fraktur font for fugacities in the superconformal index. The integrand $\cI$ is represented by the elliptic Gamma function, defined as
\begin{equation}\label{EGF}
\Gamma(z;\frakp, \frakq)=\prod_{m,n=0}^{\infty} \frac{1-\frakp^{m+1} \frakq^{n+1} / z}{1-\frakp^m \frakq^n z}=\operatorname{PE}\bigg[
    \frac{z - \frac{\frakp\frakq}{z}}{(1 - \frakq)(1-\frakp)}
    \bigg].
\end{equation}
The $\frakq$-Pochhammer symbol is defined as
\[
(x ; \frakq) :=\prod_{i=0}^{\infty}(1-x \frakq^i)\ ,
\]
and the theta function as
\[
\theta(x ; \frakq)=(x ; \frakq)(\frakq x^{-1} ; \frakq)\ .
\]

\section{ADHM contour integrals}\label{app:ADHM}

This appendix lists contributions to the instanton partition functions from various fields in 5d theories with classical gauge groups. Specifically, 5d supersymmetric theories on the $\Omega$-background effectively localize to supersymmetric quantum mechanics on instanton moduli spaces. These spaces are described by the Atiyah-Drinfeld-Hitchin-Manin (ADHM) construction \cite{Atiyah:1978ri} for classical groups. Therefore, the contributions of various fields to the instanton partition functions are derived from the ADHM descriptions of these instanton moduli spaces and their associated bundles. For a more detailed explanation, we refer readers to  \cite{Moore:1997dj,Nekrasov:2002qd,Nekrasov-Shadchin,Shadchin:2005mx,Kim:2012gu,Hwang:2014uwa}.

\subsection{\texorpdfstring{$\SU(N)$}{SU(N)} gauge group}\label{app:SU}

For SU($N$) gauge group, we list the contributions from hypermultiplets in the fundamental, (anti-)symmetric, and adjoint representations:
\begin{align}\label{SU-instanton}
\cZ_{\SU(N)_\kappa,k}^{\textrm{vec}}=&~e^{\kappa \sum_{I=1}^k \phi_I} \frac{\prod_{I \neq J} \sh( \phi_{I}-\phi_{J}) \cdot \prod_{I, J} \sh (2 \epsilon_{+}-\phi_{I}+\phi_{J})}{\prod_{I, J} \sh (\epsilon_{1,2}+\phi_{I}-\phi_{J}) \prod_{I=1}^{k}  \prod_{s=1}^{N} \sh (\epsilon_{+} \pm( \phi_{I} - a_{s}))}\cr
 \cZ_{\SU(N),k}^{N_f}(m_l)=&\prod_{l=1}^{N_f}\prod_{I=1}^k\sh(\phi_I+ m_l) \cr
\cZ_{\SU(N),k}^{\textrm{sym}}(m)=& \prod_{I=1}^{k} \sh{(2\phi_I+ m \pm \e_-)} \prod_{s=1}^{N} \sh{(\phi_I + a_s + m)} \prod_{I<J}^{k} \frac{\sh{(\phi_I + \phi_J +m\pm \e_-)}}{\sh{(-\epsilon_+ \pm(\phi_I + \phi_J+ m)) }}\cr
\cZ_{\SU(N),k}^{\textrm{anti}}(m)=& \prod_{I=1}^{k}\frac{ \prod_{s=1}^{N} \sh{(\phi_I + a_s + m)}}{ \sh{(-\epsilon_+\pm(2\phi_I+ m))}} \prod_{I<J}^{k} \frac{\sh{(\phi_I + \phi_J +m\pm \e_-)}}{\sh{(-\epsilon_+ \pm(\phi_I + \phi_J+ m))}}\cr 
\cZ_{\SU(N), k}^{\textrm{adj}}(m)=&\prod_{I=1}^{k} \frac{\sh{(\epsilon_-\pm m)}\prod_{s=1}^{N} \sh{(m\pm(\phi_I - a_s))} }{\sh{(\epsilon_+\pm m)}  } \prod_{I < J}^{k} \frac{\sh{(\epsilon_-\pm m \pm(\phi_I - \phi_J))}}{\sh{(\epsilon_+\pm m \pm(\phi_I - \phi_J))}}\cr
\end{align}

\subsection{\texorpdfstring{$\SO(n)$}{SO(n)} gauge group}\label{app:SO}
In the case of the SO($n$) gauge group, the nature of contributions depends on whether $n$ is even or odd. We represent $n=2N+\chi$, where $n\equiv \chi$ mod 2. Then, we can write the contributions from hypermultiplets in the fundamental, adjoint (antisymmetric), and symmetric representations:
\begin{align}
&\cZ_{\SO(n),k}^{\textrm{vec}}\cr=&
\frac{1}{2^{k}\,k!}\frac{\sh^{k}(2\epsilon_{+})}{\sh^{k}(\epsilon_{1,2})}
\prod_{I=1}^{k}\frac{\sh(2\epsilon_{+}\pm2\phi_{I})\,
\sh(\pm2\phi_{I})}{\sh^\chi(\epsilon_{+}\pm\phi_{I})\prod_{s=1}^{N}\sh(\epsilon_{+}\pm\phi_{I}\pm
a_{s})}  \prod_{I<J}\frac{\sh(2\epsilon_{+}\pm\phi_{I}\pm\phi_{J})\,
\sh(\pm\phi_{I}\pm\phi_{J})}{\sh(\epsilon_{1,2}\pm\phi_{I}\pm\phi_{J})}\cr
&\cZ_{\SO(n),k}^{N_f}(m_{l})=\prod_{l=1}^{N_{f}} \prod_{I=1}^{k} \operatorname{sh}\left(\pm \phi_{I}+m_{l}\right)\cr 
&\cZ_{\SO(n),k}^{\textrm{adj}}(m)=\prod_{I=1}^{k} \frac{\sh{(\pm m - \epsilon_-)} \sh^\chi(m\pm \phi_{I})  \prod_{s=1}^{\lfloor\frac{n}{2}\rfloor} \sh{(\pm\phi_I \pm a_s + m)} }{\sh{(\pm m - \epsilon_+)} \sh{(\pm2\phi_I\pm m - \epsilon_+)} } \prod_{I < J}^{k} \frac{\sh{(\pm\phi_I \pm \phi_J \pm m - \epsilon_-)}}{\sh{(\pm\phi_I \pm \phi_J \pm m - \epsilon_+)}}\cr
&\cZ_{\SO(n),k}^{\textrm{sym}}(m)=\prod_{I=1}^{k} \frac{\sh{(\pm m - \epsilon_-)} \sh{(\pm2\phi_I\pm m - \epsilon_-)}\sh^\chi(m\pm \phi_{I})\prod_{s=1}^{\lfloor\frac{n}{2}\rfloor} \sh{(\pm\phi_I \pm a_s + m)} }{\sh{(\pm m - \epsilon_+)}} \cr 
&\quad\qquad \qquad \quad \cdot\prod_{I < J}^{k} \frac{\sh{(\pm\phi_I \pm \phi_J \pm m - \epsilon_-)}}{\sh{(\pm\phi_I \pm \phi_J \pm m - \epsilon_+)}}
\end{align}

\subsection{\texorpdfstring{$\Sp(N)$}{Sp(N)} gauge group}\label{app:Sp}

For $\Sp(N)$ gauge group, the instanton partition functions depend on the discrete $\theta$-angle in general. This is rooted in the topological property that $\pi_{4}(\operatorname{Sp}(N))=\mathbb{Z}_{2}$ \cite{Morrison:1996xf,Douglas:1996xp}. Such a property is manifested in the way supersymmetric quantum mechanics is described by the $\OO(k)^\pm$ gauge group which has two disconnected components. Denoting these respective contributions as $Z^{\pm}_{k}$, a choice of taking the sum or difference of $Z^{\pm}_{k}$ corresponds to $\theta=0$ or $\theta=\pi$. Thus, the $k$-instanton partition function with trivial $\theta$-angle is given by
$$
Z^{\theta=0}_{k}=\frac{Z^{+}_{k}+Z^{-}_{k}}{2},
$$
whereas that with non-trivial $\mathbb{Z}_{2}$ element is given by
$$
Z^{\theta=\pi}_{k}=(-1)^{k} \frac{Z^{+}_{k}-Z^{-}_{k}}{2}~.
$$

Therefore, in the following, we enumerate contributions from both plus and minus sectors for hypermultiplets in the fundamental, adjoint (symmetric), antisymmetric representations.
\subsection*{Vector multiplet}
\begin{footnotesize}
\begin{align}\label{Sp-vec}
  \cZ_{\Sp(N),k=2\ell+\chi}^{\textrm{vec},+}=&\frac{1}{2^{\ell-1+\chi} \ell !}
	    \left(\frac{1}{\sh{(\epsilon_{1,2})} \, \prod_{s=1}^{N} \sh{(\epsilon_+\pm a_s)}} \cdot \prod_{I=1}^{\ell} \frac{ \sh{(\pm \phi_I)}\sh{ (2\epsilon_+\pm \phi_I )}} {\sh{ (\epsilon_{1,2}\pm \phi_I)}}\right)^{\chi}\cr
&\cdot \prod_{I=1}^{\ell} \frac{\sh{(2\epsilon_+)} }{ \sh{(\epsilon_{1,2})}  \sh(\epsilon_{1,2}\pm 2\phi_{I} ) \, \prod_{s=1}^{N} \sh{ (\epsilon_+\pm \phi_{I} \pm a_s )} }
		\prod_{I < J}^{\ell} \frac{\sh(2\epsilon_{+}\pm\phi_{I}\pm\phi_{J})\,
\sh(\pm\phi_{I}\pm\phi_{J})}{\sh(\epsilon_{1,2}\pm\phi_{I}\pm\phi_{J})}\cr
\cZ_{\Sp(N),k=2\ell+1}^{\textrm{vec},-}=& \frac{1}{2^{\ell }\ell !}\frac{1}{\sh{ (\epsilon_{1,2})} \, \prod_{s=1}^{N} \ch{ (\epsilon_+\pm a_s )}} \cdot \prod_{I=1}^{\ell} \frac{ \ch{(\pm \phi_I)}\ch{ (2\epsilon_+\pm \phi_I )}} {\ch{(\epsilon_{1,2}\pm \phi_I)}}\cr
	    &\cdot \prod_{I=1}^{\ell} \frac{\sh{(2\epsilon_+)} }{ \sh{(\epsilon_{1,2})}  \sh{ (\epsilon_{1,2}\pm 2\phi_{I} )} \, \prod_{s=1}^{N} \sh{ (\epsilon_+\pm \phi_{I} \pm a_s )} }
		\prod_{I < J}^{\ell} \frac{\sh(2\epsilon_{+}\pm\phi_{I}\pm\phi_{J})\,
\sh(\pm\phi_{I}\pm\phi_{J})}{\sh(\epsilon_{1,2}\pm\phi_{I}\pm\phi_{J})}\cr
\cZ_{\Sp(N),k=2\ell+2}^{\textrm{vec},-}
=& \frac{1}{2^{\ell}\ell!}\frac{\ch{(2\epsilon_+)}}{\sh{ (\epsilon_{1,2})} \,\sh{ (2\epsilon_{1,2})} \, \prod_{s=1}^{N} \sh{ (\pm 2a_s + 2\epsilon_+)}} \cdot \prod_{I=1}^{\ell} \frac{ \sh{(\pm 2\phi_I)}  \sh{ (4\epsilon_+\pm 2\phi_I) } } {\sh{ (2\epsilon_{1,2}\pm 2\phi_I)} } \cr
	    &\cdot \prod_{I=1}^{\ell} \frac{\sh{(2\epsilon_+)} }{ \sh{(\epsilon_{1,2} )}  \sh{ (\epsilon_{1,2}\pm 2\phi_{I} )} \, \prod_{s=1}^{N} \sh{ (\epsilon_+\pm \phi_{I} \pm a_s )} }
		\prod_{I < J}^{\ell} \frac{\sh(2\epsilon_{+}\pm\phi_{I}\pm\phi_{J})\,
\sh(\pm\phi_{I}\pm\phi_{J})}{\sh(\epsilon_{1,2}\pm\phi_{I}\pm\phi_{J})} \cr
\end{align}\end{footnotesize}

\subsection*{Fundamental hypermultiplet}
\begin{align}\label{Sp-fund}
\cZ_{\Sp(N),k=2\ell+\chi}^{N_f,+}(m_l)=&\prod_{l=1}^{N_{f}}\sh^\chi(m_{l})\prod_{I=1}^{\ell} \sh (\pm \phi_{I}+m_{l})\cr
\cZ_{\Sp(N),k=2\ell+1}^{N_f,-}(m_l)=&\prod_{l=1}^{N_{f}}\ch (m_{l}) \prod_{I=1}^{\ell} \sh( \pm \phi_{I}+m_{l})\cr
\cZ_{\Sp(N),k=2\ell+2}^{N_f,-}(m_l)=&\prod_{l=1}^{N_{f}}\sh (m_{l}) \prod_{I=1}^{\ell} \sh( \pm \phi_{I}+m_{l})\end{align}
\subsection*{Adjoint hypermultiplet}
\begin{footnotesize}
\begin{align}\label{Sp-Adj}
  \cZ^{\textrm{adj},+}_{\Sp(N),k=2\ell+\chi}=&
	   \left(  \sh{(\pm m -\epsilon_-)}\prod_{s=1}^{N} \sh{(m \pm a_s)} \prod_{I=1}^{\ell} \frac{\sh{(\pm\phi_I \pm m - \epsilon_-)}}{\sh{(\pm\phi_I \pm m - \epsilon_+)}} \right)^{\chi} \\
    &\cdot  \prod_{I=1}^{\ell} \frac{\sh{(\pm m - \epsilon_-)}  \sh{(\pm 2\phi_I \pm m - \epsilon_-)} \prod_{s=1}^{N} \sh{(\pm\phi_I \pm a_s + m)} }{\sh{(\pm m - \epsilon_+)} }\prod_{I < J}^{\ell} \frac{\sh{(\pm\phi_I \pm \phi_J \pm m - \epsilon_-)}}{\sh{(\pm\phi_I \pm \phi_J \pm m - \epsilon_+)}}\cr
\cZ^{\textrm{adj},-}_{\Sp(N),k=2\ell+1}=&
 \sh{(\pm m -\epsilon_-)}\prod_{s=1}^{N} \ch{(m \pm a_s)} \cdot \prod_{I=1}^{\ell} \frac{\ch{(\pm\phi_I \pm m - \epsilon_-)}}{\ch{(\pm\phi_I \pm m - \epsilon_+)}}  \cr
  &\cdot \prod_{I=1}^{\ell}  \frac{\sh{(\pm m - \epsilon_-)} \sh{(\pm 2\phi_I \pm m - \epsilon_+)}\prod_{s=1}^{N} \sh{(\pm\phi_I \pm a_s + m)} }{\sh{(\pm m - \epsilon_+)}} \prod_{I < J}^{\ell} \frac{\sh{(\pm\phi_I \pm \phi_J \pm m - \epsilon_-)}}{\sh{(\pm\phi_I \pm \phi_J \pm m - \epsilon_+)}}\cr
\cZ^{\textrm{adj},-}_{\Sp(N),k=2\ell+2}=& \frac{ \sh{(\pm m -\epsilon_-)}\, \sh{(\pm 2m-2\epsilon_-)}\prod_{s=1}^{N} \sh{(2m \pm 2a_s)}}{\ch{(\pm m - \epsilon_+)}} \prod_{I=1}^{\ell} \frac{\sh{(\pm2\phi_I \pm 2m - 2\epsilon_-)}}{\sh{(\pm2\phi_I \pm 2m - 2\epsilon_+)}}\cr
  &\cdot \prod_{I=1}^{\ell} \frac{\sh{(\pm m - \epsilon_-)} \sh{(\pm 2\phi_I \pm m - \epsilon_+)}  \prod_{s=1}^{N} \sh{(\pm\phi_I \pm a_s + m)} }{\sh{(\pm m - \epsilon_+)}} \cdot
    \prod_{I<J}^{\ell} \frac{\sh{(\pm\phi_I \pm \phi_J \pm m - \epsilon_-)}}{\sh{(\pm\phi_I \pm \phi_J \pm m - \epsilon_+)}}\nonumber
\end{align}\end{footnotesize}

\subsection*{Antisymmetric hypermultiplet}
\begin{footnotesize}
\begin{align}\label{Sp-anti}
  \cZ^{\textrm{anti},+}_{\Sp(N),k=2\ell+\chi}=&
	   \left( \frac{\prod_{s=1}^{N} \sh{(m \pm a_s)}}{\sh{(\pm m -\epsilon_+)}} \prod_{I=1}^{\ell} \frac{\sh{(\pm\phi_I \pm m - \epsilon_-)}}{\sh{(\pm\phi_I \pm m - \epsilon_+)}} \right)^{\chi} \prod_{I=1}^{\ell} \frac{\sh{(\pm m - \epsilon_-)} \prod_{s=1}^{N} \sh{(\pm\phi_I \pm a_s + m)} }{\sh{(\pm m - \epsilon_+)} \sh{(\pm 2\phi_I \pm m - \epsilon_+)} }\cr &\cdot \prod_{I < J}^{\ell} \frac{\sh{(\pm\phi_I \pm \phi_J \pm m - \epsilon_-)}}{\sh{(\pm\phi_I \pm \phi_J \pm m - \epsilon_+)}}\cr
\cZ^{\textrm{anti},-}_{\Sp(N),k=2\ell+1}=&
\frac{\prod_{s=1}^{N} \ch{(m \pm a_s)}}{\sh{(\pm m -\epsilon_+)}} \cdot \prod_{I=1}^{\ell} \frac{\ch{(\pm\phi_I \pm m - \epsilon_-)}}{\ch{(\pm\phi_I \pm m - \epsilon_+)}}  \frac{\sh{(\pm m - \epsilon_-)} \prod_{s=1}^{N} \sh{(\pm\phi_I \pm a_s + m)} }{\sh{(\pm m - \epsilon_+)} \sh{(\pm 2\phi_I \pm m - \epsilon_+)} } \cr
  &\cdot \prod_{I < J}^{\ell} \frac{\sh{(\pm\phi_I \pm \phi_J \pm m - \epsilon_-)}}{\sh{(\pm\phi_I \pm \phi_J \pm m - \epsilon_+)}}\cr
\cZ^{\textrm{anti},-}_{\Sp(N),k=2\ell+2}=& \frac{\ch{(\pm m - \epsilon_-)} \prod_{s=1}^{N} \sh{(2m \pm 2a_s)}}{\sh{(\pm m -\epsilon_+)}\, \sh{(\pm 2m-2\epsilon_+)}} \cr
  &\cdot \prod_{I=1}^{\ell} \frac{\sh{(\pm2\phi_I \pm 2m - 2\epsilon_-)}}{\sh{(\pm2\phi_I \pm 2m - 2\epsilon_+)}} \frac{\sh{(\pm m - \epsilon_-)} \prod_{s=1}^{N} \sh{(\pm\phi_I \pm a_s + m)} }{\sh{(\pm m - \epsilon_+)} \sh{(\pm 2\phi_I \pm m - \epsilon_+)} } \cdot
    \prod_{I<J}^{\ell} \frac{\sh{(\pm\phi_I \pm \phi_J \pm m - \epsilon_-)}}{\sh{(\pm\phi_I \pm \phi_J \pm m - \epsilon_+)}}
\end{align}\end{footnotesize}

\section{Analysis of BPS jumping}\label{app:JK}
In this supplementary appendix, we explain the method of the Jeffrey-Kirwan (JK) residue integrals  \cite{jeffrey1995localization,brion1999arrangement,szenes2003toric,Benini:2013xpa}, which is applied to instanton partition functions. The JK residue integral for an instanton partition function typically includes both non-degenerate and degenerate poles. The presence of degenerate poles leads to multiplicity coefficients and BPS jumping, as discussed in the main text. Given its importance to our study, it is beneficial to revisit the JK residue integral procedure, with a focus on handling degenerate poles. Alternatively, this appendix serves to provide detailed derivations of the multiplicity coefficients, which were not covered in \cite{Nawata:2021dlk,Chen:2023smd}.

\subsection{Lightning review of Jeffrey-Kirwan residue integrals}

An instanton partition function can be expressed as a Witten index of supersymmetric quantum mechanics on an instanton moduli space.  For supersymmetric quantum mechanics with a gauge group of rank-$k$, the instanton partition function is expressible as follows:
\begin{equation}
Z = \oint_\text{JK} \prod_{i = 1}^k \frac{d\phi_i}{2\pi i} \mathcal{Z}(\boldsymbol{\phi}) \ ,
\end{equation}
where $\phi_i$ take values of the Cartan subalgebra $\frakh$ of the gauge group, and a contour over a $k$-dimensional complex space is specified by the JK residue prescription. 
The integrand $\mathcal{Z}$ takes the form of ratios of the sine hyperbolic functions (denoted by $\sh$ here), with the poles originating from the zeros of $\sh$ in the denominator. More concretely, poles are identified by solving the equations
\be\label{JK-poles}
Q_j(\phi)+f_j(a,m,\e)=2\pi i n_j~, \qquad n_j\in\mathbb{N} \ , j= 1, 2, \ldots, k \ ,
\ee 
where $Q_j \in \frakh^*$ and $f_j(q)$ are functions dependent on  (Coulomb branch, mass, or $\Omega$-deformation) parameters. Each pole in this way corresponds to a hyperplane in $\frakh$, creating singularities in the integrand. We associate a charge $Q_j \in \frakh^*$ with each hyperplane identified by the solution of the equations \eqref{JK-poles}. A set of $k$ linearly independent charge vectors $\left\{Q_j\right\} \in \frakh^*$ constitutes a positive cone in $\mathrm{Cone}(Q) \subset \frakh^*$. Every such cone represents the intersection of $k$ hyperplanes, each associated with a non-trivial residue. Note that the range for $n_j$ in \eqref{JK-poles} is finite in $\mathbb{N}$ and is determined by the charge vectors $Q$.

Let $\mathfrak{M}_{\text{sing}}^*$ be the set of isolated points where at least $k$ linearly independent hyperplanes intersect, namely the set of $\phi_*$ solving at least $k$ linearly independent linear equations \eqref{JK-poles}. When exactly $k$ hyperplanes intersect at $\phi_*$, it is called a \emph{non-degenerate} pole. Otherwise, it is called \emph{degenerate}.
Choosing a reference vector ${\eta} \in \frakh^*$, the JK prescription chooses a contour in such a way that the integral computes the sum of residues corresponding to all cones $\mathrm{Cone}(Q)$ for which $\eta$ lies within $\mathrm{Cone}(Q)$. More concretely, we define the 
\begin{equation}
\underset{\phi=0}{\textrm{JK-Res}}\left(Q_*, \eta\right) \frac{d \phi_1 \wedge \cdots \wedge d \phi_k}{Q_{j_1}(\phi) \cdots Q_{j_k}(\phi)}=
\begin{cases}\frac{1}{\left|\det(Q)\right|} & \text { if } \eta \in \operatorname{Cone}(Q_{j_1},\ldots,Q_{j_k}) \\ 0 & \text { otherwise }\end{cases}\ .
\end{equation}
Then, the JK residues are defined by
\be 
Z=\sum_{\phi_* \in \mathfrak{M}_{\text {sing }}^*} \underset{\phi=\phi_*}{\textrm{JK-Res}}\left(Q(\phi_*), \eta\right) \cdot \cZ(\phi)~.
\ee
Note that although the nature of poles and individual JK residues may vary significantly as $\eta$ moves across different chambers, the overall integral result remains independent of the choice of $\eta$.

In the case of a non-degenerate pole, applying the above definition to evaluate the residue is quite straightforward. However, evaluating the residue at a degenerate pole is more complicated, which we will examine closely as it plays an important role in this paper. We identify an associated set of charge vectors, $Q_* = \{Q_1, \ldots, Q_n\}$, with $n > k$.

For any given $k$-sequence of linearly independent charge vectors $(Q_{j_1}, \ldots, Q_{j_k})$ from $Q_*$, we can construct a flag, denoted as $F$. This flag represents a sequence of nested subspaces within $\mathbb{R}^k$, described as:
\begin{align}
   \{0\} \subset F_1 \subset \ldots \subset F_k = \mathbb{R}^k, \quad 
   F_{\ell} = \operatorname{span}\{Q_{j_1}, \ldots, Q_{j_\ell}\}\ .
\end{align}
Although different sequences may lead to identical flags, we pick just one representative sequence. The sequence $(Q_{j_1}, \ldots, Q_{j_k})$ is commonly referred to as a basis $\mathcal{B}(F, Q_*)$ of the flag $F$ in the set $Q_*$. Note that, for a given flag $F$, the basis within $Q_*$ is not uniquely determined. Therefore, we select one basis arbitrarily for our purposes.

From each flag $F$ and its corresponding basis $\mathcal{B}(F, Q_*)$, we construct a sequence of vectors as follows:
  \begin{equation}
      \kappa(F , Q_*) := (\kappa_1, \ldots, \kappa_k)~, \qquad
      \kappa_a = \sum_{\substack{Q\in Q_*\\Q \in F_a}}Q\ .
  \end{equation}
  We also define the sign of the flag $F$, denoted as $\operatorname{sign}F$, which is the sign of the determinant of $\kappa(F, Q_*)$. For each vector sequence $\kappa(F)$, a closed cone $\textrm{Cone}(F, Q_*)$ is constructed, spanned by the vectors in $\kappa(F, Q_*)$.

With the aforementioned objects in place, the Jeffrey-Kirwan (JK) residue of a given degenerate pole $\phi_*$ is calculated as follows:
\begin{equation}\label{JK-1pole}
    \mathop{\operatorname{JK-Res}}_{\phi=\phi_*}(\eta) \cdot\mathcal{Z}
    = \sum_{F} \delta(F, \eta) \frac{\sign F}{\det \mathcal{B}(F, Q_*)} \mathop{\operatorname{Res}}_{\alpha_k = 0}\cdots
    \mathop{\operatorname{Res}}_{\alpha_1 = 0} \mathcal{Z}\bigg|_{\substack{
    Q_{j_1} (\phi) + f_{j_1} = n_{j_1}  + \alpha_1 \\
    \cdots \\
    Q_{j_k} (\phi) + f_{j_k} = n_{j_k}  + \alpha_k
    }},
\end{equation}
where the summation extends over all flags derived from $Q_*$ associated with $\phi_*$. The term $\delta(F, \eta)$ equals one if the closed-cone $\textrm{Cone}(F, Q_*)$ contains $\eta$, and zero otherwise. This definition of $\operatorname{JK-Res}$ effectively generalizes to include non-degenerate poles. Finally, for a generic choice of $\eta$, the integral is defined as:
\begin{equation}
Z=\sum_{\phi_* \in \mathfrak{M}_{\text {sing }}^*} \mathop{\operatorname{JK-Res}}_{\phi=\phi_*}(\kappa(F, Q_*),\eta) \cdot\mathcal{Z}(\phi) \ .
\end{equation}
The final result is independent of the choice of $\eta$.

\subsection{Multiplicity coefficients and BPS jumps}

To examine the emergence of degenerate poles upon tuning Coulomb branch parameters, consider the integrand of the instanton partition function. It is defined as follows for $\SU(N+8)$ with a symmetric hypermultiplet at instanton number $k$:
\begin{align}\label{SUN-1Sym}
\cZ^{\SU(N+8)_0+1\Sym}_{k}=&\frac{1}{k!}\frac{\prod_{I \neq J} \operatorname{sh}\left(\phi_I-\phi_J\right) \cdot \prod_{I, J} \operatorname{sh}\left(2 \epsilon_{+}-\phi_I+\phi_J\right)}{\prod_{I, J} \operatorname{sh}\left(\epsilon_{1,2}+\phi_I-\phi_J\right) \prod_{I=1}^k \prod_{s=1}^{N+8} \operatorname{sh}\left(\epsilon_{+} \pm\left(\phi_I-a_s\right)\right)} \\
&\times\!\prod_{I=1}^{k}\! \sh{(2\phi_I\!+ m \pm \e_-)}\! \prod_{s=1}^{N+8} \!\sh{(\phi_I\! + a_s\! + m)}\! \prod_{I<J}^{k} \!\frac{\sh{(\phi_I\! + \phi_J +m\pm \e_-)}}{\sh{(-\epsilon_+\! \pm(\phi_I\! + \phi_J\!+ m)) }} .\nonumber 
\end{align}
And for $\SU(N)$ with an antisymmetric hypermultiplet:
\begin{align}\label{SUN-1AS}
\cZ^{\SU(N)_0+1\AS}_{k}=&\frac{1}{k!}\frac{\prod_{I \neq J} \operatorname{sh}\left(\phi_I-\phi_J\right) \cdot \prod_{I, J} \operatorname{sh}\left(2 \epsilon_{+}-\phi_I+\phi_J\right)}{\prod_{I, J} \operatorname{sh}\left(\epsilon_{1,2}+\phi_I-\phi_J\right) \prod_{I=1}^k \prod_{s=1}^N \operatorname{sh}\left(\epsilon_{+} \pm\left(\phi_I-a_s\right)\right)} \cr 
& \times\prod_{I=1}^k \frac{\prod_{s=1}^N \operatorname{sh}\left(\phi_I+a_s+m\right)}{\operatorname{sh}\left(-\epsilon_{+} \pm\left(2 \phi_I+m\right)\right)} \prod_{I<J}^k \frac{\operatorname{sh}\left(\phi_I+\phi_J+m \pm \epsilon_{-}\right)}{\operatorname{sh}\left(-\epsilon_{+} \pm\left(\phi_I+\phi_J+m\right)\right)}\ .
\end{align}
For the sake of simplicity, we set the 5d Chern-Simons level to be zero $\kappa=0$ in this appendix. 
Upon taking the unrefined limit, it is straightforward to verify that the integrand of \eqref{SUN-1Sym} at the specialization \eqref{S-to-AS-unref} agrees with \eqref{SUN-1AS}.\footnote{Since the integrand includes $\sh^k(\e_+)$ in the numerator, the naive unrefined limit leads to zero. However, to derive the correct results, this term must be kept even in the unrefined limit because the JK residue procedure gives rise to a compensating $\sh^k(\e_+)$ in the denominator.}

\bigskip

Here our goal is to elucidate why we need to introduce the multiplicity coefficients in \eqref{SU-AS} when evaluating the JK residue integral of $\SU(N)_0+1\AS$ at the unrefined level. To this end, we first comment on the JK residue integral of $\SU(N+8)_0+1\Sym$, which results in the expression \eqref{SU-sym}. At the unrefined level, all poles contributing non-trivial JK residue for \eqref{SUN-1Sym} are classified by the $(N+8)$-tuple Young diagrams with a total number of boxes $|\vec\lambda|=k$ where each box $x=(i, j)\in \lambda^{(s)}$ corresponds to a pole positioned at
\begin{equation}\label{pole-location2}
\phi_s(x)=a_s+(i-j) \hbar ~.
\end{equation}
Since all the non-trivial poles at the unrefined level are non-degenerate, the JK residue can be calculated in the following manner:
\be \label{naive}
Z^{\SU(N+8)_0+1\Sym}_{\vec\lambda}=\sh^k(0)\cdot \cZ^{\SU(N+8)_0+1\Sym}_{k}(\e_1=\hbar,\e_2=-\hbar)\bigg|_{\cup_{x\in \vec\lambda}\phi_I=\phi_s(x)}~.
\ee 
In this formula, the right-hand side implies that when the pole location \eqref{pole-location2} is substituted, $k$ factors in the denominator vanish. Consequently, multiplying by $\sh^k(0)$ effectively removes these $k$ factors from the denominator, accurately determining the residue value. The expression for the resulting residue on the left-hand side is given in \eqref{SU-sym}.\footnote{At the refined level, the situation changes significantly as even degenerate poles contribute to non-trivial residues. As a result, the classification of poles cannot be solely based on sets of Young diagrams, and a closed-form expression for the refined case has not been obtained yet at present.}

However, this straightforward approach is not applicable for calculating the JK residues of $\SU(N)_0+1\AS$ \eqref{SUN-1AS}, even though the integrand can be derived from \eqref{SU-sym} by specializing the Coulomb branch parameters as in \eqref{S-to-AS-unref}. It is also necessary to incorporate the multiplicity coefficients in the formulation of the Young diagram expression, which results in the BPS jumping. In the following, we will elaborate on these aspects in the instanton partition functions of $\SU(N)_0+1\AS$.

At 1-instanton, the JK residue integral is of rank one. The evaluation is straightforward, and there is no interesting phenomenon. Therefore, our analysis begins with the 2-instanton case.

\paragraph{2-instanton} 
Consider the 2-instanton partition function for $\SU(N)_0+1\AS$, with $\eta = (1, 0.01)$ selected as the reference vector. We focus on a specific pole, described as follows
\be\label{2-poles}
\left\{\phi_1=-\frac{m}{2}-\frac{\epsilon_1}{4}+\frac{\epsilon_2}{4}, \quad \phi_2 =-\frac{m}{2}+\frac{3\epsilon_1}{4}+\frac{\epsilon_2}{4}\right\}~,
\ee 
the following two factors vanish at the pole in the denominator of the 2-instanton integrand, 
\begin{equation}\label{2-factors}
\sh(m-\epsilon_++\phi_1+\phi_2), \qquad \sh(\epsilon_1+\phi_1-\phi_2)~,
\end{equation}
whose charge vectors are
\be 
Q_*=\{(1,1),~~(1,-1) \}~.
\ee
Therefore, it is a non-degenerate pole with $|\det Q_*|=2$ and the cone formed by the charge vectors includes the reference vector $\eta$. Consequently, the JK residue at this pole can be expressed as
\begin{equation}\label{JK-2inst}
    \mathop{\operatorname{JK-Res}}_{\phi=\eqref{2-poles}}(\eta)\cdot\cZ^{\SU(N)_0+1\AS}_{k=2}
=-\frac1{4 \sh(\epsilon_1)  \sh(\epsilon_2)  \sh(2 \epsilon_2)  \sh(2\epsilon_-) \prod_i \sh(\pm\frac{\e_1}{2}-\frac{m-3\e_+}{2}-a_i) }\ ,
\end{equation}
which simplifies in the unrefined limit to
 \be\label{JK-2inst-unrefined}
-\frac1{4 \sh^2(\hbar)  \sh^2(2\hbar)  \prod_i \sh(\frac{\pm\hbar -m }{2}-a_i) }\ .
 \ee 
Furthermore, an additional pole yields an identical contribution in the unrefined limit, described as
\be\label{2-poles-2}
\left\{\phi_1=-\frac{m}{2}+\frac{\epsilon_1}{4}-\frac{\epsilon_2}{4}, \qquad \phi_2 =-\frac{m}{2}+\frac{\epsilon_1}{4}+\frac{3 \epsilon_2}{4}\right\}~.
\ee 
In the unrefined limits of \eqref{2-poles} and \eqref{2-poles-2}, the poles are located at $\frac{\pm \hbar-m}{2}$.

In our study, to perform unfreezing, we specialize the Coulomb branch parameters of the $\SU(N+8)_0+1\Sym$ instanton partition function, yielding the partition function for $\SU(N)_0+1\AS$ as \eqref{S-to-AS-unref}:
\begin{align}\label{S-to-AS-unref2}
&a_{N+1}=-\frac{m}{2}, && a_{N+2}=-\frac{m}{2}+\pi i, && a_{N+3}=-\frac{m}{2}, && a_{N+4}=-\frac{m}{2}+\pi i, \\
&a_{N+5}=\frac{ \hbar-m}{2} ,  && a_{N+6}=\frac{ \hbar-m}{2}+\pi i, && a_{N+7}=\frac{-\hbar-m}{2} ,  && a_{N+8}=\frac{- \hbar-m}{2}+\pi i~.\nonumber
\end{align}
Under this specialization, the pole locations $\phi_s(x)=\frac{\pm \hbar-m}{2}$ (see \eqref{pole-location2}) emerge from the following configurations in the ($N+8)$-tuple Young diagrams:
\begin{equation}\label{3-Young}
\renewcommand{\arraystretch}{1.2}
\begin{tabular}{|c|c|}
\hline $\vec\lambda_{*}$& $C_{\vec{\lambda},\vec{a}}^{\operatorname{anti}}$ \\
\hline 
$(\emptyset, \ldots, \emptyset, \yng(2), \emptyset, \emptyset, \emptyset ) $& 1\\\hline 
$(\emptyset, \ldots, \emptyset, \emptyset, \emptyset, \yng(1,1), \emptyset )$&1\\\hline 
$(\emptyset, \ldots, \emptyset, \yng(1), \emptyset, \yng(1), \emptyset )$&0\\\hline 
\end{tabular}
\end{equation}
Indeed, in the expression \eqref{SU-sym} over Young diagrams, each set from the above yields a contribution identical to \eqref{JK-2inst-unrefined}:
\begin{equation}
 Z_{\vec\lambda_*}^{\mathrm{SU}({N+8})_0+1\Sym} \Big|_{\eqref{S-to-AS-unref2}} = -\frac1{4 \sh^2(\hbar)  \sh^2(2\hbar)  \prod_i \sh(\frac{\pm\hbar -m }{2}-a_i) }\ .
\end{equation}
In our analysis, the accurate evaluation of JK residues reveals contributions solely from two poles, \eqref{2-poles} and \eqref{2-poles-2}. However, this contrasts with the presence of three configurations \eqref{3-Young} in the ($N+8)$-tuple Young diagrams, leading to a discrepancy in the results. This issue originates from the specialization \eqref{S-to-AS-unref2}, where the difference between $a_{N+5}=\frac{\hbar-m}{2}$ and $a_{N+7}=\frac{-\hbar-m}{2}$ is precisely $\hbar$. In the process of freezing, two D7-branes are placed by a distance of $\hbar$ near an O7${}^-$-plane. To unfreeze these D7-branes, it is necessary to position color D5-branes at an interval of $\hbar$. This positioning reinforces the $\hbar$ difference between $a_{N+5}$ and $a_{N+7}$ in specialization \eqref{S-to-AS-unref2}. To reconcile this with the JK residues and eliminate one of the three configurations in \eqref{3-Young}, we introduce multiplicity coefficients as defined in \eqref{C-anti}, which vanish at the last configuration in \eqref{3-Young}.
Then, the result is consistent with the JK residue. This can be interpreted that when color D5-branes are spaced at intervals of $\hbar$, certain instanton contributions effectively disappear.

\paragraph{3-instanton} The degenerate poles appear first at the level of 3-instanton. We pick $\eta=(1,0.01,0.001)$ as a reference vector. Let us focus on the following poles
\begin{equation}\label{3-poles}
\left\{\phi_1 =\tfrac{\epsilon_{+}-m}{2}-\epsilon_1, \quad \phi_2 = \tfrac{\epsilon_{+}-m}{2}+\epsilon_1, \quad \phi_3 = \tfrac{\epsilon_{+}-m}{2}\right\}\ .
\end{equation}
In the denominator of the 3-instanton integrand, the following four factors vanish at the poles:
\begin{equation}\label{4-factors}
\sh(m-\epsilon_++\phi_1+\phi_2), ~~\sh(\epsilon_1+\phi_1-\phi_3), ~~ \sh(\epsilon_1-\phi_2+\phi_3), ~~ \sh({m}-\epsilon_++2 \phi_3)\ .
\end{equation}
It is important to emphasize that the last pole is absent in the partition function \eqref{SUN-1Sym} of $\SU(N+8)_0+1\Sym$ at a generic value of the Coulomb branch parameters. While the other three poles are present in \eqref{SUN-1Sym}, its residue is zero due to the term $\sh(2\phi_I+ m \pm \e_-)$ in the numerator. Hence, the degenerate pole \eqref{3-poles} appears only at a specific value of the Coulomb branch parameters.

The corresponding charge vectors are
\begin{equation}
Q_*=\{(1,1, 0),~~(1, 0,-1),~~(0,-1,1),~~(0, 0, 2)\}\ .
\end{equation}
Choosing the reference vector $\eta=(1,0.01,0.002)$, only the following basis of cones contains $\eta$ among cones constructed from these charge vectors
\bea 
\cB(F, Q_*)&=((1, 0,-1),~(1,1, 0),~(0,-1,1))~,\cr
\kappa(F, Q_*)&=((1,0,-1),~(2,1,-1),~(2,0,2))~,
\eea 
where $\det \cB(F,Q_*)=2$ and $\sign F=+1$. Since the four factors \eqref{4-factors} in the denominator vanish for this triple-integral, the integral requires the evaluation of the residues at the degenerate pole \eqref{3-poles}:
\begin{multline}
\mathop{\operatorname{JK-Res}}_{\phi=\eqref{3-poles}}(\eta)\cdot\cZ^{\SU(N)_0+1\AS}_{k=3}\cr
= \bigg[12 \sh(\epsilon _1) \sh(2 \epsilon _1) \sh(3 \epsilon
   _1) \sh(\epsilon _2) \sh(2\epsilon _-)\sh(2\epsilon _1-\epsilon _2)\cr
   \times\prod_{i=1}^N \sh (2\e_+\pm\e_1-\frac{m+\e_+}{2}-a_i) \sh (2\e_+-\frac{m+\e_+}{2}-a_i) \bigg]^{-1}\ .
\end{multline}
With this choice of the reference vector, there are two additional degenerate poles
\bea 
&\left\{\phi_1=\tfrac{\epsilon_{+}-m}{2}-\epsilon_1, \quad \phi_2=\tfrac{\epsilon_{+}-m}{2}, \quad \phi_3=\tfrac{\epsilon_{+}-m}{2}+\epsilon_1\right\}\ ,\cr 
&\left\{\phi_1=\tfrac{\epsilon_{+}-m}{2}-\epsilon_1, \quad \phi_2=\tfrac{\epsilon_{+}-m}{2}+\epsilon_1, \quad \phi_3=\tfrac{\epsilon_{+}-m}{2}\right\}
\eea 
that provide the same JK residue.
Consequently, their total residue in the unrefined limit becomes
\be 
-\frac1{4\sh^2(\hbar)\sh^2(2\hbar)\sh^2(3\hbar) \prod_{i=1}^N \sh (\pm\hbar-\frac{m}{2}-a_i) \sh (-\frac{m }{2}-a_i)}\ ,
\ee 
which yields a term corresponding to $\yng(2,1)$ in \eqref{SU-AS} at one  $-m/2$ of the effective Coulomb branch parameters in \eqref{S-to-AS-unref2}. The term at the second effective Coulomb branch parameter $-m/2$ in \eqref{S-to-AS-unref2} is the total contribution from the poles
\bea 
&\left\{\phi_1 =\tfrac{\epsilon_{+}-m}{2}-\epsilon_2, \quad \phi_2 = \tfrac{\epsilon_{+}-m}{2}+\epsilon_2, \quad \phi_3 = \tfrac{\epsilon_{+}-m}{2}\right\}\ ,\cr 
&\left\{\phi_1=\tfrac{\epsilon_{+}-m}{2}-\epsilon_2, \quad \phi_2=\tfrac{\epsilon_{+}-m}{2}, \quad \phi_3=\tfrac{\epsilon_{+}-m}{2}+\epsilon_2\right\}\ ,\cr 
&\left\{\phi_1=\tfrac{\epsilon_{+}-m}{2}-\epsilon_2, \quad \phi_2=\tfrac{\epsilon_{+}-m}{2}+\epsilon_2, \quad \phi_3=\tfrac{\epsilon_{+}-m}{2}\right\}~.
\eea

\bigskip

\paragraph{4-instanton} While the evaluation of 4-instanton is quite involved, this is the first instanton number where the BPS jumping can be observed due to the presence of degenerate poles. Given its importance, it is essential to investigate the JK residues at 4-instanton in more detail. To start, we will focus on examining the following degenerate pole
\be \label{4-poles}\left\{\phi_1 = \frac{\epsilon_+-m}{2}-\epsilon_1,~~ \phi_2 = \frac{\epsilon_+-m}{2}+\epsilon_1,~~ \phi_3 = \frac{\epsilon_+-m}{2},~~ \phi_4 = \frac{\epsilon_+-m}{2}\right\}\ee
at which the following eight factors in the denominator vanish:
\begin{multline}\label{denom-8terms}
\sh(m-\epsilon_++\phi_1+\phi_2),~~ \sh(\epsilon_1+\phi_1-\phi_3),~~ \sh(\epsilon _1-\phi_2+\phi_3),~~  \sh(m-\epsilon_++2 \phi_3), \\ \sh(\epsilon _1+\phi_1-\phi_4),~~ \sh(\epsilon_1-\phi_2+\phi_4) ,~ \sh(m-\epsilon_++\phi_3+\phi_4),~ \sh(m-\epsilon_++2 \phi_4).
\end{multline}
The corresponding charge vectors are
\begin{align}\label{charge-vectors}
Q_*= \{&(1, 1, 0, 0),~~ (1, 0, -1, 0),~~ (0, -1, 1, 0),~~ (0, 0, 2, 0),\cr 
&(1, 0, 0, -1),~~ (0, -1, 0, 1), ~~(0, 0, 1, 1),~~ (0, 0, 0, 2)\}\ .
\end{align}
Note that, in the integrand \eqref{SUN-1Sym} for $\SU(N+8)_0+1\Sym$, the 4th and 8th poles are \emph{not} present at generic values of the Coulomb branch parameters. Although the remaining six poles do appear in \eqref{SUN-1Sym}, their residue becomes zero due to the term $\sh(2\phi_I+ m \pm \e_-)$ in the numerator.

Choosing $\eta=(1, 0.01, 0.002, 0.0003)$ as a reference vector, we find three flags $F_{i}$ ($i=1,2,3$) constructed from $Q_*$ such that $\eta$ lies within the Cone($F_i$, $Q_*$). The first flag $F_1$ consists of
\bea 
\cB(F_1, Q_*)=&((1, 0, -1, 0),~ (1, 1, 0, 0),~ (0, -1, 1, 0),~ (1, 0, 0, -1))\cr 
\kappa(F_1, Q_*)=&((1, 0, -1, 0),~ (2, 1, -1, 0),~ (2, 0, 2, 0),~ (3, -1, 3, 3))\ ,
\eea 
where $\det \cB(F_1, Q_*)=-2$ and $\operatorname{sign} F_1=+1$. The second flag $F_2$ consists of
\bea 
\cB(F_2, Q_*)=&((1, 0, 0, -1),~ (1, 1, 0, 0),~ (0, -1, 0, 1),~ (1, 0, -1, 0))\cr 
\kappa(F_2, Q_*)=&((1, 0, 0, -1),~ (2, 1, 0, -1),~ (2, 0, 0, 2),~ (3, -1, 3, 3))\ ,
\eea 
where $\det \cB(F_2, Q_*)=2$ and $\operatorname{sign} F_2=-1$. The third flag $F_3$ consists of
\bea 
\cB(F_3, Q_*)=&((1, 0, 0, -1), (1, 0, -1, 0), (1, 1, 0, 0), (0, -1, 1, 0))\cr 
\kappa(F_3, Q_*)=&((1, 0, 0, -1), (2, 0, -1, -1), (3, 1, -1, -1), (3, -1, 3, 3))~,
\eea 
where $\det \cB(F_3, Q_*)=2$ and $\operatorname{sign} F_3=+1$.

The contribution to the JK residue from $F_1$ can be readily computed as
\begin{multline}\label{Res-F1}
\frac{\sign F_1}{\det \cB(F_1, Q_*)}  \mathop{\operatorname{Res}}_{\alpha_4 = 0}\cdots
    \mathop{\operatorname{Res}}_{\alpha_1 = 0} \mathcal{Z}\bigg|_{\epsilon_1+\phi_1-\phi_3=\alpha_1 ,~ m-\e_++\phi_1+\phi_2=\alpha_2 ,~ \epsilon_1-\phi_2+\phi_3=\alpha_3 ,~ \epsilon_1+\phi_1-\phi_4=\alpha_4}\cr
=\sh(2 \epsilon_1+\epsilon_2)\Big[96\sh(\epsilon_1) \sh^2(2 \epsilon_1)\sh(3 \epsilon_1) \sh^2(2\epsilon_-) \sh(2 \epsilon_1-\epsilon_2)\sh^2(\epsilon_2) \cr
\times \prod_{i=1}^N \sh (2\e_+\pm\e_1-\frac{m+\e_+}{2}-a_i) \sh^2(2\e_+-\frac{m+\e_+}{2}-a_i)\Big]^{-1}\ .
\end{multline}
It is important to note that, while the eight factors \eqref{denom-8terms} in the denominator vanish at the pole \eqref{4-poles}, indicating it is a degenerate pole, four $\sh$ factors in the numerator also vanish at the pole \eqref{4-poles}, reducing the order of the pole. Consequently, the \emph{naive} (and incorrect) evaluation of the pole \eqref{4-poles} to the integrand also leads to
\begin{multline}\label{naive-res}
\sh^4(0)\cdot \mathcal{Z}^{\SU(N)_0+1\AS}_{k=4}\bigg|_{\eqref{4-poles}} \cr
=\sh(2 \epsilon_1+\epsilon_2)\Big[24\sh(\epsilon_1) \sh^2(2 \epsilon_1)\sh(3 \epsilon_1) \sh^2(2\epsilon_-) \sh(2 \epsilon_1-\epsilon_2)\sh^2(\epsilon_2) \cr
\times\prod_{i=1}^N \sh (2\e_+\pm\e_1-\frac{m+\e_+}{2}-a_i) \sh^2(2\e_+-\frac{m+\e_+}{2}-a_i)\Big]^{-1}~.
\end{multline}
Interestingly, the computed value \eqref{Res-F1} of the correct residue from $F_1$ for the degenerate pole is precisely $\frac{1}{4}$ of the value obtained from the naive evaluation mentioned above. This is indeed the origin of the multiplicity coefficient. 

Continuing to evaluate the contribution to the JK residue from $F_2$, we find 
\begin{multline}\label{Res-F2}
\frac{\sign F_2}{\det \cB(F_2, Q_*)} \mathop{\operatorname{Res}}_{\alpha_4 = 0}\cdots
    \mathop{\operatorname{Res}}_{\alpha_1 = 0} \mathcal{Z}\bigg|_{\epsilon_1+\phi_1-\phi_4=\alpha_1 ,~m-\e_++\phi_1+\phi_2=\alpha_2 ,~ \epsilon_1-\phi_2+\phi_4=\alpha_3 ,~ \epsilon_1+\phi_1-\phi_3=\alpha_4}\cr
=\sh(2 \epsilon_1+\epsilon_2)\Big[96\sh(\epsilon_1) \sh^2(2 \epsilon_1)\sh(3 \epsilon_1) \sh^2(2\epsilon_-) \sh(2 \epsilon_1-\epsilon_2)\sh^2(\epsilon_2) \cr
\times\prod_{i=1}^N \sh (2\e_+\pm\e_1-\frac{m+\e_+}{2}-a_i) \sh^2(2\e_+-\frac{m+\e_+}{2}-a_i)\Big]^{-1}\ .
\end{multline}
On the other hand, the JK residue for $F_3$ vanishes:
\begin{equation}\label{Res-F3}
\frac{\sign F_3}{\det \cB(F_3, Q_*)}  \mathop{\operatorname{Res}}_{\alpha_4 = 0}\cdots
    \mathop{\operatorname{Res}}_{\alpha_1 = 0} \mathcal{Z}\bigg|_{\epsilon_1+\phi_1-\phi_4=\alpha_1 ,~ \epsilon_1+\phi_1-\phi_3=\alpha_2 ,~m-\e_++\phi_1+\phi_2=\alpha_3 ,~ \epsilon_1-\phi_2+\phi_3=\alpha_4}=0
\end{equation}

As defined in \eqref{JK-1pole}, we sum up \eqref{Res-F1}, \eqref{Res-F2} and \eqref{Res-F3} to get the JK residue for the pole \eqref{4-poles}:
\begin{multline}
 \mathop{\operatorname{JK-Res}}_{\eqref{4-poles}}(\eta) \cdot\mathcal{Z}^{\SU(N)_0+1\AS}_{k=4}\cr    =\sh(2 \epsilon_1+\epsilon_2)\Big[48\sh(\epsilon_1) \sh^2(2 \epsilon_1)\sh(3 \epsilon_1) \sh^2(2\epsilon_-) \sh(2 \epsilon_1-\epsilon_2)\sh^2(\epsilon_2) \cr
 \times\prod_{i=1}^N \sh (2\e_+\pm\e_1-\frac{m+\e_+}{2}-a_i) \sh^2(2\e_+-\frac{m+\e_+}{2}-a_i)\Big]^{-1}~.
\end{multline}

\begin{table}[ht]
    \centering
\renewcommand{\arraystretch}{1.2}
  \begin{adjustbox}{max width=\textwidth}
\begin{tabular}{|c|c|c|}
\hline 
 poles$ :( \phi _{1} ,\phi _{2} ,\phi _{3} ,\phi _{4})$ & $ \mathcal{B}( F,Q_{*})$  & JK \\
\hline 
 $(\frac{\epsilon _{+} -m}{2} ,\frac{\epsilon _{+} -m}{2} ,\frac{\epsilon _{+} -m}{2}-\epsilon _{1} ,\frac{\epsilon _{+} -m}{2}+\epsilon _{1})$ &((2, 0, 0, 0), (1, 1, 0, 0), (0, 0, 1, 1), (-1, 0, 1, 0))  & $0$  \\
\hline 
 \multirow{2}{*}{$(\frac{\epsilon _{+} -m}{2} ,\frac{\epsilon _{+} -m}{2} -\epsilon _{1},\frac{\epsilon _{+} -m}{2} ,\frac{\epsilon _{+} -m}{2}+\epsilon _{1})$}  &((2, 0, 0,
    0), (0, 1, -1, 0), (0, 1, 0, 1), (-1, 1, 0, 0))& -$\frac14$ \\
\cline{2-3} 
  & ((2, 0, 0, 0), (0, 1, 0, 1), (-1, 1, 0, 0), (0, 1, -1, 0))  & $\frac14$ \\
  \hline 
 \multirow{2}{*}{$(\frac{\epsilon _{+} -m}{2} ,\frac{\epsilon _{+} -m}{2}-\epsilon _{1} ,\frac{\epsilon _{+} -m}{2} +\epsilon _{1},\frac{\epsilon _{+} -m}{2})$}  &  ((2, 0, 0, 0), (0,  1, 1, 0), (-1, 1, 0, 0), (0, 1, 0, -1)) & $\frac14$ \\
\cline{2-3} 
  & ((2, 0, 0, 0), (0, 1,  0, -1), (0, 1, 1, 0), (-1, 1, 0, 0)) & -$\frac14$ \\
\hline 
 &  ((1, -1, 0, 0), (1, 0, 0, 1), (0, 2, 0, 0), (1, 0, -1, 0)) & $\frac14$ \\
\cline{2-3} 
 $(\frac{\epsilon _{+} -m}{2}-\epsilon _{1} ,\frac{\epsilon _{+} -m}{2} ,\frac{\epsilon _{+} -m}{2} ,\frac{\epsilon _{+} -m}{2}+\epsilon _{1})$ &  ((1, 0, -1, 0), (1, -1, 0,    0), (1, 0, 0, 1), (0, 2, 0, 0)) & 0 \\
\cline{2-3} 
  &   ((1, 0, -1, 0), (1, 0, 0, 1), (0,  0, 2, 0), (1, -1, 0, 0)) &$\frac14$  \\
\hline 
   &  ((1, -1, 0, 0), (1, 0, 1, 0), (0, 2, 0,  0), (1, 0, 0, -1)) & $\frac14$ \\
\cline{2-3} 
 $(\frac{\epsilon _{+} -m}{2}-\epsilon _{1} ,\frac{\epsilon _{+} -m}{2} ,\frac{\epsilon _{+} -m}{2}+\epsilon _{1} ,\frac{\epsilon _{+} -m}{2})$  &   ((1, 0, 0, -1), (1, -1, 0, 0), (1, 0, 1, 0), (0, 2, 0, 0)) & 0 \\
\cline{2-3} 
  &   ((1, 0, 0, -1), (1, 0, 1, 0), (0, 0, -1, 1), (1, -1, 0, 0)) & $\frac14$ \\
\hline 
  &  ((1, 0, -1, 0), (1, 1, 0, 0), (0, -1, 1,  0), (1, 0, 0, -1)) & $\frac14$ \\
\cline{2-3} 
  $(\frac{\epsilon _{+} -m}{2} -\epsilon _{1},\frac{\epsilon _{+} -m}{2} +\epsilon _{1},\frac{\epsilon _{+} -m}{2} ,\frac{\epsilon _{+} -m}{2})$   &   ((1, 0, 0, -1), (1, 1, 0, 0), (0, -1, 0,  1), (1, 0, -1, 0))) & $\frac14$ \\
\cline{2-3} 
  &    ((1, 0, 0, -1), (1, 0, -1, 0), (1, 1, 0,  0), (0, -1, 1, 0)) & 0 \\
 \hline
\end{tabular}
\end{adjustbox}
 \caption{The first column displays the location of each pole which is a permutation of \eqref{4-poles}. The second column identifies the basis vector of a flag $F$, relevant to the pole, whose cone contains $\eta$. The third column specifies the coefficient which, when multiplied by the value in \eqref{naive-res}, yields the JK residue associated with the basis $\cB(F,Q_*)$. Notably, the last three rows of the table are associated with the flags $F_i$ ($i=1,2,3$) for the pole \eqref{4-poles}, as elaborated above.}
    \label{tab:1}
\end{table}

To identify additional cones containing $\eta=(1, 0.01, 0.002, 0.0003)$, we will explore permutations of the poles in \eqref{JK-1pole}. These permutations as well as their associated flags and residues are summarized in table \ref{tab:1}. It is noteworthy that all JK residues are proportional to the naive evaluation in \eqref{naive-res}. Accordingly, the third column of table \ref{tab:1} specifies coefficients which, upon being multiplied by the value in \eqref{naive-res}, calculate the corresponding JK residues. Consequently, the total contribution from these poles is 
\begin{multline}
    \sh(2 \epsilon_1+\epsilon_2)\Big[16\sh(\epsilon_1) \sh^2(2 \epsilon_1)\sh(3 \epsilon_1) \sh^2(2\epsilon_-) \sh(2 \epsilon_1-\epsilon_2)\sh^2(\epsilon_2) \cr\times\prod_{i=1}^N \sh (2\e_+\pm\e_1-\frac{m+\e_+}{2}-a_i) \sh^2(2\e_+-\frac{m+\e_+}{2}-a_i)\Big]^{-1}\ ,
\end{multline}
whose unrefined limit is
\begin{equation}\label{unrefine-1}
\frac1{16\sh^2(\hbar) \sh^4(2 \hbar)\sh^2(3 \hbar) \prod_{i=1}^N \sh (\pm\hbar-\frac{m}{2}-a_i) \sh^2(-\frac{m}{2}-a_i)}~.
\end{equation}

Similarly, we can evaluate the contributions from the pole 
\be\label{4-poles-2}
\left\{\phi_1 = \frac{\epsilon_+-m}{2}-\epsilon_2,~~ \phi_2 = \frac{\epsilon_+-m}{2}+\epsilon_2,~~ \phi_3 = \frac{\epsilon_+-m}{2},~~ \phi_4 = \frac{\epsilon_+-m}{2}\right\}
\ee
and its permutations. Although these poles are all degenerate, a similar cancellation of the zeros in the denominator and numerator occurs so that the pole order is still four. Consequently, the naive (and incorrect) evaluation of the residue at the pole \eqref{4-poles-2} is
\begin{multline}\label{naive-res-2}
\sh^4(0)\cdot\mathcal{Z}^{\SU(N)_0+1\AS}_{k=4}\bigg|_{\eqref{4-poles-2}} \cr 
=\sh(2 \epsilon_{2}+\epsilon_{1})\Big[24\sh(\epsilon_{2}) \sh^2(2 \epsilon_{2})\sh(3 \epsilon_{2}) \sh^2(2\epsilon_-) \sh(2 \epsilon_{2}-\epsilon_{1})\sh^2(\epsilon_{1}) \cr
\times\prod_{i=1}^N \sh (2\e_+\pm\e_2-\frac{m+\e_+}{2}-a_i) \sh^2(2\e_+-\frac{m+\e_+}{2}-a_i)\Big]^{-1}~.
\end{multline}
Since the correct evaluation procedure for these degenerate poles is the same as before, we omit the detail. Nonetheless, the total contribution from these poles is proportional to \eqref{naive-res-2}, which is
\begin{multline}\label{total-2}
\sh(2 \epsilon_{2}+\epsilon_{1})\Big[16\sh(\epsilon_{2}) \sh^2(2 \epsilon_{2})\sh(3 \epsilon_{2}) \sh^2(2\epsilon_-) \sh(2 \epsilon_{2}-\epsilon_{1})\sh^2(\epsilon_{1}) \cr
\times\prod_{i=1}^N \sh (2\e_+\pm\e_2-\frac{m+\e_+}{2}-a_i) \sh^2(2\e_+-\frac{m+\e_+}{2}-a_i)\Big]^{-1}~,
\end{multline}
whose unrefined limit is the same as \eqref{unrefine-1}:
\begin{equation}\label{unrefine-2}
\frac1{16\sh^2(\hbar) \sh^4(2 \hbar)\sh^2(3 \hbar) \prod_{i=1}^N \sh (\pm\hbar-\frac{m}{2}-a_i) \sh^2(-\frac{m}{2}-a_i)}~.
\end{equation}

We continue to proceed the evaluation of residues for poles of a similar kind. Consider the following pole and its permutations:
\be\label{4-poles-3}
\left\{\phi_1 =- \frac{\epsilon_++m}{2}, ~~\phi_2 =2\epsilon_+\! -\frac{\epsilon_++m}{2}, ~~\phi_3 = \epsilon_1- \frac{\epsilon_++m}{2}, ~~\phi_4 =\epsilon_2- \frac{\epsilon_++m}{2}\right\}\ .
\ee
This pole is also degenerate due to the vanishing of these seven factors in the denominator:
\bea 
\Big\{\sh(-m-\!\epsilon _+\!-2 \phi
   _1),~\sh(m-\epsilon _+\!+\phi _1+\phi
   _2),~\sh(\epsilon _1+\phi _1-\phi _3), ~
   \sh(\epsilon _2-\phi _2+\phi _3),\cr
   \sh(\epsilon _2+\phi _1-\phi
   _4),~~\sh(\epsilon _1-\phi _2+\phi _4),~~\sh(m-\epsilon _++\phi _3+\phi _4)\Big\}\ .
   \eea
The associated charge vectors are given by
\begin{align}
    Q_*=(~(-2, 0, 0, 0),\quad (1, 1, 0, 0),\quad  (1, 0, -1, 0),\quad  (0, -1, 1, 0),\cr
(1, 0,  0, -1),\quad  (0, -1, 0, 1),\quad  (0, 0, 1, 1)~) \ .
\end{align} 
With our chosen $\eta$, among these vectors, only one flag's cone contains $\eta$:
\bea 
\cB(F,Q_*)=~&((1, 1, 0, 0),~~ (1, 0, 0, -1),~~ (1, 0, -1, 0),~~ (-2, 0, 0, 0))\ ,\cr 
\kappa(F,Q_*)=~&((1, 1, 0, 0),~~ (2, 1, 0, -1), ~~(3, 1, -1, -1),~~ (1, -1, 1, 1))\ ,
\eea 
where $\det \mathcal{B}(F, Q_*)=-2$ and $\sign F=+1$. The JK residue for this flag is
\begin{align}\label{correct-res-3}
 &\mathop{\operatorname{JK-Res}}_{\eqref{4-poles-3}}(\eta) \cdot\mathcal{Z}^{\SU(N)_0+1\AS}_{k=4}\cr    =
    &\frac{\sign F}{\det \cB(F, Q_*)} \mathop{\operatorname{Res}}_{\alpha_4 = 0}\cdots
    \mathop{\operatorname{Res}}_{\alpha_1 = 0} \mathcal{Z}\bigg|_{m-\epsilon _++\phi _1+\phi
   _2=\alpha_1 ,~\epsilon _2+\phi _1-\phi_4=\alpha_2 ,~ \epsilon _1+\phi _1-\phi _3=\alpha_3 ,~ -m-\epsilon _+-2 \phi_1=\alpha_4}\cr
 =&-\Big[48 \sh(\epsilon _{1,2}) \sh(2 \epsilon _{1,2}) \sh(2 \epsilon_2-\epsilon _1) \sh^2(2\epsilon_-)\sh(2 \epsilon _1-\epsilon _2)\cr
   & \quad \quad \times\prod_{i=1}^N \sh(\frac{\epsilon_{+}-m}{2}-a_i)\sh(\epsilon _{1,2}+\frac{\epsilon_{+}-m}{2}-a_i)\sh(2\epsilon _{+}+\frac{\epsilon_{+}-m}{2}-a_i)\Big]^{-1}.
\end{align}
As before, a cancellation of zeros in both the denominator and numerator occurs, maintaining the pole order at four, and the naive (and incorrect) evaluation of the residue at the pole \eqref{4-poles-3} is 
\begin{multline}\label{naive-res-3}
\sh^4(0)\cdot \mathcal{Z}^{\SU(N)_0+1\AS}_{k=4} \bigg|_{\eqref{4-poles-3}}
=\Big[24 \sh(\epsilon _{1,2}) \sh(2 \epsilon _{1,2}) \sh(2 \epsilon_2-\epsilon _1) \sh^2(2\epsilon_-)\sh(2 \epsilon _1-\epsilon _2)\\
\times \prod_{i=1}^N \sh(\frac{\epsilon_{+}-m}{2}-a_i)\sh(\epsilon _{1,2}+\frac{\epsilon_{+}-m}{2}-a_i)\sh(2\epsilon _{+}+\frac{\epsilon_{+}-m}{2}-a_i)\Big]^{-1}\ . 
\end{multline}
However, the correct residue evaluation \eqref{correct-res-3} yields a value that is $-\frac12$ of this naive estimate.  As in table \ref{tab:2}, each permutation of the pole \eqref{4-poles-3} has one flag that contributes to the JK residue, and their total contribution is  
\begin{multline}
-\Big[8 \sh(\epsilon _{1,2}) \sh(2 \epsilon _{1,2}) \sh(2 \epsilon_2-\epsilon _1) \sh^2(2\epsilon_-)\sh(2 \epsilon _1-\epsilon _2)\cr
\times  \prod_{i=1}^N \sh(\frac{\epsilon_{+}-m}{2}-a_i)\sh(\epsilon _{1,2}+\frac{\epsilon_{+}-m}{2}-a_i)\sh(2\epsilon _{+}+\frac{\epsilon_{+}-m}{2}-a_i)\Big]^{-1}\ ,
\end{multline}
where the unrefined limit is
\begin{equation}\label{unrefine-3}
\frac1{8\sh^2(\hbar) \sh^4(2 \hbar)\sh^2(3 \hbar) \prod_{i=1}^N \sh (\pm\hbar-\frac{m}{2}-a_i) \sh^2(-\frac{m}{2}-a_i)}~.
\end{equation}

\begin{table}[ht]
    \centering
\renewcommand{\arraystretch}{1.2}
  \begin{adjustbox}{max width=\textwidth}
    \begin{tabular}{|c|c|c|}
\hline 
 poles$ :( \phi _{1} ,\phi _{2} ,\phi _{3} ,\phi _{4})$ & $ \mathcal{B}( F,Q_{*})$  & JK \\
\hline 
 $(-\frac{\epsilon _{+} +m}{2} ,2\epsilon _+-\frac{\epsilon _{+} +m}{2},\epsilon _{2}-\frac{\epsilon _{+} +m}{2} ,\epsilon _{1}-\frac{\epsilon _{+} +m}{2})$ &((1, 1, 0, 0), (1, 0, 0, -1), (1, 0, -1, 0), (-2, 0, 0, 0)) & -$\frac12$  \\
\hline 
 $(-\frac{\epsilon _{+} +m}{2} ,2\epsilon _+-\frac{\epsilon _{+} +m}{2},\epsilon _{1}-\frac{\epsilon _{+} +m}{2} ,\epsilon _{2}-\frac{\epsilon _{+} +m}{2})$ &     ((1, 1, 0, 0), (1, 0, 0, -1), (1, 0, -1, 0), (-2, 0, 0, 0))& -$\frac12$ \\
\hline
  $(-\frac{\epsilon _{+} +m}{2} ,\epsilon _{2}-\frac{\epsilon _{+} +m}{2} ,2\epsilon _{+}-\frac{\epsilon _{+} +m}{2},\epsilon _{1}-\frac{\epsilon _{+} +m}{2} )$ &      ((1, 0, 1, 0), (1, 0, 0, -1), (1, -1, 0, 0), (-2, 0, 0, 0))  & -$\frac12$ \\
  \hline 
  $(-\frac{\epsilon _{+} +m}{2} ,\epsilon _{2}-\frac{\epsilon _{+} +m}{2} ,\epsilon _{1}-\frac{\epsilon _{+} +m}{2} ,2\epsilon _{+}-\frac{\epsilon _{+} +m}{2})$  &       ((1, 0, 0, 1), (1, 0, -1, 0), (1, -1, 0, 0), (-2, 0, 0, 0))& -$\frac12$ \\
\hline
  $(-\frac{\epsilon _{+} +m}{2} ,\epsilon _{1}-\frac{\epsilon _{+} +m}{2} ,2\epsilon _{+}-\frac{\epsilon _{+} +m}{2},\epsilon _{2}-\frac{\epsilon _{+} +m}{2} )$&      ((1, 0, 1, 0), (1, 0, 0, -1), (1, -1, 0, 0), (-2, 0, 0, 0)) & -$\frac12$ \\
\hline 
  $(-\frac{\epsilon _{+} +m}{2} ,\epsilon _{1}-\frac{\epsilon _{+} +m}{2} ,\epsilon _{2}-\frac{\epsilon _{+} +m}{2} ,2\epsilon _{+}-\frac{\epsilon _{+} +m}{2})$ &       ((1, 0, 0, 1), (1, 0, -1, 0), (1, -1, 0, 0), (-2, 0, 0, 0)) & -$\frac12$ \\
\hline
\end{tabular}
\end{adjustbox}
 \caption{The first column displays the location of each pole that is a permutation of \eqref{4-poles-3}. The second column identifies the basis vector of a flag $F$, relevant to the pole, whose cone contains $\eta$. The third column specifies the coefficient which, when multiplied by the value in \eqref{naive-res-3}, yields the JK residue associated with the basis $\cB(F,Q_*)$. Notably, the first row of the table is associated with the pole \eqref{4-poles-3}, as discussed above.}
    \label{tab:2}
\end{table}

In the unrefined limits, the permutations of \eqref{4-poles}, \eqref{4-poles-2} and \eqref{4-poles-3} fall into a class of poles at
\be\label{4-poles-pos}
\left\{-\frac{m}{2},~\ -\frac{m}{2}+\hbar,~ -\frac{m}{2}-\hbar,~ -\frac{m}{2} \right\}~.
\ee 
 Summing up \eqref{unrefine-1}, \eqref{unrefine-2} and \eqref{unrefine-3}, we find the total contribution to the unrefined instanton partition function associated with these poles to be
\begin{equation}\label{unrefine-total}
\frac1{4\sh^2(\hbar) \sh^4(2 \hbar)\sh^2(3 \hbar) \prod_{i=1}^N \sh (\pm\hbar-\frac{m}{2}-a_i) \sh^2(-\frac{m}{2}-a_i)}~.
\end{equation}
Note that although the choice of a different reference vector for $\eta$ changes the pole structure, the final outcome remains unchanged.

\bigskip

On the other hand, examining \eqref{pole-location2}, the following configurations in the $(N+8)$-tuple Young diagrams correspond to \eqref{4-poles-pos}:
\begin{equation}\label{6-Young}
\renewcommand{\arraystretch}{1.2}
\begin{tabular}{|c|c|}
\hline $\vec\lambda_{**}$& $C_{\vec{\lambda},\vec{a}}^{\operatorname{anti}}$ \\
\hline 
$( \emptyset, \ldots, \emptyset, \yng(2,2), \emptyset, \emptyset, \emptyset,  \emptyset, \emptyset, \emptyset, \emptyset) $& -1\\\hline 
$( \emptyset, \ldots, \emptyset, \emptyset, \emptyset,\yng(2,2),  \emptyset,  \emptyset, \emptyset, \emptyset, \emptyset)$&1\\\hline 
$( \emptyset, \ldots, \emptyset, \yng(2,1), \emptyset, \yng(1), \emptyset,  \emptyset, \emptyset, \emptyset, \emptyset) $&1\\\hline $( \emptyset, \ldots, \emptyset, \yng(1), \emptyset,\yng(2,1),  \emptyset,  \emptyset, \emptyset, \emptyset, \emptyset)$&1\\\hline 
$( \emptyset, \ldots, \emptyset, \yng(2), \emptyset, \yng(1,1), \emptyset,  \emptyset, \emptyset, \emptyset, \emptyset) $&1\\ \hline $( \emptyset, \ldots, \emptyset, \yng(1,1), \emptyset,\yng(2),  \emptyset,  \emptyset, \emptyset, \emptyset, \emptyset)$&1\\ \hline
\end{tabular}
\end{equation}
Note that non-trivial Young diagrams show up at the positions 
associated to the effective Coulomb branch parameter $-\frac{m}{2}$.  When evaluating the expression \eqref{SU-sym} over these Young diagrams at the specialization \eqref{S-to-AS-unref2}, each configuration contributes:
\begin{equation}
 Z_{\vec\lambda_{**}}^{\mathrm{SU}({N+8})_0+1\Sym} \Big|_{\eqref{S-to-AS-unref2}} =\frac1{16\sh^2(\hbar) \sh^4(2 \hbar)\sh^2(3 \hbar) \prod_{i=1}^N \sh (\pm\hbar-\frac{m}{2}-a_i) \sh^2(-\frac{m}{2}-a_i)}~.
\end{equation}
Thus, since there are six configurations in \eqref{6-Young}, simply specializing the Coulomb branch parameters as in \eqref{S-to-AS-unref2} does not bring the expression \eqref{SU-sym} for $\SU(N+8)_0+1\Sym$ to the partition function of $\SU(N)_0+1\AS$. Nevertheless, once we incorporate the multiplicity coefficients, the top two configurations in \eqref{6-Young} effectively cancel each other out, and only the remaining four consequently contribute, yielding the correct value \eqref{unrefine-total}.

\paragraph{Conclusion}
In summary, the reason why the BPS jumping occurs at a specific value of Coulomb branch parameters can be attributed to the following two reasons. 

First, when there is a difference of $\hbar$ in the Coulomb branch parameters in $\SU(N+8)_0+1\Sym$, there may be a mismatch between the number of poles in the JK residue integral and those represented in the Young diagram sum expression. This requires the introduction of multiplicity coefficients, as we have observed in the 2-instanton level analysis. However, it is worth noting that multiplicity coefficients of this kind can often be mitigated by massaging the formula, like 
Eq. (2.12) in \cite{Chen:2023smd}.

Second, degenerate poles contribute even at the unrefined level in the instanton partition functions of $\SU(N)_0+1\AS$. Namely, the presence of degenerate poles at specific Coulomb branch values in $\SU(N+8)_0+1\Sym$ is a more fundamental reason behind the BPS jumping.  Despite being degenerate, these poles maintain an order equal to the instanton number due to the cancellation of zeros in both the numerator and the denominator of the integrand. Consequently, the JK residue at a degenerate pole is proportional to the naive estimate of the residue as done in \eqref{naive} (or \eqref{naive-res}, \eqref{naive-res-2}, \eqref{naive-res-3}). Additionally, a degenerate pole may give rise to multiple flags whose cone contains a reference vector.  The resulting multiplicity coefficients depend on these proportional coefficients and the number of flags. As the instanton number increases, the changing structure of flags within a degenerate pole affects these coefficients. This explains why \eqref{C-anti} depends on Young diagrams through $\alpha(\lambda)$ and $\beta(\lambda)$. 

While our primary focus has been on the relation between $\SU(N+8)_0+1\Sym$  and $\SU(N)_0+1\AS$, similar considerations apply to the relation between $\SO(2N+8)$ and $\Sp(N)_\theta$ pure Yang-Mills theories \cite{Nawata:2021dlk}. In the $\Sp(N)_\theta$ theory, degenerate poles also make non-trivial contributions, leading to the presence of multiplicity coefficients \eqref{C-Sp}. The analysis, albeit tedious, follows the same principles outlined above. To keep the discussion concise and focused, we omit the details.

As observed, this phenomenon manifests not only at the point of ``unfreezing'' but also where degenerate poles appear within JK integrals. Thus, if chemical potentials (or fugacities) admit physical interpretation and can be varied in JK integrals, it would be insightful to conduct a systematic investigation into the values of these chemical potentials and their corresponding JK residues at which new degenerate poles emerge.

%%%%%%%%%%%%%%%%%%%%%%%%%%%%%
%%%%%%%%%%%%%%%%%%%%%%%%%%
\bibliographystyle{JHEP}
\bibliography{bibli.bib}

\end{document}